\documentclass[aps,prb,twocolumn,superscriptaddress,floatfix,letterpaper,nofootinbib]{revtex4-1}

\usepackage[all,arrow,matrix]{xy}

\allowdisplaybreaks

\newcommand{\gSq}{\mathbb{Sq}}
\newcommand{\Sq}{\text{Sq}}
\newcommand{\Bs}{\beta}

\newcommand{\se}[1]{\overset{\scriptscriptstyle #1}{=}}

\newcommand{\hcup}[1]{\underset{{\scriptscriptstyle #1}}{\smile}}
\newcommand{\toZ}[1]{\lfloor #1 \rceil}

\usepackage[normalem]{ulem}

\newcommand{\Fig}[1]{Fig.~\ref{#1}}
\newcommand{\nn}{\nonumber\\}

\newcommand{\ket}[1]{|{#1}\rangle}
\newcommand{\bra}[1]{\langle{#1}|}
\newcommand{\p}{\partial}
\newcommand{\ra}{\rightarrow}
\newcommand{\lra}{\leftrightarrow}

\newcommand{\str}[1]{\overset{\scriptscriptstyle *}{#1}}

\begin{document}

\begin{titlepage}

\title{Lattice models that realize $\Z_n$-1-symmetry protected topological 
states for even $n$}

\author{Lokman Tsui}
\affiliation{Department of Physics, Massachusetts Institute of Technology, Cambridge, MA 02139, USA}

\author{Xiao-Gang Wen}
\affiliation{Department of Physics, Massachusetts Institute of Technology, Cambridge, MA 02139, USA}

\begin{abstract} 

Higher symmetries can emerge at low energies in a topologically ordered state
with no symmetry, when some topological excitations have very high energy
scales while other topological excitations have low energies.  The low energy
properties of topological orders in this limit, with the emergent higher
symmetries, may be described by higher symmetry protected topological order.
This motivates us, as a simplest example, to study a lattice model of
$\Z_n$-1-symmetry protected topological (1-SPT) states in 3+1D for even $n$.
We write down an exactly solvable lattice model and study its boundary
transformation. On the boundary, we show the existence of anyons with
non-trivial self-statistics.  For the $n=2$ case, where the bulk
classification is given by an integer $m$ mod $4$, we show that the boundary
can be gapped with double semion topological order for $m=1$ and toric code
for $m=2$. The bulk ground state wavefunction amplitude is given in terms of
the linking numbers of loops in the dual lattice.  Our construction can be
generalized to arbitrary 1-SPT protected by finite unitary symmetry.

\end{abstract}

\pacs{}

\maketitle

\end{titlepage}

{\small \setcounter{tocdepth}{1} \tableofcontents }

\section{Introduction}

In the last few decades, there has been rapid progress in understanding
``topological phases" of matter, which despite sharing the same symmetry, must
undergo a phase transition to reach one phase from another. Some famous
examples are the topological ordered states with no symmetry
\cite{W8987,W9039} which have degenerate ground states on topological
non-trivial closed manifolds, as well as symmetry protected topological (SPT)
states with symmetry\cite{GW0931,CLW1141,CGL1172,CGW1038}, which does not have
topological order and have a unique gapped ground state in closed manifolds. 

A 3+1D topological order can have point-like and string-like topological
excitations \cite{KW1458,LW170404221,LW180108530}.  For example, a 3+1D
topological order described by $\Z_n$ gauge theory has $\Z_n$ charges (the
point-like topological excitations) and $\Z_n$ flux-lines (the string-like
topological excitations).  If the $\Z_n$ charges have very large energy gap,
then the theory for low energy $\Z_n$ flux-lines will have an emergent higher
symmetry -- a $\Z_n$ 1-symmetry \cite{W181202517}. In other words the low
energy effective Hamiltonian is invariant under the symmetry transformations
that act on all closed 2-dimension subspaces of the 3-dimensional space.  Thus
to understand the topological orders in such a limit, we can study
Hamiltonians with a 1-symmetry. This motivates us to study 1-symmetry in this
paper, such as the lattice Hamiltonian that realize 1-symmetry and the
associated symmetry protected topological order, as well as their boundaries.

We will refer the transformations that act closed 2-dimension subspaces as the
transformation membrane.  If the 3d space have a boundary, the transformation
membrane may intersect with the boundary.  Such an intersection will be called
transformation string.

\subsection{Statement of results}

In this paper, we will study lattice systems with  higher symmetries
\cite{K032,W0303,LW0316,HW0541,NOc0605316,NOc0702377,Y10074601,B11072707,KT13094721,GW14125148,TK151102929,BM170200868,KR180505367,Yoshida2016}.
Like the usual symmetry (0-symmetry) that can have SPT order
\cite{GW0931,CLW1141,CGL1172,CGW1038}, higher symmetry can also have higher
SPT order \cite{KT13094721,TK151102929,W181202517}.  In this paper, we will
concentrate on 3+1D systems with $\Z_n$ 1-symmetry and the associated
associated 3+1D $\Z_n$ 1-SPT states.  Those systems can appear  as low energy
effective theories for 3+1D $\Z_n$ topological order where the $\Z_n$ charges
have a large energy gap.

The 3+1D $\Z_n$ 1-SPT states are known to have a $\Z_{2n}$
classification\cite{ZW180809394,WW181211967}, labeled by $m\in\Z_{2n}$. We study them in the Hamiltonian formalism and write down an exactly solvable bulk Hamiltonian, which has a compact expression when $m$ is even.

The boundary of our system can also have the $\Z_n$ 1-symmetry, but such a
$\Z_n$ 1-symmetry is anomalous \cite{W1313,KT13094721,TK151102929}.  We find
that on the boundary, the transformation strings can carry non-trivial
self-statistics, as a reflection of the anomaly. This predicts the gapped
boundary of the 1-SPT to have emergent anyons. We also find that it is
possible for its surface state to be a gapped topological ordered state. The
topological ordered boundary state has degenerate ground states if the surface
manifold has non-zero genus. These degenerate states exhibit the spontaneous
breaking of 1-symmetry. We also give a geometric interpretation of the ground
state wave function, by writing the wave function amplitude in terms of the
linking numbers of loops in the dual lattice.

\subsection{Notations and conventions}\label{notation}

In some part of this paper, we will use the Lagrangian formalism to describe
quantum lattice systems. This allows us to use extensively the notion of
cochain, cocycle, and coboundary, as well as their higher cup product
$\hcup{k}$ and Steenrod square $\gSq^k$, to construct exactly solvable
Lagrangian that realize topological orders and (higher) SPT orders.  The
reason to use modern mathematical formalisms is that they allow us to see the
features of topological order and (higher) SPT order easily and quickly.  

But the modern mathematical formalisms are not widely used in condensed matter
theory. So we provide a brief introduction in Appendix \ref{cochain}.  Also,
the Lagrangian formalism does not give us a lattice Hamiltonian explicitly.
So in this paper, we present a systematic and direct way to obtain a lattice
Hamiltonian from the those exactly solvable Lagrangian.

We will abbreviate the cup product of cochains $a\smile b$ as $ab$ by dropping
$\smile$.  We will use $\se{n}$ to mean equal up to a multiple of $n$, and use
$\se{\dd }$ to mean equal up to $\dd f$ (\ie up to a coboundary).  We will use
$\<l,m\>$ to denote the greatest common divisor of $l$ and $m$ ($\<0,m\>\equiv
m$).  We will also use $\toZ{x}$ to denote the integer that is closest to $x$.
(If two integers have the same distance to $x$, we will choose the smaller
one, \eg.  $\toZ{\frac12}=0$.)

In this paper, we will deal with $\Z_n$-value quantities. We will denote them
as: 
\begin{align*}
a^{\Z_n}:= a - n \toZ{\frac{a}{n}},
\end{align*}
so the value of $a^{\Z_n}$ has a range from $ -\toZ{\frac {n-1}2}$ to $
\toZ{\frac n2}$. We will sometimes lift a $\Z_n$-value to $\Z$-value, and when
we do so we omit the superscript, \eg. $a^{\Z_n}\ra a^{\Z} = a$, so we can
make sense of expressions like $a^{\Z_n}+a'^{\Z}$, which means
$a^{\Z}+a'^{\Z}$. Since $(a+nu^{\Z})^{\Z_n} = a^{\Z_n}$, whenever we lift a
$\Z_n$-value to $\Z$-value we need to take care whether the final result is
independent of choice of lifting, \ie choice of $u^{\Z}$.

We will also use $D$ to denote spacetime dimensions and $d$ to denote space
dimensions.

\subsection{Overview of paper} 

The structure of the paper and a road map for
reading is presented as follows.

In section \ref{rev}, we review some background information connecting the
cohomology models we studied to the standard many-body theory.  We explained
what are those cohomology models, and some simple examples of those model that
realize simple topological orders and (higher) SPT orders.

In section \ref{overview} we present an intuitive, informal argument for one of our
major results, the self and mutual statistics of boundary transformation
strings, without using the mathematical machinery of cochains and cocycles. 
The formal argument begins from section \ref{model}, where we cite from the literature that the $\Z_n$-1 SPT has
$\Z_{2n}$ classification from cohomology, such that each phase is labeled by
$m\in \Z_{2n}$. We write down the exactly solvable Lagrangian, the expression
for $\om_4$ in \eqref{om4}. We also show that it changes by a boundary term
under gauge transformation via \eqref{sqB1}$\se{d}$\eqref{sqB5}, 
\begin{align*} 
\om_4[\hat{B}^{\Z_n}+\dd a^{\Z_n}]
=\om_4[\hat{B}^{\Z_n}]+\dd \phi_{3}[\hat{B}^{\Z_n},a^{\Z_n}]
\end{align*} 
for some function $\phi_{3}$. This implies $\om_4[\hat{B}^{\Z_n}+\dd
a^{\Z_n}]$ and $\om_4[\hat{B}^{\Z_n}]$ gives the same answer when summed over
a closed manifold, which is expected from gauge invariance \eqref{gaugeinv}. 

In section \ref{exact} we specialize to the case $\hat{B}^{\Z_n}=0$ and give
the explicit form of $\phi_3[a^{\Z_n}]$ in \eqref{phi3}. In Appendix
\ref{apdx:h} and \ref{apdx:gs}, we argue that on a closed spatial manifold
$\cM^3$, $\ee^{2\pi \ii \int_{\cM^3}\phi_3[a^{\Z_n}]}$ is the amplitude of the
ground state wavefunction. We achieve this by examining the time-evolution
operator $\ee^{-T \hat{H}_{\infty}}$ whose matrix elements are given in
\eqref{eth}. We show that it is a projection operator(hence an infinite gap)
and has trace 1(hence a unique ground state). We further argue this transfer
matrix can be decomposed into local commuting projection operators $P_{ij}$
\eqref{eq:Pelem}. We then build our exactly solvable Hamiltonian with a finite
gap by summing over the $-P_{ij}$'s. We then verify that the ground state
wavefunction is indeed given in terms of $\phi_3[a^{\Z_n}]$.

To write down the exactly solvable Hamiltonian, we consider a particular
triangulation of $\cM^3$, given in Appendix \ref{apdx:lat}. We compute the
explicit form for $P_{ij}$ for the even $m$ case in Appendix \ref{apdx:Peven}
and the odd $m$ case in Appendix \ref{apdx:Podd}. Unfortunately, we are unable
to further simplify the expression in the odd $m$ case. The results are
summarized and presented in section \ref{exact}.

In section \ref{gsbd} we consider the case when $\cM^3$ has a boundary. We
introduced the notion of a ``boundary state"\eqref{bdstatedef}, which is
obtained by fixing the degrees of freedom on the boundary and relaxing the
bulk degrees of freedom to their ground state. As a result, the originally
non-anomalous 1-symmetry transformation from the bulk now transform the
boundary states with an additional phase  $\ee^{2\pi \ii \int_{\p
\cM^3}\phi_2[a^{\Z_n},h^{\Z_n}]}$. This phase captures the 't Hooft anomaly of
1-symmetry in the boundary. Any boundary Hamiltonian must be symmetric under
this anomalous 1-symmetry in order to cancel the 't Hooft anomaly. We show that
$\phi_2[a^{\Z_n}]$ is related to the ground state wavefunction
$\phi_3[a^{\Z_n}]$ by \eqref{delphi3}:
\begin{align*}
\phi_3[(a+\dd h)^{\Z_n}]-\phi_3[a^{\Z_n}] = -\dd \phi_2[a^{\Z_n},h^{\Z_n}]
\end{align*}
which states that under the 1-symmetry, the ground state wavefunction changes
by a boundary term. We write down the explicit form of
$\phi_2[a^{\Z_n},h^{\Z_n}]$ in \eqref{phi2alzn}. Using this explicit form, we
are able to compute the self\eqref{selfstat} and mutual\eqref{mutstat}
statistics of the transformation strings. Details of the computation are given
in \ref{apdx:stat}. The boundary transformation strings may be interpreted as
hopping operators for anyons residing on the end of the strings. This predicts
the emergence of such anyon on the boundary theory and is the main result of
the paper.

In section \ref{gapbd} we test our prediction by writing down some gapped
boundary Hamiltonians which obeys the anomalous 1-symmetry. We specialize to
$n=2$ and check the cases $m=2$ and $m=1$. We show that the gapped boundary is
identical to the toric code model (for $m=2$) and the double semion model (for
$m=1$). We verify in both cases that the boundary indeed contains an anyon
with the predicted statistics. Details of the computation for the boundary
Hamiltonian are given in \ref{wi}.

In section \ref{gswfmeaning} we return to examine the ground state
wavefunction. We present the geometric interpretation of the bulk wave
function amplitude as a knot invariant (linking number) of loops dual to $\dd
a^{\Z_n}$.

In section \ref{nonzerobg} we extend our study to the case with a non-zero
background gauge field. In the even $m$ case, we find a line charge with
charge $-m$ is attached to the dual line of the background gauge field.
Details are presented in Appendix \ref{apdx:hnzB} and \ref{apdx:gsdzB}.

In Appendix \ref{apdx:gen} we go deeper into the origin of the connections
between $\om_4$, $\phi_3$, $\phi_2$, and show that they are members of a
series of algebraic objects $\phi_k$ which encodes the same cocycle $\om_4$ at
sub-manifolds of dimension $k$.

In Appendix \ref{apdx:gen2} we present the result of generalizing the
computation of boundary string statistics to other unitary groups.

\section{A brief review of topological order, SPT states, and higher SPT
states}

\label{rev}

A large class of topological orders can be realized by exactly solvable
Lagrangian model.  To write down the Lagrangian model, we first triangulate
the spacetime to obtain a spacetime lattice $\cM^D$, whose vertices are
labeled by $i,j,\cdots$.  The physical degrees of freedom $\cB_{ij}$ live on
the link $ij$, and takes value in a group $G$, \ie $\cB_{ij} \in G$.  In this
paper, we always assume $G$ to be Abelian.  The collection of those values
$\cB_{ij}$ give us a field $\cB$ on spacetime, which, in this case, is also
called a gauge configuration.  A quantum system in Lagrangian formulation is
described by a path integral with an action amplitude.  For our model, the
action amplitude assigns a $U(1)$ phase $\ee^{2\pi \ii S^\text{top}[\cM^D,
\cB]}$ to a gauge configuration $\cB$ on a $D$-dimensional spacetime lattice
$\cM^D$.  The gauge field $\cB$ satisfies the ``flatness condition'' $\dd
\cB=0$ which is enforced by an energy penalty term $U |\dd \cB|^2$ in $U\to
\infty$ limit.  The model is exactly solvable if the $U(1)$ phase is a
topological invariant, meaning it remains unchanged under ``deformations" of
the lattice $\cM^D$ (change of triangulation), and is also invariant under gauge transformations $ \cB
\ra  \cB + \dd a$, \ie (in this paper we will assume the underlying group $G$
is an Abelian finite group.)
\begin{align}
S^\text{top}[\cM^D, \cB+\dd a]\se{1}S^\text{top}[\cM^D, \cB],\ \ \
a\in G
\label{gaugeinv}
\end{align}
Here $\se{1}$ means equal up to 1.  The partition function , after summing all
the degrees of freedom (\ie the $G$ values in all the links), is given by
\begin{align*}
Z(\cM^D)=\sum_{\{\cB\}} \ee^{2\pi \ii S^\text{top}[\cM^D, \cB]}.
\end{align*}
Up to a volume term,\cite{KW1458,WW180109938} the partition function
$Z(\cM^D)$ is a topological invariant of manifold $\cM^D$, that characterize a
topological order.  When $S^\text{top}[\cM^D, \cB] =0$, our model realize a
$G$ topological order described by a $G$-gauge theory.  When
$S^\text{top}[\cM^D, \cB] \neq 0$, our exactly solvable model realizes a
topological order described by a twisted $G$-gauge theory, which is also known
as Dijkgraaf-Witten model\cite{DW9093}.

The action amplitude $\ee^{2\pi \ii S^\text{top}[\cM^D, \cB]}$ of the exactly
solvable model can also be viewed as an SPT invariant\cite{W1447,HW1339,K1459}
that characterizes an SPT order protected by symmetry $G$, if we view $\cB$ as
the background gauge field $\hat B$ that describes the symmetry twist on the
space-time $\cM^D$.  Such a relation is also referred to as ``ungauging" a topological
order, which results in a SPT order.  The SPT invariant $\ee^{2\pi \ii
S^\text{top}[\cM^D, \hat B]}$ characterizes a large class of SPT orders.

To realize the SPT states characterized by the above SPT invariant, we write $
\cB=\hat B+\dd a$, fix a background gauge configuration $\hat B$, and treat
the different gauge transformations $a$ as distinct physical fields. The
partition function, after summing all the degrees of freedom $a$, reproduces
the SPT invariant, up to a space-time volume term:\cite{KW1458,WW180109938}
\begin{align*}
Z[\cM^D,\hat B] =
\int\cD a \ee^{2\pi \ii S^\text{top}[\cM^D,\hat B+\dd a]}
\sim \ee^{2\pi \ii S^\text{top}[\cM^D,\hat B]}
\end{align*}
Note that the action is invariant under the symmetry $a\ra a+ \al$ for $\al$
satisfying $\dd \al=0$. An SPT is trivial if $\ee^{2\pi \ii
S^\text{top}[\cM^D,\hat B]}=1$ for all closed manifolds and background gauge
fields $\hat B$. SPTs also form an Abelian group under stacking. The
topological action $S^\text{top}$ for the stacked SPT is the sum of the
topological actions of its layers. The trivial SPT is the identity element
under stacking and describes a direct product state.

``Group cohomology construction"\cite{CGL1172} is one way to write down
$S^\text{top}[\cM^D,\hat B]$. In this construction, we assume that
$S^\text{top}[\cM^D,\hat B]$ can be written as a sum over all the
$D$-simplices $\Delta^D$: 
\begin{align}
\label{Stop}
S^\text{top}[\cM^D,\hat B]=\int_{\cM^D}\om_D[\hat B]
=\sum_{\Delta^D}\om_D[\hat B]
\end{align}
where $\om_D[\hat B]$ assigns a number to each $D$-simplex. The requirement
that $S^\text{top}[\cM^D,\hat B]$ is invariant under triangulation leads to
the following constraint on $\om_D[\hat B]$, known as the ``cocycle
condition":
\begin{align*}
\dd \om_D[\hat B] \se{1} 0,
\end{align*}
whose solutions are called cocycles. (The left hand side is evaluated on a
$D+1$-simplex and $\dd$ is called the coboundary operator analogous to the
exterior derivative for differential forms. See Appendix \ref{cochain} for
further details.) Distinct solutions of the cocycle condition do not
necessarily correspond to distinct topological phases, since two solutions
$\om_D$, $\om_D'$ may give the same $S^\text{top}$ on closed manifolds if
$\om_D=\om_D'+\dd \bt_{D-1}$ for some function $\bt_{D-1}$. Defining an
equivalence relation $\om_D\sim\om_D'+\dd \bt_{D-1}$ on cocycles and solving
for the equivalence classes of cocycles, the resulting algebraic object is
known as a cohomology group, which also provide a way to classify SPTs.

In the traditional SPT, the gauge field $\hat B$ assigns a group element of
$G$ to every 1-dimensional simplex(\ie links), and are thus called 1-cochain.
(A $G$-valued $m$-cochain is an assignment of a group element of $G$ to each
$m$-simplex.) Gauge transformations are parameterized by a 0-cochain $a$ which
assigns a group element to every 0-dimensional simplex(\ie vertices). Symmetry
is parameterized by 0-cochain $\al$. The condition $\dd \al=0$ implies $\al$
is a constant function on every connected component. Physically this
corresponds to a global symmetry acting on a connected component of the
spatial slice. An example is the $\Z_2$-protected SPT in
$D=2+1$.\cite{CLW1141} The $\Z_2$ symmetric ground state wavefunction can be
constructed as the superposition of domain walls in the $\Z_2$ symmetry
breaking state, with $(-1)^{\text{no. of domain walls}}$ as its
amplitude\cite{Levin2012}. 

With the above description of usual SPT states, we can now describe higher
SPTs.  Higher SPT states, or ``$k$-symmetry protected topological states"
($k$-SPT)\cite{KT13094721,TK151102929,W181202517}, is a generalization of
traditional SPTs. They have symmetry acting on closed sub-lattices of
codimension
$k$.\cite{K032,W0303,LW0316,HW0541,NOc0605316,NOc0702377,Y10074601,B11072707,KT13094721,GW14125148,TK151102929,BM170200868,KR180505367,Yoshida2016}.
The 1-cochain (\ie the vector field) $\cB$ is promoted to $(k+1)$-cochain.
The gauge transformation is now described by a  $k$-cochain $a$:
\begin{align}
 \cB \to \cB +\dd a.
\end{align}

The path integral on spacetime lattice $\cM^D$ that realize
a higher SPT state is given by
\begin{align}
\label{ZMB}
 Z[\cM^D,\hat B] =
\sum_{\{a\}} \ee^{2\pi \ii \int_{\cM^D} \om_D(\hat B+\dd a)}
\end{align} 
where the dynamical field $a$ is now a $k$-cochain (a field which takes values
on the $k$ simplices), and $S^\text{top}[\cM^D,\cB]$ is given by \eqn{Stop}.
In such a lattice model, the higher symmetry is generated by a $k$-cocycle
$\al$:
\begin{align}
 a \to a +\al.
\end{align}
We see that the symmetry acts on $k$-simplices where $\al\neq 0$. Such
$k$-simplices are dual to a $(D-k)$-dimensional manifold $\tilde{\al}$ on the
dual lattice. The condition $\dd \al=0$ implies $\tilde{\al}$ has no boundary
within the space-time manifold. $\tilde{\al}$ may have a non-empty boundary if
it intersects the boundary of the space-time manifold $\p \cM^D$. 

When $\om_D=0$, \eqn{ZMB} describes a state with trivial $k$-SPT order. When
$\om_D$ is a non-trivial cocycle, \eqn{ZMB} realizes a state with a
non-trivial $k$-SPT order.  The traditional SPT corresponds to $k=0$ case.

The above Lagrangian is a realization of higher SPT states.  In this paper, we
show how to convert the above Lagrangian realization into a Hamiltonian
realization.  In the Hamiltonian formalism, a $k$-symmetry operator acts on
codimension $k$ sub-lattices in the spatial manifold. For example in a 3 space
dimensions, a 1-symmetry operator acts on closed membranes. These membranes
may intersect the boundary as strings. We show in Section \ref{gsbd} that the
$1$-symmetry membrane operators in the bulk corresponds to $1$-symmetry string
operators on the boundary.

The hallmark of non-trivial SPT is that its boundary cannot be gapped with a
unique ground state on all manifolds. If it were the case, we could start from
a trivial SPT, nucleate a small bubble of the non-trivial SPT, and expand the
bubble to fill up the entire space. This would have provided a path connecting
the trivial and the non-trivial SPTs without closing the energy gap, leading
to a contradiction. Generically the boundary of non-trivial SPT is gapless,
breaks symmetry spontaneously, or has topological order. The inability for the
boundary to achieve a uniquely gapped state on all manifolds is encoded by the
't Hooft anomaly of the $k$-symmetry on the boundary. Therefore studying such
anomaly is a way to probe the non-trivial nature of the topological bulk.

\section{Intuitive argument for boundary transformation string
statistics}\label{overview} In this section we present an informal argument
for the self and mutual statistics of the boundary strings.

The 't Hooft anomaly of the boundary transformation of SPTs may be interpreted
via symmetry fractionalization\cite{Barkeshli_2019,Chen_2017}: when the
symmetry acts on the entire boundary manifold, and hence can be extended into
the bulk, the group representation structure is preserved. But when we attempt
to examine the symmetry acting only on a local patch of the boundary manifold,
various group representation structures may be spoiled.

Take for example the non-trivial $G$-protected 1d 0-SPT\cite{CGW100837451},
for which the AKLT\cite{AKLT} chain with $G=SO(3)$ is a well-known instance.
The boundary of a 1d segment are its two endpoints, indexed by $L$ and $R$
respectively. When the bulk is gapped, the low energy effective theory are
described in terms of its boundary degrees of freedom, and the Hilbert space
may be expressed as a tensor product $\cH_L \otimes \cH_R$ of the local
Hilbert spaces $\cH_L$ and $\cH_R$ for the two ends. For two group elements
$g,h$ acting on the tensor product space, we have
\begin{align*}
\cR_{L+R}(g)\cR_{L+R}(h)=\cR_{L+R}(gh)
\end{align*}
which says $\cR_{L+R}(g)=\cR_{L}(g)\otimes\cR_{R}(g)$ is a linear
representation of $G$. This is because when the same $g\in G$ acts on both
boundaries, it may be extended into a symmetry acting globally in the bulk,
where the group is represented linearly.  When localizing on the left end,
$\cR_{L}$ becomes a projective representations of $G$:
\begin{align*}
\cR_{L}(g)\cR_{L}(h)=\om(g,h)\cR_{L}(gh).
\end{align*}
The 't Hooft anomaly is expressed as the non-trivial phase $\om(g,h)$, which
spoils the linearity of the representation.

In the same spirit, for our case with $\Z_n$-1-SPT, we may expect 't Hooft
anomaly to appear as the spoiling of some group representation structure when
localizing to a part of the boundary. If the boundary symmetry can be extended
into the bulk, the group representation structure is expected to be preserved.

Consider the case where we have two 1-symmetries, $W_1$ and $W_2$, associated
with group elements $q_1^{\Z_n}$ and $q_2^{\Z_n}$ respectively. They act on
two contractible loops, as shown below:
\begin{eqnarray*}
\includegraphics{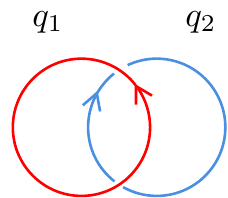}
\end{eqnarray*}

Each loop can be extended into the bulk as a 1-symmetry acting on a hemisphere. In the bulk, the 1-symmetries commute. We therefore expect that on the boundary, the two loop operators also commute. This is represented by the diagrammatic equation:
\begin{eqnarray*}
\includegraphics{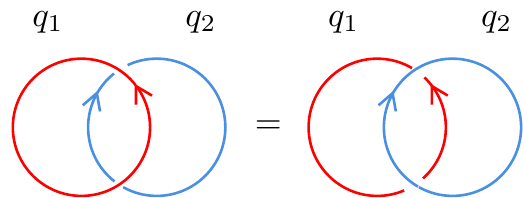}
\end{eqnarray*}

There are two intersections of the loops. Motivated by the symmetry fractionalization picture, we may guess that when localizing to one of the intersections, the commutativity is spoiled by a $U(1)$ phase $\ee^{2\pi \ii \th_{q_1q_2}}$:
\begin{eqnarray}
\includegraphics{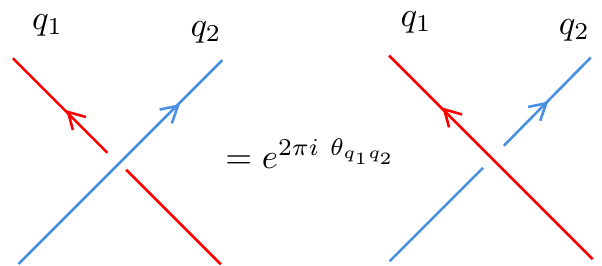}
\label{q1q2crossing}
\end{eqnarray}

If we further assume that two parallel lines associated with group elements $p_1^{\Z_n}$ and $q_1^{\Z_n}$ could stack into a single line $(p_1+q_1)^{\Z_n}$ without incurring any phase, we can deduce that $\th_{q_1q_2}$ is a linear function of $q_1$ by the following manipulations:
\begin{eqnarray*}
\includegraphics{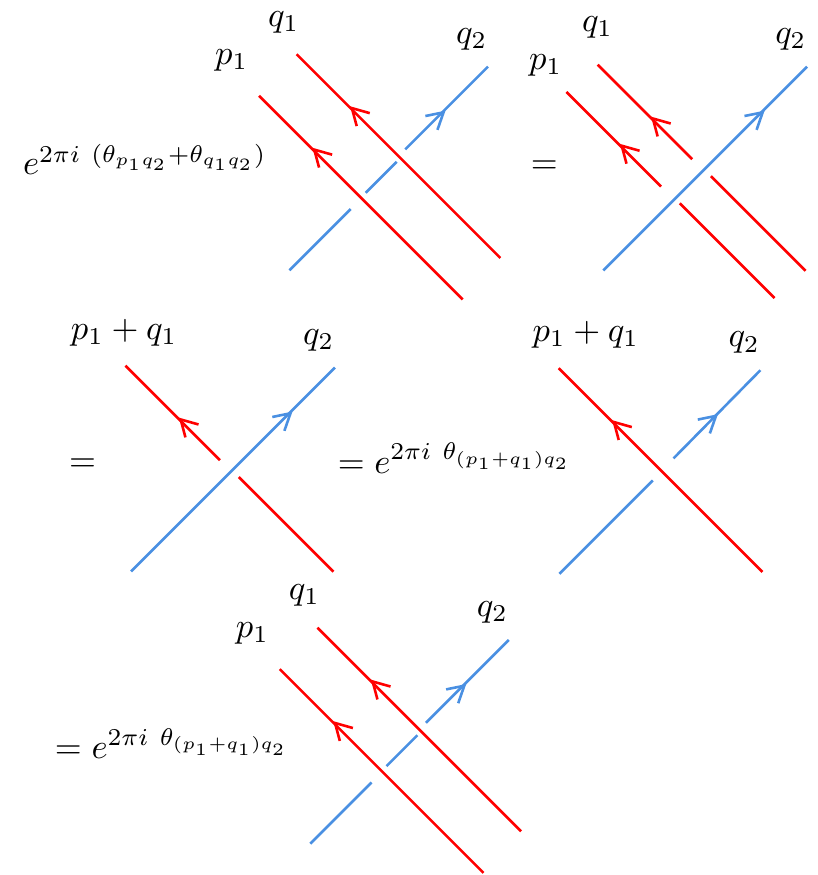}
\end{eqnarray*}
A similar argument shows $\th_{q_1q_2}$ is linear in $q_2$. We conclude
\begin{align*}
\th_{q_1q_2}\propto q_1q_2
\end{align*}
Since $q_1^{\Z_n}$, $q_2^{\Z_n}$ are defined up to multiples of $n$, we expect $\ee^{2\pi \ii \th_{q_1q_2}}$ to be invariant under $q_1\ra q_1+n$. Thus the coefficient should be a fraction $\frac{m}{n}$ for some integer $m$.
\begin{align}
\th_{q_1q_2}= \frac{m}{n} q_1q_2 \label{thq1q2guess}
\end{align}

When $q_1=q_2=q$, we may also entertain the possibility that at an intersection, the transformation string may ``change track" and incur a $U(1)$ phase $\ee^{2\pi \ii \th_q}$ or $\ee^{2\pi \ii \bar{\th}_q}$, depending on the orientation of the crossing:
\begin{eqnarray*}
\includegraphics{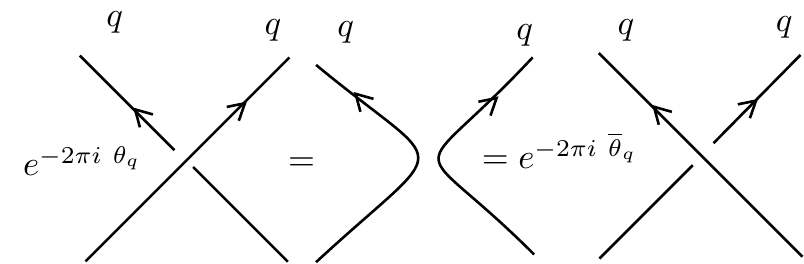}
\end{eqnarray*}
Comparing to \eqref{q1q2crossing} with $q_1=q_2=q$, we observe that
\begin{align}
\th_{q}-\bar{\th}_{q}\se{1}\th_{qq}.\label{thqq}
\end{align}

On the other hand, we also have the equality of these two diagrams:
\begin{eqnarray*}
\includegraphics{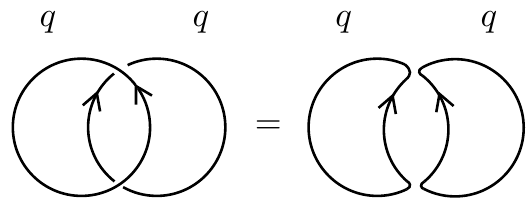}
\end{eqnarray*}
The reason is as follows, imagine when the left hand side extends into the bulk as two hemispheres which intersect on a line. On the intersection line, the 1-symmetry acts trivially and may be removed by reconnecting the membranes near the line. Thus we may reconnect the two intersecting hemispheres into two non-intersecting membranes which terminates on the surface as shown on the right hand side. This implies
\begin{align*}
\th_q+\bar{\th}_q\se{1}0.
\end{align*}
Adding this equation to \eqref{thqq}, we get $2\th_{q}\se{1}\th_{qq}$. Thus
\begin{align*}
\th_{q}\se{1/2}\frac{\th_{qq}}{2}\se{1/2}\frac{m}{2n} q^2.
\end{align*}
so
\begin{align*}
\th_{q}\se{1}\frac{m}{2n} q^2+\frac{1}{2}f(q,m,n)
\end{align*}
For some integer-valued function $f$. An argument similar to that for the linearity of $\th_{q_1q_2}$ implies $\th_q$ is proportional to $q^2$. So $f(q,m,n)\propto q^2$ and the coefficient of $\frac{q^{2}}{2n}$ in $\th_{q}$ is an integer. Let's redefine $m$ to be this integer coefficient. Thus we have
\begin{align*}
\th_{q}=\frac{m}{2n} q^2
\end{align*}
Upon $q\ra q+n$, the above equation transform as
\begin{align*}
\th_{q}&=\frac{m}{2n} q^2 \ra \frac{m}{2n} (q^2+2qn+n^2) \nn
&\se{1}\th_{q} + \frac{nm}{2}
\end{align*}
so in order for $\th_q$ to be invariant mod 1, $nm$ must be even. In the case $n$ is even, $m$ can be chosen from $0,1,\dots,2n-1$. With $n$ odd, $m$ must be an even number chosen from $0,2,4,\dots,2n-2$. Each choice gives a distinct set of $\th_{q}$ and $\th_{q_1q_2}$.

Assuming the set of $\th_{q}$ and $\th_{q_1q_2}$ is bijective to the set of 't Hooft anomalies, which is bijective to the set of 1-SPT phases in the bulk, we would expect the bulk $\Z_n$-1-SPT to have $\Z_{2n}$ classification for even $n$, and $\Z_{n}$ classification for odd $n$. This agrees with the classification results in Ref. \cite{ZW180809394} derived from cohomology and Ref. \cite{,WW181211967} from cobordism group.

We stress that the odd $n$ case and the even $n$ case differs only in that the odd $m$'s are forbidden for odd $n$. In fact, all the results in our paper for even $m$ also applies to odd $n$ case.

In the rest of the paper, we will re-derive the expressions for $\th_{q_1q_2}$ and $\th_{q}$ formally, using the language of group cohomology.

\section{A 3+1D model to realize a $\Z_n$-1-SPT phase for even $n$}\label{model}

\begin{figure}[t]
\begin{center}
\includegraphics[scale=0.8]{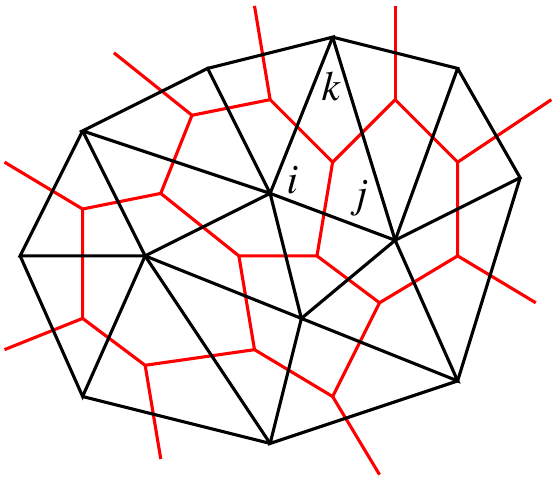}
\end{center}
\caption{(Color online) 
The black lines describe a 2-dimensional spacetime complex $\cM^2$.
The red lines describe the dual complex $\t \cM^2$.
}
\label{TriDTri}
\end{figure}

To construct lattice models with higher symmetries, it is
convenient to do so in the spacetime Lagrangian formalism.  We construct a
spacetime lattice by first triangulating a $D$-dimensional spacetime manifold
$M^D$. So a spacetime lattice is a $D$-complex $\cM^D$
with vertices labeled by $i$, links labeled by $ij$, triangles labeled by
$ijk$, \etc (see Fig. \ref{TriDTri}).  The $D$-complex $\cM^D$ also has a dual
complex denoted as $\t \cM^D$.  The  vertices of $\cM^D$ correspond to the
$D$-cells in $\t \cM^D$, The links of $\cM^D$ correspond to the $(D-1)$-cells
in $\t \cM^D$, \etc

Our spacetime lattice model may have a field living on the vertices, $g_i$.
Such a field is called a 0-cochain.  The model may also have a field living on
the links, $a_{ij}$. Such a field is called a 1-cochain, \etc.  To construct
spacetime lattice models, in particular, the topological spacetime lattice
models,\cite{KT13094721,TK151102929,W161201418,LW180901112} we will use
extensively the mathematical formalism of cochains, coboundaries, and cocycles
(see Appendix \ref{cochain}).

\subsection{The bulk exactly solvable Lagrangian} \label{SecZn1SPT}

We consider a 3+1D bosonic model on a spacetime complex
$\cM^4$, with $\Z_n$-valued dynamic field $ a^{\Z_n}_{ij}$ on the links $ij$ of
the complex $\cM^4$.  
Here $n$ is even.
We also have a $\Z_n$-valued non-dynamical background
field $\hat B^{\Z_n}_{ijk}$  on the triangles $ijk$ of the complex $\cM^4$. $\hat B^{\Z_n}$ is a $\Z_n$-valued 2-cocycle
\begin{align}
\dd  \hat B^{\Z_n}\se{n}0. \label{dB}
\end{align}
The path integral of our bosonic model is given by
\begin{align}
&Z = \sum_{\{a^{\Z_n}\} } 
\ee^{2\pi \ii \int_{\cM^4} \omega_4[\cB] } \label{Zn1SPT} \\
&\omega_4[\cB]:=\frac{m}{2n} \gSq^2\cB^{\Z_n},\label{om4}\\
&\cB := \hat B+\dd a \label{cB} \\
&\cB^{\Z_n} = \hat B+\dd a - n \toZ{\frac{\hat B+\dd a}{n}}\nonumber
\end{align}
where $m,n=$ integers, $\sum_{\{a^{\Z_n}\} }$ sums over $\Z_n$-valued
1-cochains $a^{\Z_n}$.
We have lifted the $\Z_n$-valued quantities $\hat B^{\Z_n}$ and $a^{\Z_n}$ to $\Z$-valued quantities  $\hat B$ and $a$. Also $\gSq^2$ is the generalized Steenrod square defined by \eqn{Sqdef}.
We will show that the above model realizes a $\Z_n$-1-SPT phase.

Since $\omega_4[\cB]=\omega_4[\cB^{\Z_n}]$ and $\cB^{\Z_n}$ is invariant under the transformation
\begin{align}
\label{gaugeZn}
 \hat B \to \hat B + n b^{\Z},\ \ \ \
 a \to a + n u^{\Z},
\end{align}
where $b^{\Z}$ and $u^{\Z}$ are any $\Z$ valued 2-cochain and 1-cochain, the action amplitude in \eqn{Zn1SPT} is invariant, even when $\cM^4$ has a boundary.
The above result also implies
that the model has a $\Z_n$-1-symmetry generated by
\begin{align}
\label{Zn1symm}
 a \to a +  \al^{\Z_n}, \ \ \ \dd  \al^{\Z_n} \se{n} 0,
\end{align}
even when $\cM^4$ has a boundary.  

Also it can be checked that $\ee^{2\pi\ii \om_4}$ is a $U(1)$-valued cocycle: Using \eqref{Sqdef}, \eqref{cupkrel} and $\dd \cB^{\Z_n}\se{n}0$ which follows from \eqref{dB}, and remembering that $n$ is even, we have
\begin{align*}
&\dd \omega_4[\cB^{\Z_n}]\nn
&=\frac{m}{2n} \dd \gSq^2 \cB^{\Z_n}\nn
&=\frac{m}{2n} \dd\left(\cB^{\Z_n}\cB^{\Z_n}+\cB^{\Z_n}\hcup{1}\dd \cB^{\Z_n}\right)\nn
&=\frac{m}{2n} \big(\dd \cB^{\Z_n}\cB^{\Z_n}+ \cB^{\Z_n}\dd \cB^{\Z_n} \nn
&\ \ \ + \dd \cB^{\Z_n} \hcup{1}\dd \cB^{\Z_n} + \cB^{\Z_n}\dd \cB^{\Z_n} -\dd \cB^{\Z_n} \cB^{\Z_n}\big)\nn
&=\frac{m}{2n} \left(2\cB^{\Z_n}\dd \cB^{\Z_n} + \dd \cB^{\Z_n} \hcup{1}\dd \cB^{\Z_n}\right)\se{1} 0.
\end{align*}

In \eqn{cB}, $\hat B^{\Z_n}$ is the $\Z_n$ background 2-connection to
describe the twist of the $\Z_n$-1-symmetry.  The model has a $\Z_n$ gauge
symmetry:
\begin{align}
\label{Zngauge}
 a \to a +  \hat{a}^{\Z_n},  \ \ \ \ \
 \hat B \to \hat B -  \dd\hat{a}^{\Z_n}. 
\end{align}
Also, using $\dd \cB\se{n}0$, \eqref{Sqdef} and \eqref{cupkrel},
\begin{align}
&\frac{m}{2n} \gSq^2\cB^{\Z_n} \label{sqB1} \\
&= \frac{m}{2n} \gSq^2(\cB - n \toZ{\frac{\cB}{n}})\nn
&=\frac{m}{2n}\big[\big(\cB- n \toZ{\frac{\cB}{n}}\big)\big(\cB- n \toZ{\frac{\cB}{n}}\big)\nn
& \ \ \ +\big(\cB- n \toZ{\frac{\cB}{n}}\big)\hcup{1}\dd \big(\cB- n \toZ{\frac{\cB}{n}}\big) \big]\nn
&\se{1} \frac{m}{2n} \gSq^2\cB + \frac{m}{2}\big(\toZ{\frac{\cB}{n}}\cB + \cB \toZ{\frac{\cB}{n}}+\cB\hcup{1}\dd \toZ{\frac{\cB}{n}}\big)\nn
&\se{1} \frac{m}{2n} \gSq^2\cB + \frac{m}{2}\dd\big( \cB\hcup{1} \toZ{\frac{\cB}{n}}\big)\label{sqB2} \\
&\se{\dd} \frac{m}{2n} \gSq^2\cB
=\frac{m}{2n} \gSq^2(\hat{B}+\dd a)\label{sqB3} \\
&=\frac{m}{2n}\big[\big(\hat{B}+\dd a\big)\big(\hat{B}+\dd a\big)
+ \big(\hat{B}+\dd a\big)\hcup{1}\dd\big(\hat{B}+\dd a\big)\big]\nn
&\se{1}\frac{m}{2n}\big[\gSq^2\hat{B}+\hat{B}\dd a + \dd a \hat{B} +\dd a \dd a + \dd a \hcup{1} \dd \hat{B}\big] \nn
&\se{1}\frac{m}{2n}\big[\gSq^2\hat{B} + \dd \big( a \dd a +\dd a \hcup{1} \hat{B} + 2 a \hat{B} \big)\big]\label{sqB4} \\
&\se{\dd}\frac{m}{2n}\gSq^2\hat{B}\se{\dd}\frac{m}{2n}\gSq^2\hat{B}^{\Z_n}.\label{sqB5}
\end{align}
In the last step we reused $\eqref{sqB1}\se{1,d}\eqref{sqB3}$ with $\cB$ replaced by $\hat{B}$. Therefore
\begin{align}
\ee^{2\pi \ii \int_{\cM^4} \frac{m}{2n} \gSq^2 (\hat B+\dd a)^{\Z_n} }
=\ee^{2\pi \ii \int_{\cM^4} \frac{m}{2n} \gSq^2 \hat B^{\Z_n} }\label{gaugeinv2}
\end{align}
for closed spacetime $\cM^4$. This is expected from gauge invariance \eqref{gaugeinv}. The model is exactly solvable and gapped for
closed spacetime $\cM^4$.

Eqn. \eq{Zn1SPT} has no topological order
since on closed spacetime and for $\hat B^{\Z_n}=0$
\begin{align}
 Z(\cM^4) &= \sum_{\{a^{\Z_n}\} } \ee^{2\pi \ii \int_{\cM^4} \frac{m}{2n}
\gSq^2 (\dd a)^{\Z_n} } = \sum_{\{a^{\Z_n}\} } 1 =
n^{N_l},\label{Zclosedeq1}
\end{align}
where $N_l$ is the number of links in the spacetime complex $\cM^4$.
$n^{N_l}$ is the so called the volume term that is linear in the spacetime
volume.  The topological partition function $Z^\text{top}$ is given by removing the volume term:\cite{KW1458,WW180109938}
\begin{align}
Z^\text{top}(\cM^4)
 =Z(\cM^4)/ n^{N_l},
\end{align}
which is equal to $1$ for all closed
4-complex $\cM^4$.  Thus the above model has no topological order.  After we
turn on the flat $\Z_n$-2-connection $\hat B^{\Z_n}$, the topological partition
function of the model \eq{Zn1SPT} becomes
\begin{align}
&Z^\text{top}(\cM^4, \hat B)= \ee^{2\pi \ii \int_{\cM^4} \frac{m}{2n}
[\hat B^{\Z_n}\hat B^{\Z_n} + \hat B^{\Z_n} \hcup{1} \dd \hat B^{\Z_n}]},\label{Ztop}\\
&\ \ \ 
\dd \hat B^{\Z_n} \se{n}0. \nonumber
\end{align}
In \Ref{ZW180809394}, it was shown that $H^4(\cB(\Z_n,2);\R/\Z)= \Z_{2n}$ for
$n=\text{even}$. Furthermore, the classification of higher-SPTs based on a generalized cobordism theory approach also obtains a $\Z_{2n}$ for $n$ is even. See Table 7 of \Ref{WW181211967}. Thus the above 1-SPT invariant is non-trivial.  There are
$2n$ distinct $\Z_n$-1-SPT phases labeled by $m=0,\cdots,2n-1$. 

\section{Exactly solvable Hamiltonian}\label{exact}
In this section we derive the exactly solvable $\Z_n$-1-SPT Hamiltonian. For simplicity we focus on the untwisted theory and set the non-dynamical background 2-connection $\hat{B}^{\Z_n}=0$ so $Z^\text{top}$ depends on $a^{\Z_n}$ only. In section \ref{nonzerobg} we will examine the case with non-zero $\hat{B}^{\Z_n}$.

The action \eqref{Zn1SPT} is:
\begin{align}
Z^\text{top} =\frac{1}{n^{N_l}} \sum_{\{a^{\Z_n}\}} \ee^{2\pi \ii \int_{\cM^4} \om_4[\dd a^{\Z_n}]}, \label{Ztop1}
\end{align}
using $\eqref{sqB1}\se{1}\eqref{sqB2}$ and $\eqref{sqB3}\se{1}\eqref{sqB4}$ with $\hat{B}=0$, 
\begin{align}
\om_4[\dd a^{\Z_n}] &= \frac{m}{2n} \gSq^2(\dd a)^{\Z_n} \nn
&\se{1} \frac{m}{2n} \gSq^2\dd a^{\Z_n} + \dd\big(\frac{m}{2} \dd a^{\Z_n} \hcup{1} \toZ{\frac{\dd a^{\Z_n}}{n}}\big)\nn
&=\frac{m}{2n} \gSq^2\dd a^{\Z_n} + \dd\xi_3[a^{\Z_n}]\nn
&= \dd\big(\frac{m}{2n}a^{\Z_n}\dd a^{\Z_n} + \xi_3[a^{\Z_n}]\big)\nn
&=\dd \phi_3 [a],\label{om4da}
\end{align}
(note that $\dd a^{\Z_n}=\dd (a^{\Z_n})\neq (\dd a)^{\Z_n}$) where
\begin{align}
\xi_3[a]:&=\frac{m}{2} \dd a \hcup{1} \toZ{\frac{\dd a}{n}}\label{xi3}\\
\phi_3 [a] :&= \frac{m}{2n}a^{\Z_n} \dd a^{\Z_n} + \xi_3[a^{\Z_n}] \nn
&\se{1} \frac{m}{2n}a \dd a + \xi_3[a] + \dd \xi_2[a]\label{phi3}\\
\xi_2[a] :&= \frac{m}{2}\left(a\toZ{\frac{a}{n}}+\dd a\hcup{1}\toZ{\frac{a}{n}}\right).\label{xi2}
\end{align}
\eqref{phi3} and \eqref{xi2} are obtained from the previous line by writing out $a^{\Z_n}\ra a-n\toZ{\frac{a}{n}}$. By construction we have $\phi_3[a]=\phi_3[a + n u^{\Z}] = \phi_3[a^{\Z_n}]$ for any $\Z$-valued 1-cochain $u^{\Z}$. However, $\xi_3$ and $\xi_2$ do not enjoy this property.
(See Appendix \ref{apdx:gen} for relationship between $\om_4$ and $\phi_3$ in general.)

We will analyze the cases for even and odd $m$ separately. For each case we write down the Hamiltonian \begin{align*}
H=-\sum_{ij} P_{ij},
\end{align*}
which is the sum over links $ij$ of projections $P_{ij}$, as described in Appendix \ref{apdx:h}. We can compute $P_{ij}$ by assuming a hypercubic lattice for the space-time $\cM^4=\R^4$ triangulated as in Appendix \ref{apdx:lat}. The Hilbert space is spanned by $\ket{\{a^{\Z_n}_{ij}\}}$ for links $ij$ in the 3D cubic lattice.

\subsection{Even $m$ case}\label{evenm}

When $m$ is even, \eqref{om4da} and \eqref{phi3} are simplified considerably. The result is
\begin{align}
\om_4[\dd a^{\Z_n}] &\se{1} \frac{m}{2n}\gSq^2(\dd a^{\Z_n})\label{om4even}\\
&\se{1}\dd \phi_3[a]\nn
\phi_3[a]&\se{1}\frac{m}{2n} a \dd a.
\end{align}
We will also triangulate $\R^3$ as described in Appendix \ref{apdx:lat}. The variables in our lattice model lives on links. There are three types of links: 1-diagonal, 2-diagonal or 3-diagonal. A link $ij$ is defined to be $k$-diagonal if the displacement vector from $i$ to $j$ differs by $k$ distinct unit vectors $\in \{\hat{x}_1,\hat{x}_2,\hat{x}_3\}$. In the even $m$ case, as shown in Appendix \ref{apdx:Peven}, the 2-diagonal and 3-diagonal links form product states and can be ignored.

\begin{figure}
\centering
\includegraphics[width=1\linewidth]{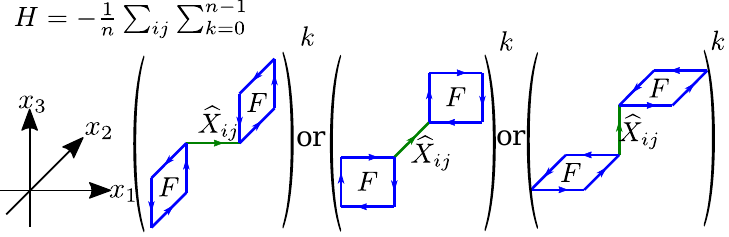}
\caption{(Color online) The lattice Hamiltonian for even $m$ 1-symmetric SPT consists of commuting projections $P_{ij}$ summed over all links in the cubic lattice. The projection consists of an operator $\widehat{X}_{ij}$(depicted in green) which increments the link value $a^{\Z_n}\ra (a+1)^{\Z_n}$, and the operators $\ee^{2\pi \ii \frac{m}{2n} F}$ that multiplies the state by a phase proportional to the fluxes $F=\sum_{\square}\dd a^{\Z_n}$ through the two squares(depicted in blue) touching $ij$.
}
\label{fig:latH}
\end{figure}

For the 1-diagonal links, the topological action
\begin{align}
Z^\text{top} =\frac{1}{n^{N_l}} \sum_{\{a^{\Z_n}\}} \ee^{2\pi \ii \int_{\cM^4} \frac{m}{2n} \gSq^2(\dd a)^{\Z_n} }\label{eq:Ztopeven}
\end{align}
leads to mutually commuting projections \eqref{eq:Peven}
\begin{align}
P_{ij} =  \frac{1}{n} \sum_{k=0}^{n} \widehat{X}_{ij}^{k} \ee^{2\pi \ii \frac{m k}{2 n} \frac{\eps^{\al \bt \ga}}2 [F_{\bt \ga}(\vec{r}_{ij}+\vec{\frac 12})+F_{\bt \ga}(\vec{r}_{ij}-\vec{\frac 12})]},\label{eq:Peven2}
\end{align}
where the sum is carried over 1-diagonal links $ij=(\vec{n},\vec{n}+\hat{\al})$, $\widehat{X}_{ij}\ket{a^{\Z_n}_{ij}} = \ket{(a_{ij}+1)^{\Z_n}}$, $\vec{r}_{ij} = \vec{n}+\frac{\hat{\al}}{2}$ is the mid-point of the link. $\vec{\frac 12}=(\frac 12,\frac 12,\frac 12)$ and $F_{\bt \ga}(\vec{r})$ reads off the ``flux" through the square centered at $\vec{r}$, or more specifically
\begin{align}
F_{\bt\ga}(\vec{r})&:= \<\dd a^{\Z_n},(\vec{0},\hat{\bt},\hat{\bt}+\hat{\ga})_{\vec{r}-\frac{\hat{\bt}}{2}-\frac{\hat{\ga}}{2}}\> - (\bt\lra\ga),\label{F}
\end{align}
where $(\vec{a},\vec{b},\dots,\vec{c})_{\vec{n}}$ is a shorthand for $(\vec{a}+\vec{n},\vec{b}+\vec{n},\dots,\vec{c}+\vec{n})$. The Hamiltonian is illustrated in \Fig{fig:latH}

\subsection{General $m$ case}



In appendix \ref{apdx:Podd} we show for general $m$ the corresponding projections are given by \eqref{eq:Podd}:
\begin{align*}
P_{ij} = & \frac{1}{n} \sum_{k=0}^{n}  \widehat{X}_{ij}^{k} \ee^{2\pi \ii \int_{\R^3}\del_k \phi_3[a^{\Z_n}]}.
\end{align*}

Here $\del_k \phi_3[a^{\Z_n}]$ is the change in $\phi_3[a^{\Z_n}]$ when a single link $ij$ changes as $a_{ij}^{\Z_n}\ra(a_{ij}+k)^{\Z_n}$. Under our triangulation, it is evaluated for 1-, 2-, 3- diagonal links in Appendix \ref{apdx:Podd}.

\section{Ground state wavefunctions and boundary transformations}\label{gsbd}

By Appendix \ref{apdx:gs}, the ground state wavefunction in closed space 3-manifold $\cM^3$ is given by
\begin{align}
\ket{\psi_0} = \sum_{\{a^{\Z_n}\}} \ee^{2 \pi i \int_{\cM^3} \phi_3[a^{\Z_n}]} \ket{\{a^{\Z_n}\}}. \label{gswf}
\end{align}

For physical interpretation of these wavefunctions, see Section \ref{gswfmeaning}.

\subsection{Boundary States and their 1-symmetry transformations}
Suppose we are interested in space 3-manifold which has a boundary. We may write down a ``boundary state" by separating $\{a^{\Z_n}\}=\{a^{\Z_n}_{bulk},a^{\Z_n}_{\p}\}$ into boundary and bulk links, fixing the values of $a_{\p}^{\Z_n}$ at the boundary in \eqref{gswf} and only sum over links $a_{bulk}^{\Z_n}$ inside the bulk.
\begin{align}
\ket{\{a^{\Z_n}_{\p}\}}_{\p}:=\sum_{\{a^{\Z_n}_{bulk}\}} \ee^{2 \pi i \int_{\cM^3} \phi_3[a^{\Z_n}_{bulk},a^{\Z_n}_{\p}]} \ket{\{a^{\Z_n}_{bulk},a^{\Z_n}_{\p}\}}.\label{bdstatedef}
\end{align}

Consider a 1-symmetry transformation
\begin{align}
\ket{a^{\Z_n}} \ra \ket{a'^{\Z_n}} = \ket{(a + \al)^{\Z_n}}, \label{transform}
\end{align}
where $\al^{\Z_n}$ is a $\Z_n$-valued 1-cochain. We have
\begin{align}
&\ket{\{a^{\Z_n}_{\p}\}}_{\p} \nn
&\ra \sum_{\{a^{\Z_n}_{bulk}\}} \ee^{2 \pi i \int_{\cM^3} \phi_3[a^{\Z_n}]} \ket{\{(a+ \al)^{\Z_n}\}}\nn
&= \sum_{\{a'^{\Z_n}_{bulk}\}} \ee^{2 \pi i \int_{\cM^3} \phi_3[a'^{\Z_n}] - \delta_{\al} \phi_3[a^{\Z_n}]} \ket{\{a'^{\Z_n}\}} \label{bdstate0}
\end{align}
with
\begin{align*}
\del_{\al} \Phi[a] &:= \Phi[a+\al] - \Phi[a].
\end{align*}
for any function $\Phi[a]$.

For 1-symmetry, we have $\dd \al^{\Z_n} \se{n} 0$, then
\begin{align}
-\del_{\al} \phi_3[a^{\Z_n}] &\se{1} -\frac{m}{2n}  \al^{\Z_n} \dd a^{\Z_n} + \frac{m}{2}\big(a^{\Z_n} \frac{\dd \al^{\Z_n}}{n} + \al^{\Z_n} \frac{\dd \al^{\Z_n}}{n} \big) \nn
&\ \ \ + \frac{m}{2} \dd a^{\Z_n}\hcup{1}  \frac{\dd \al^{\Z_n}}{n} +\del_{\al^{\Z_n}}\dd \xi_2[a^{\Z_n}]\nn
&\se{1}\dd\big[\frac{m}{2n}\al^{\Z_n} a^{\Z_n}+\frac{m}{2} \big(a^{\Z_n} \hcup{1} \frac{\dd \al^{\Z_n}}{n}   \big)\nn
&\ \ \ +\del_{\al^{\Z_n}}\xi_2[a^{\Z_n}]\big]+\frac{m}{2}\al^{\Z_n} \frac{\dd \al^{\Z_n}}{n}.\label{delphi3al}
\end{align}

Assuming $\al=\dd h^{\Z_n}$ for a $\Z_n$ valued 0-cochain $h^{\Z_n}$, then the last term can be made into a total derivative:
\begin{align*}
\frac{m}{2} \al^{\Z_n} \frac{\dd \al^{\Z_n}}{n} &\se{1}\frac{m}{2} \left(\al \frac{\dd \al}{n} + \al \dd \toZ{\frac{\al}{n}}\right)\nn
&=\frac{m}{2} \dd h^{\Z_n}\dd \toZ{\frac{\dd h^{\Z_n}}{n}}\nn
&=\dd \big(\frac{m}{2}h^{\Z_n}\dd \toZ{\frac{\dd h^{\Z_n}}{n}}\big)\nn
&\se{1}\dd \big(\frac{m}{2}h\dd \toZ{\frac{\dd h}{n}}\big).
\end{align*}

So
\begin{align}
-\del_{\al} \phi_3[a^{\Z_n}] &=\dd \phi_2[a,h]\label{delphi3}\\
\phi_2[a,h]:&=\frac{m}{2n} \al^{\Z_n} a^{\Z_n}+\frac{m}{2} \big(a^{\Z_n} \hcup{1} \frac{\dd \al^{\Z_n}}{n}   \big)\nn
&\ \ \ +\del_{\al^{\Z_n}}\xi_2[a^{\Z_n}]+\frac{m}{2} \big(h^{\Z_n}\dd \toZ{\frac{\dd h^{\Z_n}}{n}}\big)\label{phi2alzn}\\
&\se{1}\frac{m}{2n} \al a +  \frac{m}{2}\big(a \hcup{1} \frac{\dd \al}{n}   \big)+ \del_{\al}\xi_2[a]\nn
&+ \dd \big(\frac{m}{2}\toZ{\frac{\al}{n}}\hcup{1}a\big)+\frac{m}{2} \al \toZ{\frac{\al}{n}} + \frac{m}{2} \big(h\dd \toZ{\frac{\dd h}{n}}\big)\nn
&\se{1}\frac{m}{2n}\dd h a+ \del_{\dd h}\xi_2[a] + \dd \xi_1[a,h]\label{phi2h} \\
\xi_1[a,h] :&= \frac{m}{2}\big(\toZ{\frac{\dd h}{n}}\hcup{1}a + h \toZ{\frac{\dd h}{n}}\big). \nonumber
\end{align}

By construction \eqref{phi2alzn} we have 
\begin{align}
\phi_2[a,h]=\phi_2[a^{\Z_n},h^{\Z_n}].\label{phi2inv}
\end{align}
(See Appendix \ref{apdx:gen} for relationship between $\om_4$, $\phi_3$ and $\phi_2$ in general.)

We also see that $\int_{\cM^3} \delta_{\al} \phi_3[\{a^{\Z_n}\}]$ is independent of $a^{\Z_n}_{bulk}$ or $a'^{\Z_n}_{bulk}$, so we may take it out of the sum in the last line of \eqref{bdstate0} and write:
\begin{align}
&\ket{\{a^{\Z_n}_{\p}\}}_{\p} \nn
&\ra \ee^{-2 \pi i \int_{\cM^3}  \delta_{\al} \phi_3[a_{\p}^{\Z_n}]} \ket{\{a'^{\Z_n}_{\p}\}}_{\p}\nn
&=\ee^{2 \pi i \int_{\p\cM^3}  \phi_2[a_{\p}^{\Z_n},h^{\Z_n}]} \ket{\{a'^{\Z_n}_{\p}\}}_{\p}. \label{bdstate}
\end{align}


In the even $m$ case, \eqref{phi2alzn} simplifies to
\begin{align*}
\phi_2[a,h]\se{1}\frac{m}{2n} \al^{\Z_n} a^{\Z_n},
\end{align*}
so the non-onsite phase for the anomalous 1-symmetry is
\begin{align}
\int_{\p \cM^3} \phi_2[a^{\Z_n},h^{\Z_n}]
&\se{1} \frac{m}{2n} \int_{\p \cM^3} \al^{\Z_n} a^{\Z_n} \nn
&=\frac{m}{2n} \int_{\p \cM^3 \cap \al^{\Z_n}} a^{\Z_n}. 
\label{bdphase}
\end{align}
Here $\cap$ is the cap product\cite{hatcher}, which takes as input a $q$-cochain $\phi_q$ and $n$-chain $(0\ra n)$, and outputs a $(n-q)$-chain given by:
\begin{align}
(0\ra n) \cap \phi_q := \<\phi_q,(0\ra q)\> (q \ra n). \label{cap}
\end{align}

In the more general case, by \eqref{phi2alzn} and \eqref{xi2}, the non-onsite phase is:
\begin{align}
&\int_{\p \cM^3} \phi_2[a^{\Z_n},h^{\Z_n}]\nn
&\se{1}\int_{\p \cM^3}\frac{m}{2n} \al^{\Z_n} a^{\Z_n}   \nn
&\ \ \ +\frac{m}{2} \big(a^{\Z_n} \hcup{1} \frac{\dd \al^{\Z_n}}{n}+(a^{\Z_n}+\al^{\Z_n})\toZ{\frac{a^{\Z_n}+\al^{\Z_n}}{n}}\nn
&\ \ \ + \dd a^{\Z_n} \hcup{1} \toZ{\frac{a^{\Z_n}+\al^{\Z_n}}{n}} +h^{\Z_n}\dd \toZ{\frac{\dd h^{\Z_n}}{n}}\big),
\label{bdphaseodd}
\end{align} 
where $\al^{\Z_n}=(\dd h^{\Z_n})^{\Z_n}$

\subsection{Boundary transformation strings}
On the boundary $\p \cM^3$, the 1-cocycle $\al^{\Z_n}$ is Poincar\'e dual to closed loops $\p \cM^3 \cap \al^{\Z_n}$. These loops are the boundary of a 2-manifold $- \cM^3 \cap \al^{\Z_n}$ in the bulk. While the 1-symmetry is on-site in the bulk, it is non-onsite on the boundary, accompanied by the phase $\int_{\p \cM^3} \phi_2$. Since the bulk is a non-trivial SPT with 1-symmetry, we expect that its boundary cannot be uniquely gapped without breaking the 1-symmetry.

The 1-symmetry acts on the boundary as string operators. These string operators can be thought of as hopping operators for some emergent flux anyons. We measure the statistics of these anyons in the following subsection. 

\subsection{Self and mutual statistics of boundary transformation strings}
\begin{figure}
\centering
\includegraphics[width=1\linewidth]{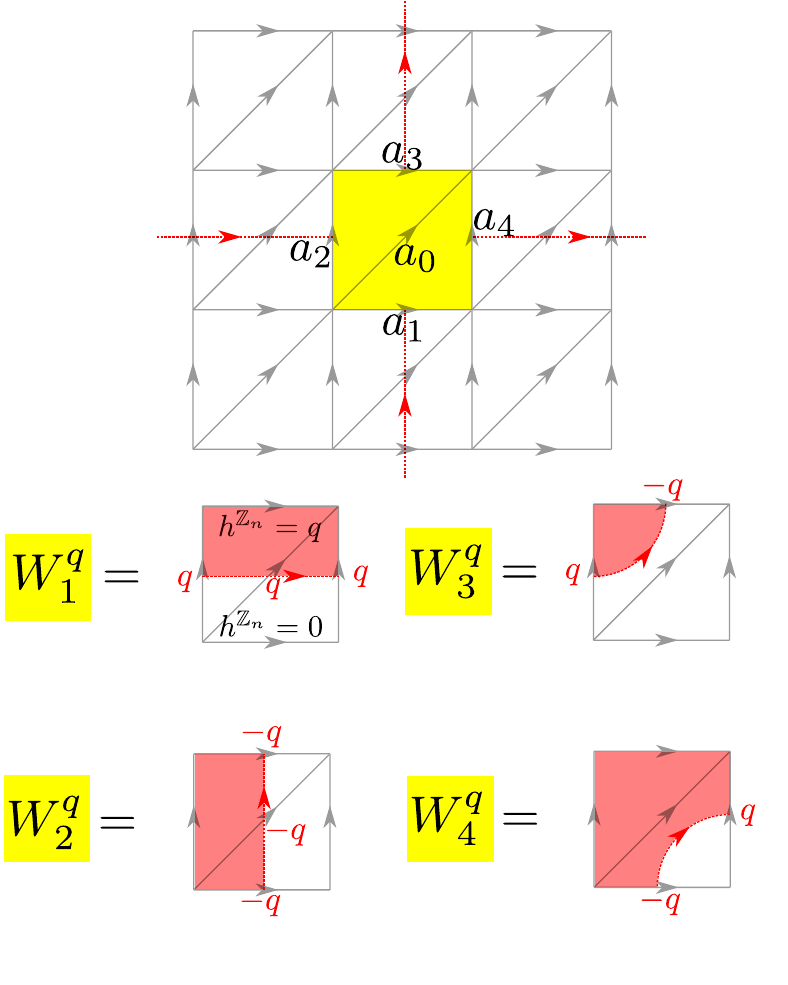}
\caption{(Color online)  Measurement of self and mutual statistics of flux anyons. Top: A 2D region of $\p \cM^3$ is shown. Here the Hilbert space is spanned by the boundary states given in \eqref{bdstatedef} and the degrees of freedom are $a^{\Z_n}$ living on gray links. Red lines depict the boundary 1-symmetry $W^q_{i}$ which can be regarded of as an anyon hopping operator. The corresponding $\al^{\Z_n}$ is non-zero on the gray links intersecting the red dotted lines, and change these links by $\ket{a^{\Z_n}}_{\p}\ra \exp(2\pi \ii \int \phi_2)\ket{(a+\al)^{\Z_n}}_{\p}$, where $\phi_2$ is given by \eqref{bdphase} or \eqref{bdphaseodd}. It turns out that, due to $\phi_2[a,h]=0$ when $\dd h=0$, in our calculation for self and mutual statistics it is only necessary to keep track of links $a^{\Z_n}_{i}$ in the central square, highlighted in yellow. Bottom: the configurations of four different 1-symmetries in the central square. Here $h^{\Z_n}=q$ in the region shaded in pink and $h^{\Z_n}=0$ in the unshaded regions. The non-zero values of $\al = \dd h^{\Z_n}$ are shown in red on gray links intersected by the red dotted lines.
}
\label{fig:scatter}
\end{figure}

We triangulate the 2-dimensional boundary $\p \cM^3$ as shown in \Fig{fig:scatter}. We only focus on a yellow central square, whose links are labeled as $a^{\Z_n}_{i}$, $i=0,1,2,3,4$. We define string operators: $W_{i}^q$, $i=1,2,3,4$, to be the hopping operator depicted in the bottom of \Fig{fig:scatter}.

Each string operator $W_{i}^q$ is represented by an oriented red line in the figure. The red line intersects links in the lattice (colored in gray). Every lattice link intersecting the red string is being updated as in \eqref{transform} with $\al = \dd h^{\Z_n} $. 
$h^{\Z_n}=q$ in the pink shaded region and  $h^{\Z_n}=0$ in the other unshaded regions.
The operator $W_{i}^q$ acts on the boundary Hilbert space as described in \eqref{bdstate}, with $\phi_2$ given by \eqref{bdphase} or \eqref{bdphaseodd}. 

\begin{figure}
\centering
\includegraphics[width=1\linewidth]{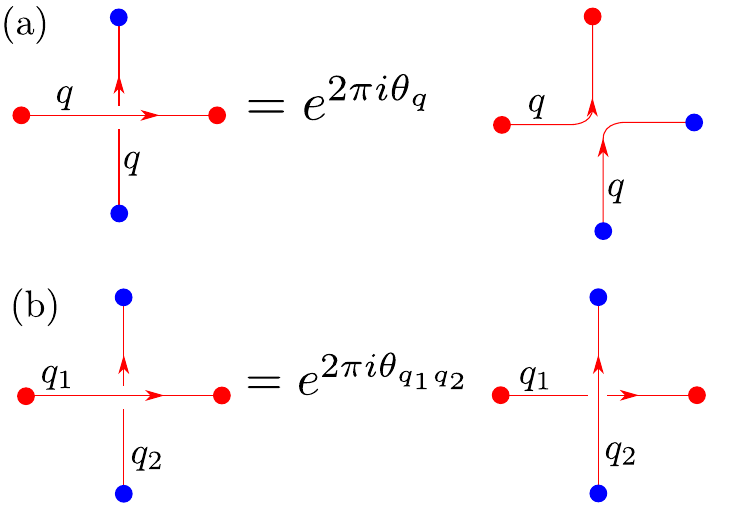}
\caption{(Color online) (a) To measure the self statistics, we compare the outcome of two processes. The first hops an anyon (shown in red) from left to right, and then hops an anyon (shown in blue) from bottom to top; the second hops an anyon from left to top and then hops an anyon from bottom to right. The results differ by exchange of two anyons at their final positions. (b) To measure the mutual braiding statistics, we compare the outcome of two processes: the first hops a red anyon (with flux $q_1$) from left to right, followed by hopping a blue anyon (with flux $q_2$) from bottom to top; the second process do these two operations in the different order. The results differ by a change of linking number $\Delta Lk(q_1,q_2)=1$ between the world lines of the two anyons.}
\label{fig:exchange}
\end{figure}

\subsubsection{Self-statistics}

To compute the self statistics for anyon with flux $q$, we compare the result
of hopping an anyon from bottom to top, then another anyon from left to right,
versus the result of hopping an anyon from bottom to right, and another anyon
from left to top.\cite{LW0316} As shown in \Fig{fig:exchange}a, the resulting
positions of the two final anyons are exchanged in the two processes. More
explicitly the self-statistic is given by $\th_{q}$, where
\begin{align}
W_{1}^q \circ W_{2}^q = \ee^{2 \pi \ii \th_q} W_{4}^q \circ W_{3}^q.\label{thq}
\end{align}

Using \eqref{bdphaseodd} to compute the actions of $W^q_i$, the result is (derivation details in Appendix \ref{apdx:stat})
\begin{align}
\th_{q} = q^2\frac{m}{2n}, \label{selfstat}
\end{align}
which is consistent with \Ref{W161201418}: The 3+1D bulk state that we have
constructed is a $\Z_n$-1-SPT state labeled by $m \in \{0,1,\cdots,2n-1\}$,
protected by an on-site (anomaly-free) $\Z_n$-1-symmetry.  Its boundary has an
anomalous (non-on-site) $\Z_n$-1-symmetry generated by closed string operators
(see \eqn{bdphase} or \eqn{bdphaseodd}).  The corresponding open string
operators will created topological excitations on the boundary.  The anomaly of
the boundary 1-symmetry is encoded in the fractional statistics of those
topological excitations. For instance if $n=2$, $q=1$, then for $m=2$, the
anyon is an emergent fermion. For $m=1$ the anyon is an emergent semion.

\subsubsection{Mutual-statistics}
Similarly to compute the mutual statistics for two anyons with flux $q_1$ and $q_2$, we compare the result of hopping a flux $q_1$ anyon from left to right, then the flux $q_2$ anyon from bottom to top, versus the result of doing the two processes in a different order, as illustrated in \Fig{fig:exchange}b. The mutual-statistic is given by $\th_{q_1 q_2}$, where (derivation details in Appendix \ref{apdx:stat})
\begin{align}
W_{1}^{q_1} \circ W_{2}^{q_2} = \ee^{2 \pi \ii \th_{q_1 q_2}} W_{2}^{q_2} \circ W_{1}^{q_1} ,\label{thq1q2}
\end{align}
and the result is
\begin{align}
\th_{q_1 q_2} = q_1q_2\frac{m}{n}.\label{mutstat}
\end{align}

\section{Gapped symmetric boundaries}\label{gapbd}
In this section we attempt to write down boundary Hamiltonians which are symmetric under the non-onsite transformation \eqref{bdstate}, and contain emergent anyons with self-statistics predicted by \eqref{selfstat}. We will show that it is possible to gap out the boundary by realizing a topological order, which in the $(n,m)=(2,1)$ case is the double semion (DS) topological order, which contains an emergent semion. In the $(n,m)=(2,2)$ case the toric code is realized on the boundary, which contains an emergent fermion. 
The degenerate ground states for these systems on a manifold with non-trivial cycles spontaneously break the 1-symmetry.

An easy way to see this is as follows. From $\omega_4=\dd \phi_3[a]$, if $\cM^4$ has a boundary,
\begin{align}
Z^\text{top} &=\frac{1}{n^{N_l}} \sum_{\{a^{\Z_n}\}} \ee^{2\pi \ii \int_{\cM^4} \om_4[\dd a^{\Z_n}]}\nn
&=\frac{1}{n^{N_{l\p}}} \sum_{\{a_{\p}^{\Z_n}\}} \ee^{2\pi \ii \int_{\p\cM^4} \phi_3[a^{\Z_n}]},\label{Ztopbdd}
\end{align}
where $N_{l\p}$ is the number of links in the space-time triangulation of the boundary.

If we impose the constraint $\dd a^{\Z_n}\se{n}0$ by hand (the constraint doesn't violate 1-symmetry since it is invariant under \eqref{Zn1symm}), then from the expression for $\phi_3$ \eqref{phi3}, we have
\begin{align*}
\left.\phi_3[a^{\Z_n}]\right\vert_{\dd a\se{n}0} &\se{1} \frac{m}{2n}a^{\Z_n}\dd a^{\Z_n}=\frac{m}{2}a^{\Z_2}\frac{\dd a^{\Z_2}}{2},
\end{align*}
where in the last step we specialized to the case $n=2$. This can be recognized as the Lagrangian for the surface topological order. To recast it into a more familiar form, we have
\begin{align*}
\frac{\dd a^{\Z_2}}{2} = \bt_2 a^{\Z_2} \se{2} \gSq^1(a^{\Z_2}) \se{2} a^{\Z_2}  a^{\Z_2},
\end{align*}
where $\bt_2$ is the Bockstein homomorphism and the second equality follows from \eqref{Sq1Bs2}, and the third equality is by definition of Steenrod square \eqref{cupkrel1} and $\dd a^{\Z_2}\se{2}0$.
So, \eqref{Ztopbdd} becomes
\begin{align*}
\left.Z^\text{top}\right\vert_{\dd a\se{2}0} &=\frac{1}{n^{N_{l\p}}} \sum_{\{\dd a \se{2}0\}} \ee^{2\pi \ii \int_{\p\cM^4} \frac{m}{2}a^{\Z_2}\frac{\dd a^{\Z_2}}{2}} \nn
&=\frac{1}{n^{N_{l\p}}}\sum_{\{\dd a\se{2}0\}} \ee^{2\pi \ii \int_{\p\cM^4} \frac{m}{2} a^{\Z_2} a^{\Z_2} a^{\Z_2}},
\end{align*}
which for $m=1$ is (up to a volume term) the partition function for double semion topological order (see for instance \Ref{Wang2018}). For $m=2$ the Lagrangian$\se{2}0$ and describes the $\Z_2$ gauge theory, \ie toric code.

\subsection{Engineering boundary gapped Hamiltonian}

Alternatively, we can explicitly engineer a gapped Hamiltonian consisting of
mutually commuting terms on the boundary Hilbert space respecting the anomalous
1-symmetry and realizing the DS topological order.

The following boundary Hamiltonian is proposed:
\begin{align}
H_{\p} &= -\sum_{i}H_{s,i}-\sum_{\Delta}H_{p,\Delta} \label{hp} \\
H_{s,i} &:=W_{\bigodot i}\nn
H_{p,\Delta} &:=\delta_{\<\dd a,\Delta\>^{\Z_n},0}.\nonumber
\end{align}
Here $i$ is summed over all sites and $\Delta$ is summed over all 2-simplices
(\ie triangles) in the boundary. $\delta$ is the Kronecker delta function.
$\<\dd a,\Delta\>$ is evaluating the 2-cochain $\dd a$ on the 2-simplex $\Delta$.
Hence $H_{p,\Delta}$ enforces the ``no flux" constraint $\dd a\se{n}0$ on every
2-simplices. $W_{\bigodot i}$ is the 1-symmetry operator corresponding to a
tiny loop surrounding site $i$ (see \Fig{fig:wi}). 

\subsubsection{$H_{\p}$ is exactly solvable and has 1-symmetry}
To show that $H_{\p}$ consists of mutually commuting terms, which also commutes with the boundary 1-symmetry operators, it suffices to check the following commutators vanishes:
\begin{align}
[H_{p,\Delta},W(h^{\Z_n})]&=0 \label{comm1}\\
[W_{\bigodot i},W(h^{\Z_n})]&=0 \label{comm2}
\end{align}
for any $\Z_n$-valued 0-cochain $h^{\Z_n}$, where $W(h^{\Z_n})$ denotes a 1-symmetry operator parameterized by $h^{\Z_n}$, whose action is described by \eqref{bdstate} with $a'^{\Z_n}_{\p}=(a_{\p}+\dd h)^{\Z_n}$.\\

To show \eqref{comm1}, notice that 
\begin{align*}
&W(h^{\Z_n})^{-1}H_{p,\Delta}W(h^{\Z_n})\nn
&= W(h^{\Z_n})^{-1}\delta_{\<da,\Delta\>^{\Z_n},0}W(h^{\Z_n})\nn
&=\delta_{\<\dd (a+\dd h),\Delta\>^{\Z_n},0}=\delta_{\<\dd a,\Delta\>^{\Z_n},0}=H_{p,\Delta},
\end{align*}
where we used the fact that the non-onsite phases from $W(h^{\Z_n})$ and $W(h^{\Z_n})^{-1}$ cancels, since $\delta_{\<da,\Delta\>^{\Z_n},0}$ does not change the value of $a^{\Z_n}$ in the ket.

To show \eqref{comm2}, notice that for any two $\Z_n$-valued 0-cochain $h_1^{\Z_n}$ and $h_2^{\Z_n}$, we have
\begin{align}
&W(h_2^{\Z_n})^{-1}W(h_1^{\Z_n})^{-1}W(h_2^{\Z_n})W(h_1^{\Z_n})\ket{\{a^{\Z_n}\}}_{\p}\nn
&=\exp \Big[2\pi \ii \int_{\p \cM^3} \big(-\phi_2[(a+\dd h_2)^{\Z_n},h_1^{\Z_n}] - \phi_2[a^{\Z_n},h_2^{\Z_n}]\nn
&\ \ \ \ \ \ +\phi_2[(a+\dd h_1)^{\Z_n},h_2^{\Z_n}] + \phi_2[a^{\Z_n},h_1^{\Z_n}] \big) \Big]\ket{\{a^{\Z_n}\}}_{\p}\nn
&=\exp \Big[2\pi \ii \int_{\p \cM^3} \dd \phi_1[a^{\Z_n},h_1^{\Z_n},h_2^{\Z_n}]\Big]\ket{\{a^{\Z_n}\}}_{\p},\label{wh1h2}
\end{align}
where we applied \eqref{phi2inv} and \eqref{phi2h} in the last step to show that the integrand in the exponent is a total derivative:
\begin{align*}
&\phi_2[(a+\dd h_1)^{\Z_n},h_2^{\Z_n}] - \phi_2[a^{\Z_n},h_2^{\Z_n}]- (h_1\lra h_2)\nn
&=\phi_2[a+\dd h_1,h_2] - \phi_2[a,h_2]- (h_1\lra h_2)\nn
&=\frac{m}{2n}\dd h_2 (a+\dd h_1)+ \del_{\dd h_2}\xi_2[a+\dd h_1] + \dd \xi_1[a+\dd h_1,h_2]\nn
&\ \ \ -\frac{m}{2n}\dd h_2 a- \del_{\dd h_2}\xi_2[a] - \dd \xi_1[a,h_2]- (h_1\lra h_2)\nn
&=\frac{m}{2n}\dd h_2 \dd h_1+\dd \del_{\dd h_1}\xi_1[a,h_2]- (h_1\lra h_2)\nn
&=\dd \phi_1[a^{\Z_n},h_1^{\Z_n},h_2^{\Z_n}]
\end{align*}
where
\begin{align}
\phi_1[a,h_1,h_2]:&=\frac{m}{2n}h_2^{\Z_n} \dd h_1^{\Z_n}+ \xi_1[a^{\Z_n}+\dd h_1^{\Z_n},h_2^{\Z_n}]\nn
& \ \ \ - \xi_1[a^{\Z_n},h_2^{\Z_n}]-(h_1\lra h_2)\label{phi1def} \\
&\se{1}\frac{m}{2n} h_2 \dd h_1+ \del_{\dd h_1}\xi_1[a,h_2]+\dd \xi_0[h_1,h_2]\nn
&\ \ \ - (h_1\lra h_2) \nn
\xi_0[h_1,h_2]:&=\frac{m}{2}\big(\toZ{\frac{h_2}{n}}\hcup{1}\dd h_1 + h_1 \toZ{\frac{h_2}{n}}\big).\nonumber
\end{align}
(See Appendix \ref{apdx:gen} for relationship between $\om_4$, $\phi_3$, $\phi_2$ and $\phi_1$ in general.)

\begin{figure}[tb]
\centering
\includegraphics[width=1\linewidth]{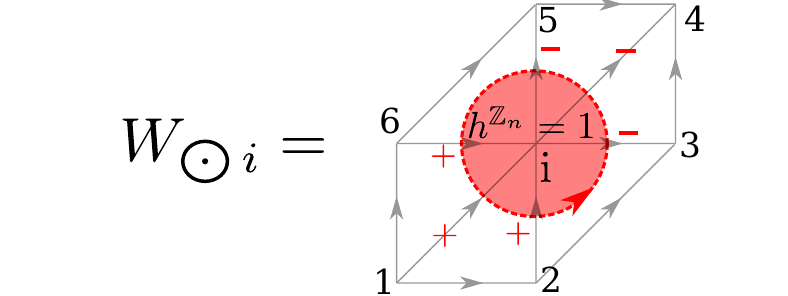}
\caption{(Color online) $W_{\bigodot i}$ is the 1-symmetry operator on the boundary Hilbert space for a tiny loop surrounding site $i$. Here the pink shaded region has $h^{\Z_n}=1$. The non-zero values of $\al=\dd h^{\Z_n}=\pm 1$ are drawn in red. The neighboring sites of $i$ are labeled $j=1,\dots,6$.}
\label{fig:wi}
\end{figure}

Thus by Stoke's theorem, \eqref{wh1h2} implies that when evaluated on the closed manifold $\p\cM^3$, $[W(h_1^{\Z_n}),W(h_2^{\Z_n})]=0$ for any $\Z_n$-valued 0-cochains $h_1$, $h_2$. For the case of our interest \eqref{comm2}, we may take $h_1^{\Z_n}=h_{\bigodot i}^{\Z_n}$ where $W_{\bigodot i}=W(h_{\bigodot i}^{\Z_n})$ as depicted in \Fig{fig:wi}, and $h_2^{\Z_n}=h^{\Z_n}$ to be an arbitrary 1-symmetry. Alternatively we can evaluate \eqref{wh1h2} by integrating the exponent over a patch covering the region where $h_{\bigodot i}^{\Z_n}\neq 0$ and use $\phi_1[a^{\Z_n},h_1^{\Z_n}=0,h_2^{\Z_n}]=0$.\\


\subsubsection{Topological ordered surface states for $n=2$}
We can specialize to the case $(n,m)=(2,1)$ and evaluate $W_{\bigodot i}$. Assuming ``no flux" constraint is enforced, we have (see Appendix \ref{wi} for details)
\begin{align}
&\left.W_{\bigodot i}\ket{\{a^{\Z_2}_{ij},a^{\Z_2}_{jj'}\}}\right\vert_{\dd a\se{2}0}\nn
&=\prod_{\<j,j'\>}(-)^{a_{ij}a_{ij'}}\ket{\{(a_{ij}+1)^{\Z_2},a^{\Z_2}_{jj'}\}},\label{eq:wi}
\end{align}
where $j,j'\in\{1,\dots,6\}$ are neighboring sites of $i$ (see \Fig{fig:wi}). The product is taken over six links with neighboring $j,j'$. The resulting $H_{\p}$ gives rise to DS topological order.

For the $m=2$ case, we have
\begin{align*}
&\left.W_{\bigodot i}\ket{\{a^{\Z_2}_{ij},a^{\Z_2}_{jj'}\}}\right\vert_{\dd a\se{2}0}=\ket{\{(a_{ij}+1)^{\Z_2},a^{\Z_2}_{jj'}\}}
\end{align*}
So $h_{s,i}$ is the usual star term and $h_{p,\Delta}$ is the usual plaque term for the toric code model. Thus $H_{\p}$ gives rise to the toric code topological order.

\subsubsection{Connection to Works of Wan and Wang}
A general theory of gapped symmetric boundaries of higher SPT is presented in Section III of \Ref{WW181211955}, which is a generalization of \Ref{Wang2018}. It was then applied in Section IX of \Ref{WW181211968}, and Section 8 of \Ref{WW190400994}, which also contains a lattice Hamiltonian description for a 4+1D bulk/3+1D boundary. 

We give a rough review of their result in the following. In general, a 1-SPT protected by 1-form finite symmetry $\Pi_2$(which is Abelian) and 0-form finite symmetry $G$, may be associated with a ``2-group" $\mathbb{G}$, such that its classifying space $B\mathbb{G}=B(G,\Pi_2)$ has $\pi_1 = G$, $\pi_2 = \Pi_2$ and $\pi_{0}=\pi_{k>2}=0$. (In addition, $\mathbb{G}$ also contains the data\cite{ZW180809394} $\al_2 : G \ra \text{Aut}(\Pi_2) $ and $\bar{n}_3 \in H^3(BG,\Pi_2^{\al_2})$ describing the interplay between $G$ and $\Pi_2$.) 
A space-time field configuration is a map $\phi:~\cM^D \ra B\mathbb{G}$, and the cocycle $\om_D$ describing the 1-SPT is the pullback: $\om_D=\phi^*\bar{\om}_D$ for a topological term $\bar{\om}_D$, which can be an element of $H^D(B\mathbb{G},U(1))$, or a bordism invariant in general. Section III of \Ref{WW181211955} claimed that the gapped boundary of 1-SPT corresponds to a fibration:
\begin{align*}
B\mathbb{K}\ra B \mathbb{H} \ra B \mathbb{G}
\end{align*}
such that the topological invariant $\bar{\om}_D$ in $ B\mathbb{G}$ is pulled back to a trivial topological invariant in $ B\mathbb{H}$. Here $\mathbb{H}$ is a 2-group, viewed as an extension of $\mathbb{G}$. $B\mathbb{K}$ is the total space of a fibration $B^2K_{[1]} \ra B\mathbb{K} \ra BK_{[0]}$, where $K_{[0]}$, $K_{[1]}$ are some 0-form and 1-form symmetries respectively. For a finite group $G$, $B^2G=K(G,2)$ is a Eilensberg-MacLane space for which the only non-trivial homotopy group is $\pi_2=G$. To be precise, the topological invariants of the classifying spaces $B\mathbb{G}$, $B\mathbb{H}$ are bordism invariants of the bordism groups\cite{Kapustin2014,Kapustin2015} $\Omega^{S_G}_{D}(B\mathbb{G}) $, $\Omega^{S_H}_{D}(B\mathbb{H}) $, computed\cite{WW181211967} with respect to an ``$S$-structure", $S_{G,H}=\text{SO}/\text{O}/\text{Spin}/\text{Pin}^{\pm}$, corresponding to unitary bosonic SPT/time-reversal invariant bosonic SPT/unitary fermionic SPT/time-reversal invariant fermionic SPT with $T^2=(\mp)^{F}$, respectively.

In this framework, our $\Z_2$-1-SPT has $D=4$, $(G,\Pi_2)=(0,\Z_2)$ and $S_G=\text{SO}$. Gapping out its 2+1D boundary for $m=2$ with toric code topological order corresponds to the fibration:
\begin{align*}
B\Z_2 \ra B\text{Spin}(4)\times B^2\Z_2 \ra B\text{SO}(4)\times B^2\Z_2
\end{align*}
where the pullback of $\bar{\om}_4$ is trivial because of a relation between $\Z_2$-valued 2-cocycle $\cB^{\Z_2}$ and the Stiefel-Whitney classes $\text{w}_1,\text{w}_2$ (which is derived using Wu formula, \eg. in Appendix D.5 of \Ref{W161201418}):
\begin{align*}
\gSq^2(\cB^{\Z_2})=(\text{w}_1^2+\text{w}_2)\cB^{\Z_2},
\end{align*}
where $\text{w}_1,\text{w}_2$ vanishes when pulled back to $B\text{Spin}(4)\times B^2\Z_2$, which is a spin manifold. The emergent fermion is due to the emergent spin structure.

\section{Geometric interpretation of ground state wavefunction}\label{gswfmeaning}
In this section we attempt to provide an intuitive interpretation of the ground state wavefunction \eqref{gswf} on a closed 3-manifold.

Recall from \eqref{gswf} and \eqref{phi3}, the ground state wavefunction is
\begin{align*}
\ket{\psi_0} &= \sum_{\{a^{\Z_n}\}} \ee^{2 \pi i \int_{\cM^3} \phi_3[a^{\Z_n}]} \ket{\{a^{\Z_n}\}}\nn
\phi_3 [a] &= \frac{m}{2n}a \dd a + \frac{m}{2} \dd a \hcup{1} \toZ{\frac{\dd a}{n}} + \dd \xi_2[a].
\end{align*}
In a closed 3-manifold $\cM^3$, we can ignore the $\dd \xi_2$ term. In the dual manifold $\tilde{\cM}^3$, $a$ is dual to 2-chains $\tilde{a}$, and $\dd a$ is dual to $\p \tilde{a}$, which is a 1-cycle.

If we focus on the term $\frac{m}{2n}a \dd a$, which only depends on 1-diagonal links, we can imagine the dual 2-chains and 1-cycles as living on the dual faces and links of a simple cubic lattice. Geometrically, $\frac{m}{2n}a \dd a$ is contributed from the intersections of $\tilde{a}$ and $\p \tilde{a}'$, which is $\p \tilde{a}$ displaced by the framing vector $-\vec{\frac 12}=(-1/2,-1/2,-1/2)$.
\begin{align*}
\int_{\cM^3} \frac{m}{2n}a \dd a=\sum_{p\in \tilde{a} \cap \p \tilde{a}'} \frac{m}{2n} q_{a,p} q_{\p a',p},
\end{align*}
where $q_{a,p}, q_{\p a',p} \in \Z $ denote the integer coefficients of the 2-chain $\tilde{a}$ and 1-cycle $\p \tilde{a}'$ at the intersection point $p$.

\begin{figure}[tb]
\centering
\includegraphics[width=1\linewidth]{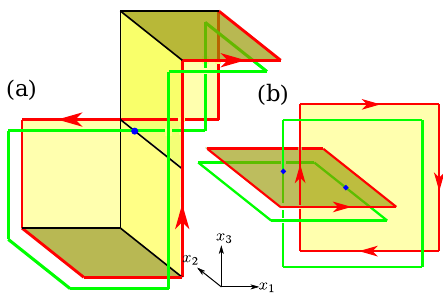}
\caption{(Color online) Evaluation of $\int_{\cM^3} \frac{m}{2n}a \dd a$ in the dual lattice. Yellow squares, red links and green links represent $\tilde{a}$, $\p \tilde{a}$ and $\p \tilde{a}'$ respectively. Blue dot represents the intersection of green links and yellow square, where $a \dd a\neq 0$. The configuration of $\p \tilde{a}$ is: (a) an unknot with self-linking number +1, where the coefficient of the 2-chain $\tilde{a}$ on every yellow square is $q$; (b) a Hopf link with linking number +1, where the coefficient of the 2-chain $\tilde{a}$ on the two yellow squares are $q_1$ and $q_2$.}
\label{fig:gswfmeaning}
\end{figure}

If the 1-cycle $\p \tilde{a}$ can be resolved into non-intersecting loops $K_i$, then $\tilde{a}$ are the Seifert surfaces $S_i$ for these loops. A Seifert surface of loop $K_i$ is an oriented surface with $K_i$ as its boundary. It is known that the signed intersection number between $K_i$ and a Seifert surface of $K_j'$ is the sum of signed crossings between $K_i$ and $K_j'$ (viewed from the $-\vec{\frac 12}$ direction), which is the linking number $Lk(K_i,K_j')$ \footnote{\url{https://www.math.ias.edu/files/wam/LicataLecture3.pdf}}. Thus
\begin{align}
&\int_{\cM^3} \frac{m}{2n}a \dd a=\sum_{i,j} \sum_{S_i\cap K'_j} \frac{m}{2n} q_i q_j \nn
&\ \ \ =\frac{m}{2n}\sum_{i,j}  q_i q_j Lk(K_i,K'_j)\nn
&\ \ \ = \frac{m}{2n}\sum_{i} q_{i}^{2} w(K_i) + \frac{m}{n} \sum_{i<j} q_i q_j Lk(K_i,K_j), \label{gswf=lk}
\end{align}
where $w(K_i)=Lk(K_i,K_i')$ is the self-linking number of $K_i$, and for $i\neq j$, $Lk(K_i,K_j')=Lk(K_i,K_j)$ is the linking number between $K_i$ and $K_j$. $q_i \in \Z$ denote the ``strength" of each loop $K_i$. Note the result \eqref{gswf=lk} is invariant (mod 1) under $q_i\ra q_i + n u_i$ for any integers $\{u_i\}$ for general $m$. For example in Figure \ref{fig:gswfmeaning}(a), we see that for an unknot with self linking number $+1$ carrying flux $\dd a =q$, $\int_{\cM^3} \frac{m}{2n}a \dd a = \frac{m}{2n} q^{2}$. In Figure \ref{fig:gswfmeaning}(b), for the Hopf link with linking number 1, with two loops carrying flux $\dd a =q_1$ and $q_2$, we have $\int_{\cM^3} \frac{m}{2n}a \dd a = \frac{m}{n} q_1 q_2$. This could be regarded as an alternative way to derive the self-statistics \eqref{selfstat} and mutual-statistics \eqref{mutstat} of the boundary transformation strings, from the 3d bulk space perspective instead of the 2+1d boundary spacetime perspective.

However, when multiple lines intersect at a point, we need to carefully resolve the 1-cycle $\p\tilde{a}$ into non-intersecting loops. We will consider the even $m$ case and the odd $m$ case separately.

\subsection{Even $m$ case}
\begin{figure}[tb]
\centering
\includegraphics[width=1\linewidth]{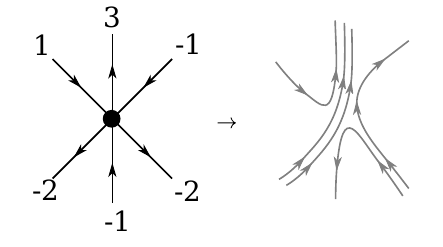}
\caption{(Color online) ``No-crossing" resolution at an intersection. In the simple cubic dual lattice, every vertex is connected to six oriented links. Each link carries an integer $q$ which is the coefficient of the 1-cycle $\p\tilde{a}$. The figure on the left shows one such configuration, viewed from $\vec{\frac{1}{2}}$. In the ``no-crossing" resolution, each link is resolved into $q$ parallel strands away from the original vertex. Near the vertex the strands are connected such that there is no crossing when viewed from $\vec{\frac{1}{2}}$. Such resolution may not be unique, but can be fixed by choosing some convention.}
\label{fig:crossing}
\end{figure}

In the case of even $m$, \eqref{phi3} is
\begin{align*}
\phi_3 [a] \se{1} \frac{m}{2n}a \dd a.
\end{align*}
Each lattice point of the dual cubic lattice has six connecting dual links. Given a dual cycle configuration $\p \tilde{a}$, we project these six links onto the plane perpendicular to the $-\vec{\frac{1}{2}}$ framing vector. Then we resolve the intersection into disjoint loops with ``no-crossing" resolution: requiring that no crossing occurs in this intersection. An example is shown in \Fig{fig:crossing}.

Since all the crossings are contributed away from intersections, the wavefunction amplitude \eqref{gswf=lk} is
\begin{align*}
\int_{\cM^3} \phi_3[a] =\frac{m}{2n}\sum_{i} q_{i}^{2} w(K_i) + \frac{m}{n} \sum_{i<j} q_i q_j Lk(K_i,K_j).
\end{align*}
with $K_i$ obtained from $\p\tilde{a}$ by ``no-crossing" resolution at each vertex.

\subsection{Odd $m$ case}
In the case of odd $m$, \eqref{phi3} is
\begin{align*}
\phi_3 [a] \se{\dd,1} \frac{m}{2n}a \dd a + \frac{m}{2} \dd a \hcup{1} \toZ{\frac{\dd a}{n}}.
\end{align*}
As explained previously, 
\begin{align*}
\int_{\cM^3}\frac{m}{2n}a \dd a=\frac{m}{2n}\times \sum
\begin{pmatrix}
&\text{signed crossings}\\
&\text{ away from intersections}
\end{pmatrix}.
\end{align*}
In the following we will also interpret the second term as $\frac{m}{2n}\times$the sum of signed crossings under a ``quotient-remainder" resolution at intersections.

\begin{figure}[tb]
\centering
\includegraphics[width=1\linewidth]{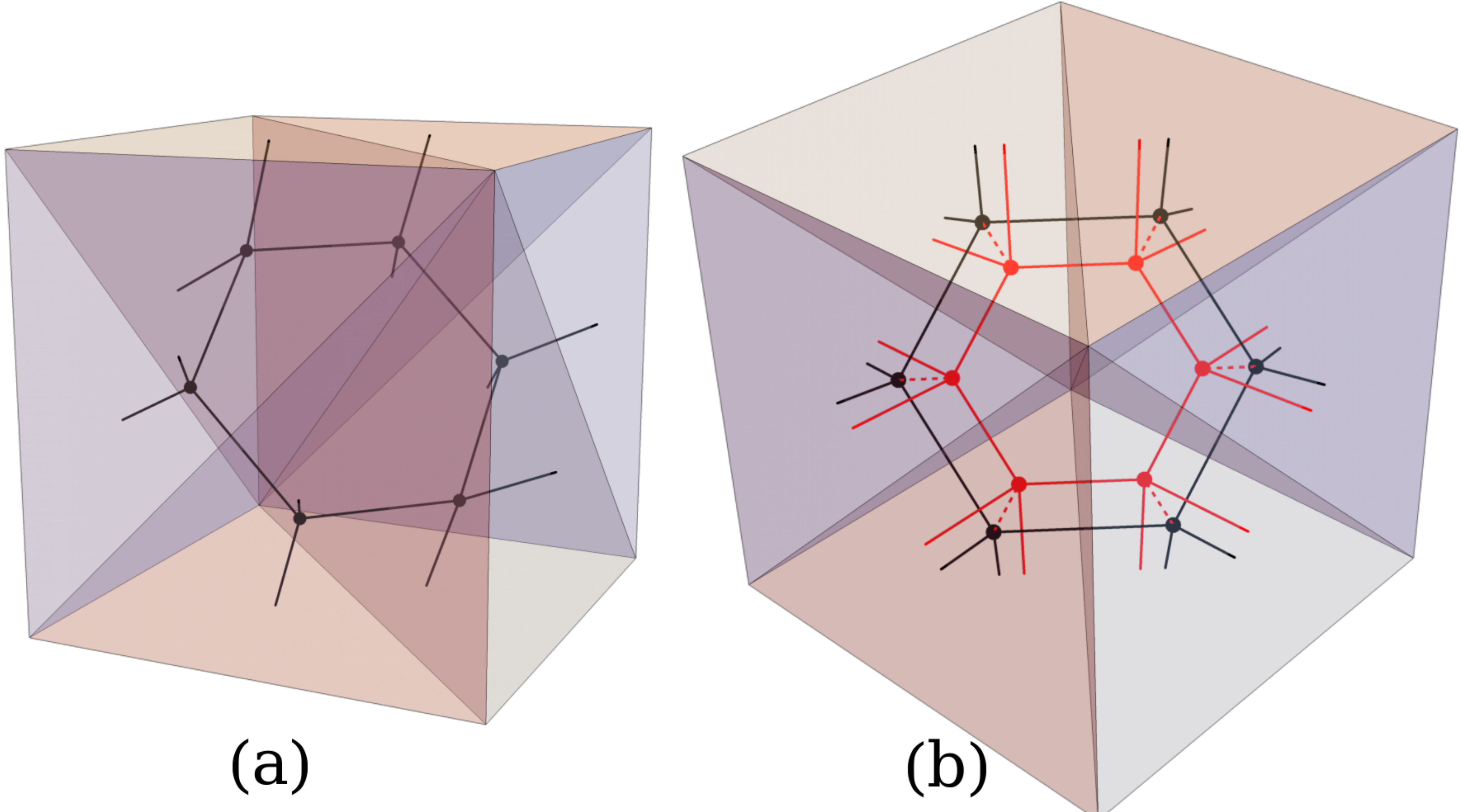}
\caption{(Color online) (a)A cubic cell is triangulated with six tetrahedrons. In the dual lattice they corresponds to six vertices (shown in black), which forms a hexagon. There are two external links at each face which connects to vertices of a neighboring cube. (b) Each link is separated into a ``remainder" channel (black) dual to $(\dd a)^{\Z_n}$ and a ``quotient" channel (red) dual to $n\toZ{\frac{\dd a}{n}}$. In each tetrahedron, the intersections for ``quotient" channels (red dots) were displaced slightly towards the center of the cube, away from intersections for ``remainder" channels (black dots). The resulting crossings between ``remainder" and ``quotient" channels viewed from $\vec{\frac{1}{2}}$ contribute to $\dd a \hcup{1} \toZ{\frac{\dd a}{n}}$. The two channels are coupled within each tetrahedron by a red dashed line, dual to $n\dd \toZ{\frac{\dd a}{n}}$.}
\label{fig:qr}
\end{figure}

Since the term $\frac{m}{2} \dd a \hcup{1} \toZ{\frac{\dd a}{n}}\se{1}\frac{m}{2} (\dd a)^{\Z_n} \hcup{1} \toZ{\frac{\dd a}{n}}$ also depends on 2- and 3-diagonal links, we need to use the full triangulation described in Appendix \ref{apdx:lat} with six tetrahedrons per unit cubic cell. The dual lattice is a cubic lattice with six sites forming a hexagon in each unit cell, depicted in \Fig{fig:qr}(a).

The ``quotient-remainder" resolution is the following: write $\dd a=(\dd a)^{\Z_n}+n \toZ{\frac{\dd a}{n}}$. These two terms are called the ``remainder" and ``quotient" respectively. The 1-cycles dual to $\dd a$ live on the links of the dual lattice. We split each link in the dual lattice into two channels: the ``remainder" channel dual to $(\dd a)^{\Z_n}$, and the ``quotient" channel dual to $n \toZ{\frac{\dd a}{n}}$. They are depicted as black and red links respectively in \Fig{fig:qr}(b). If we detach the `quotient" intersections from the ``remainder" intersections by displacing them slightly towards the center of each cube, then $\dd a \hcup{1} n\toZ{\frac{\dd a}{n}}$ is the sum of signed crossings (mod $2n$) between the ``remainder" channels and the ``quotient" channels, viewed from $\vec{\frac{1}{2}}$, as depicted in \Fig{fig:qr}(b). All other intersections (black and red dots in \Fig{fig:qr}(b)) are resolved with the ``no-crossing" resolution. 


Thus $\frac{m}{2} \dd a \hcup{1} \toZ{\frac{\dd a}{n}}$ is $\frac{m}{2n}\times$sum of signed crossings between quotient channels and remainder channels at a vertex. As before, the sum of signed crossings is the sum of linking numbers between resolved loops. Hence the wavefunction amplitude \eqref{gswf=lk} is
\begin{align*}
\int_{\cM^3} \phi_3[a] =\frac{m}{2n}\sum_{i} q_{i}^{2} w(K_i) + \frac{m}{n} \sum_{i<j} q_i q_j Lk(K_i,K_j).
\end{align*}
with $K_i$ obtained from $\p\tilde{a}$ by ``quotient-remainder" resolution at each vertex.

The term $\frac{m}{2} \dd a \hcup{1} \toZ{\frac{\dd a}{n}}$ is necessary to ensure that $\phi_3 [a]$ is invariant mod 1 under $a\ra a+nu$ for $\Z$-valued 1-chain $u$. Indeed under $a\ra a+nu$, all changes occur only in the quotient channel $n\toZ{\frac{\dd a}{n}}\ra n\toZ{\frac{\dd a}{n}}+ n\dd u$. The change to $\phi_3 [a]$ mod 1 is $\frac{1}{2}\times$the sum of signed crossings between the dual of $(\dd a)^{\Z_2}$ in remainder channel and the dual of $\dd u$ in the quotient channel. Since both of them are closed loops living in separate channels, the total number of signed crossing is even. Hence $\phi_3 [a]$ is invariant mod 1.

\section{Non-zero background gauge field}\label{nonzerobg}
We may also extend our derivations to the case where the background gauge field $\hat{B}$ in \eqref{cB} is non-zero. By keeping track of the coboundary terms in \eqref{sqB1}$\se{d}$\eqref{sqB5}, it can be shown that \eqref{om4da},\eqref{xi3},\eqref{phi3},\eqref{xi2} become
\begin{align*}
\om_4[(\hat{B}+\dd a)^{\Z_n}] &= \frac{m}{2n} \gSq^2\hat{B}^{\Z_n} +\dd \phi_3 [a,\hat{B}]\nn
&= \om_4[\hat{B}^{\Z_n}] +\dd \phi_3 [a,\hat{B}]
\end{align*}
where
\begin{align}
\phi_3 [a,\hat{B}] :&= \frac{m}{2n}(a^{\Z_n} \dd a^{\Z_n} +a^{\Z_n}\hat{B}^{\Z_n}+\hat{B}^{\Z_n}a^{\Z_n})\nn
&\ \ \ + \xi_3[a^{\Z_n},\hat{B}^{\Z_n}] \label{phi3nzB} \\
&\se{1} \frac{m}{2n}(a \dd a+a\hat{B}+\hat{B}a) + \xi_3[a,\hat{B}] + \dd \xi_2[a,\hat{B}]\nn
\xi_3[a,\hat{B}]:&=\frac{m}{2} \big[(\hat{B}+\dd a) \hcup{1} \toZ{\frac{\hat{B}+\dd a}{n}}+a\hcup{1}\frac{\dd \hat{B}}{n}\nn
&\ \ \ +\hat{B}\hcup{1}\toZ{\frac{\hat{B}}{n}}\big]\nn
\xi_2[a,\hat{B}] :&= \frac{m}{2}(a\toZ{\frac{a}{n}}+\dd a\hcup{1}\toZ{\frac{a}{n}}+a\hcup{1}\toZ{\frac{\hat{B}}{n}}+\hat{B}\hcup{1}\toZ{\frac{a	}{n}}).\nonumber
\end{align}
and \eqref{delphi3al},\eqref{delphi3},\eqref{phi2alzn},\eqref{phi2h} become
\begin{align}
&-\del_{\al} \phi_3[a^{\Z_n},\hat{B}^{\Z_n}] =\dd\big[\frac{m}{2n} \big(\al a-\al\hcup{1}\hat{B}\big)\nn
& \ \ \ + \frac{m}{2}\big(a \hcup{1} \frac{\dd \al}{n}+\del_{\al}\xi_2[a,\hat{B}]+\hat{B}\hcup{2}\frac{\dd \al}{n}\big)\big]\nn
&\ \ \ -\frac{m}{n}\hat{B}\al+\frac{m}{2}\al\frac{\dd\al}{n}
=\dd \phi_2[a,h,\hat{B}]\label{delphi3alnzB}
\end{align}
where
\begin{align}
&\phi_2[a,h,\hat{B}]:=\frac{m}{2n}\big( \al^{\Z_n} a^{\Z_n}-\al^{\Z_n}\hcup{1}\hat{B}^{\Z_n}\big)\nn
&+\frac{m}{2} \big(a^{\Z_n} \hcup{1} \frac{\dd \al^{\Z_n}}{n}
 +\del_{\al^{\Z_n}}\xi_2[a^{\Z_n},\hat{B}^{\Z_n}]+\hat{B}^{\Z_n}\hcup{2}\frac{\dd \al^{\Z_n}}{n}\big)\nn
&\ \ \  -\frac{m}{n}\hat{B}^{\Z_n}h^{\Z_n}+\frac{m}{2} h^{\Z_n}\dd \toZ{\frac{\dd h^{\Z_n}}{n}}\nn
&\se{1}\frac{m}{2n}(\dd h a-\dd h \hcup{1}\hat{B})-\frac{m}{n}\hat{B}h
 + \del_{\dd h}\xi_2[a,\hat{B}] + \dd \xi_1[a,h] \label{phi2nzB}\\
&\xi_1[a,h] := \frac{m}{2}\big(\toZ{\frac{\dd h}{n}}\hcup{1}a + h \toZ{\frac{\dd h}{n}}\big). \nonumber
\end{align}

In Appendix \ref{apdx:hnzB}, we generalize the construction of an exactly solvable Hamiltonian with a unique ground state to the case of non-zero $\hat{B}$. Also in Appendix \ref{apdx:gsdzB} we generalize the expression of the ground state wavefunction \eqref{gswfnzB} in terms of $\phi_3$:
\begin{align}
\ket{\psi_0[\hat{B}]} = \frac{1}{N_{\psi}}\sum_{\{a\}}\ee^{2\pi i \int_{\cM^3} \phi_3[a,\hat{B}]-\phi_3[0,\hat{B}]} \ket{\{a\}}.
\end{align}

In the following we consider the case $m$ is even, where \eqref{phi3nzB}, \eqref{phi2nzB} simplifies to
\begin{align}
\phi_3 [a,\hat{B}]&\se{1}\frac{m}{2n}(a \dd a+a\hat{B}+\hat{B}a) \label{phi3evennzB}\\
\phi_2[a,h,\hat{B}]&\se{1}\frac{m}{2n}(\dd h a-\dd h \hcup{1}\hat{B})-\frac{m}{n}\hat{B}h\label{phi2evennzB}
\end{align}
\subsection{Exactly Solvable Hamiltonian}
The bulk Hamiltonian is given by
\begin{align*}
H=-\sum P_{ij}[\hat{B}]
\end{align*}
By using \eqref{PelemnzB} to write down matrix elements of $P_{ij}[\hat{B}]$ It can be shown that $P_{ij}[\hat{B}]$ are the same as \eqref{eq:Peven2},
\begin{align*}
P_{ij}[\hat{B}] =  \frac{1}{n} \sum_{k=0}^{n} \widehat{X}_{ij}^{k} \ee^{2\pi \ii \frac{m k}{2 n} \frac{\eps^{\al \bt \ga}}2 [F_{\bt \ga}(\vec{r}_{ij}+\vec{\frac 12})+F_{\bt \ga}(\vec{r}_{ij}-\vec{\frac 12})]},
\end{align*}
except that in the definition \eqref{F} of the flux $F$, $\dd a$ is replaced by $\hat{B}+\dd a$:
\begin{align}
F_{\bt\ga}(\vec{r})&:= \<(\hat{B}+\dd a)^{\Z_n},(\vec{0},\hat{\bt},\hat{\bt}+\hat{\ga})_{\vec{r}-\frac{\hat{\bt}}{2}-\frac{\hat{\ga}}{2}}\> - (\bt\lra\ga)
\end{align}

\subsection{Geometric interpretation of wavefunction}
In \eqref{phi3evennzB}, the background gauge field $\hat{B}$ is coupled to $a$ through the extra terms 
\begin{align}
\frac{m}{2n}(a\hat{B}+\hat{B}a)\label{aBBa}
\end{align}
Geometrically, in 3d space, the 2-cocycle gauge field $\hat{B}$ is dual to a 1d line $\tilde{B}$. we may shift these lines in the $\pm\vec{\frac{1}{2}}$ directions to obtain $\tilde{B}_{\pm}$. Then the extra terms \eqref{aBBa} contributes a phase $\ee^{2\pi \ii \frac{m}{2n}}$ to every signed intersections between $\tilde{B}_{\pm}$ and $\tilde{a}$ (Recall $\tilde{a}$ is the surface dual to $a$).

For simplicity let's pretend $\tilde{a}$ will not fluctuate too wildly near the intersection, and so $\tilde{B}_{\pm}$ gives the same number of signed intersections as $\tilde{B}$, then the extra terms contribute a phase $\ee^{2\pi \ii \frac{m}{n}}$ for every such intersection.
\begin{align*}
\int_{\cM^3}\frac{m}{2n}(a\hat{B}+\hat{B}a)\approx\frac{m}{n}\times \sum
\begin{pmatrix}
&\text{signed intersections}\\
&\text{between $\tilde{B}$ and $\tilde{a}$}
\end{pmatrix}.
\end{align*}

We may interpret such phase as a charge attachment to $\tilde{B}$. In 1-SPT, charged objects are 1-dimensional: a charge $k$ line pick up a phase $\ee^{2\pi \ii \frac{k}{n}}$ for every intersection with a unit-shift(\ie acting by the generator of $\Z_n$) 1-symmetry membrane operator. Charge lines live on the original lattice.

Thus in a $\Z_n$-1-SPT labeled by $m$, consider the 1-symmetry transformation $\ket{\{a\}}\ra \ket{\{a+\al\}}=\ket{\{a'\}}$, where $\al$ is a unit-shift acting on a membrane intersecting $\tilde{B}$ once. We have(as in \eqref{bdstate0})
\begin{align*}
\ket{\psi_0[\hat{B}]}&=\frac{1}{N_{\psi}}\sum_{\{a\}}\ee^{2\pi i \int_{\cM^3} \phi_3[a,\hat{B}]-\phi_3[0,\hat{B}]} \ket{\{a\}}\nn
&\ra \frac{1}{N_{\psi}}\sum_{\{a'\}}\ee^{2\pi i \int_{\cM^3} \phi_3[a',\hat{B}]-\phi_3[0,\hat{B}]-\del_{\al} \phi_3[a,\hat{B}]} \ket{\{a'\}}\nn
&=\ee^{2\pi i \int_{\cM^3}\frac{-m}{n}\hat{B}\al}\ket{\psi_0[\hat{B}]}.
\end{align*}
using \eqref{delphi3alnzB} and assuming $\p\cM^3=\emptyset$. Hence the ground state wavefunction picks up a phase $ \ee^{-2\pi \ii \frac{m}{n}} $ due to the background gauge field. Thus the dual of the background gauge fields $\tilde{B}$(with unit gauge strength) is attached a charge $-m$ line(located at $\tilde{B}_+$, to be exact).

\begin{figure}
\centering
\includegraphics{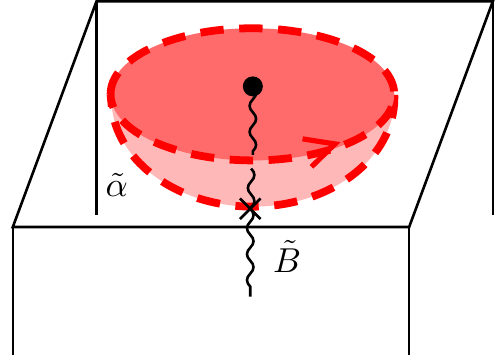}
\caption{A non-zero background gauge field $\hat{B}$, whose dual $\tilde{B}$ is depicted as the wiggly line in the figure, intersects the 1-symmetry membrane $\al$ acting in the bulk(whose dual $\tilde{\al}$ is depicted as dotted red lines). The intersection in the bulk is denoted by a cross $\times$. From the boundary, the endpoint of $\tilde{B}$(depicted as a black dot) is enclosed in a region $h^{\Z_n}=1$(shaded in pink), where $\al=\dd h$, and $h=0$ outside the pink region. The line $\tilde{B}$ with unit gauge strength acquires a phase $\ee^{-2\pi \ii \frac{m}{n}}$ under a unit shift 1-symmetry $\al$.}
\label{bulkbddnzB}
\end{figure}

\subsection{Boundary perspective}

We can alternatively consider the effect of background gauge field from a boundary perspective. Consider the arrangement depicted in \Fig{bulkbddnzB}. In the bulk, the unit strength background gauge $\hat{B}$ intersects the 1-symmetry $\al$ once. On the boundary, the 0d endpoint of background gauge field is enclosed in a region where $h^{\Z_n}=1$, and the endpoint is far away from the 1-symmetry operator in the boundary (so $\dd h=0$ near the end point). From \eqref{phi2evennzB}, under this 1-symmetry there is an extra term 
\begin{align}
&\int_{\p\cM^3}\phi_2[a_{\p},h,\hat{B}]-\int_{\p\cM^3}\phi_2[a_{\p},h,0]\nn
&=\int_{\p\cM^3}-\frac{m}{n}\hat{B}h = -\frac{m}{n}\label{nzBextraphi2}
\end{align}
contributed to the boundary transformation \eqref{bdstate}, compared to the case without background gauge fields. The boundary state hence acquires a phase $\ee^{-2\pi \ii \frac{m}{n}}$ due to the background gauge field. \ie the endpoint of $\tilde{B}$ has charge $-m$ under the boundary 1-symmetry.

We observe that the above boundary argument extends to the odd $m$ case as well. \eqref{nzBextraphi2} still holds by inspecting \eqref{phi2nzB} and again assuming $\dd h=0$ near the end point where $\hat{B}\neq0$. So we expect the same charge attachment also occurs for odd $m$.

\section{Conclusions}

In this paper we studied the $\Z_n$-1-symmetry protected topological states
in 3+1-dimensions, which is labeled by $m\in \{0,1,\cdots,2n-1\}$.  The
$\Z_n$-1-symmetry is generated by closed membrane operators.  We presented an
exactly solvable Hamiltonian which commutes with the closed membrane operators,
and wrote down the ground state wavefunction.  We also studied the effective
boundary theory in 2+1-dimensions.  The effective boundary theory has an
anomalous $\Z_n$-1-symmetry generated by closed string operators.  We showed
that those boundary string operators create topological excitations at the
string ends, which may have non-trivial self-statistics. In particular for the
$n=2$ case, they have self-semionic (for $m=1$) or fermionic statistics (for
$m=2$). In these cases we can gap out the boundary with an engineered boundary
Hamiltonian \emph{with the anomalous $\Z_n$-1-symmetry}, which gives the same ground
state as the toric code model (for $m=2$) and double-semion model (for $m=1$)
on the boundary. We interpreted the wavefunction amplitudes of the bulk
grounds states as linking numbers of strings in the dual lattice. Finally we extend to the case of non-zero background gauge field and find the lines dual to the background gauge field is attached with line charge $-m$.

In the future, we would like to study the nature of the gapless boundary
states.  It is also interesting to see whether other knot invariants can be
derived from the wavefunction amplitude for other 1-SPT's.

LT thanks Yuan-Ming Lu and Juven Wang for helpful discussions. LT is supported by the Croucher
Fellowship for Postdoctoral Research.  XGW is partially
supported by NSF Grant No.  DMS-1664412 and by the Simons Collaboration on
Ultra-Quantum Matter, which is a grant from the Simons Foundation (651440)

\appendix
\allowdisplaybreaks

\section{Space-time complex, cochains, and cocycles} 

\label{cochain}

In this paper, we consider models defined on a spacetime lattice.  A
spacetime lattice is a triangulation of the $D$-dimensional spacetime $M^D$,
which is denoted by $\cM^D$.  We will also call the triangulation $\cM^D$ as a
spacetime complex, which is formed by simplices -- the vertices, links,
triangles, \etc.  We will use $i,j,\cdots$ to label vertices of the spacetime
complex.  The links of the complex (the 1-simplices) will be labeled by
$(i,j),(j,k),\cdots$.  Similarly, the triangles of the complex  (the
2-simplices)  will be labeled by $(i,j,k),(j,k,l),\cdots$.

\begin{figure}[t]
\begin{center}
\includegraphics[scale=0.5]{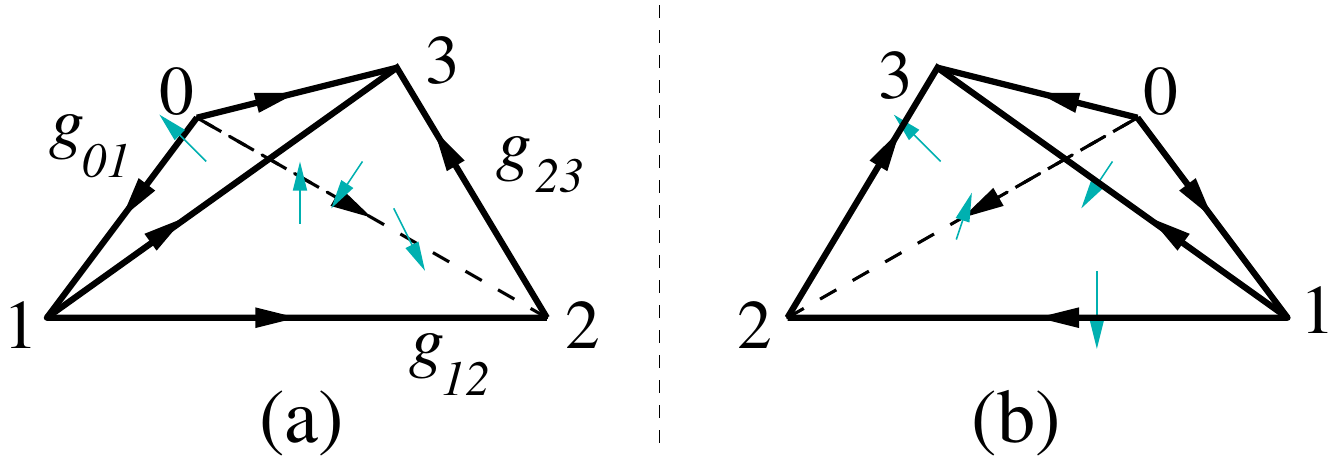} 
\end{center}
\caption{ (Color online) Two branched simplices with opposite orientations.
(a) A branched simplex with positive orientation and (b) a branched simplex
with negative orientation.  }
\label{mir}
\end{figure}

In order to define a generic lattice theory on the spacetime complex
$\cM^D$ using local Lagrangian term on each simplex, it is important
to give the vertices of each simplex a local order.  A nice local scheme to
order  the vertices is given by a branching
structure.\cite{C0527,CGL1172,CGL1204} A branching structure is a choice of
orientation of each link in the $d$-dimensional complex so that there is no
oriented loop on any triangle (see Fig. \ref{mir}).

The branching structure induces a \emph{local order} of the vertices on each
simplex.  The first vertex of a simplex is the vertex with no incoming links,
and the second vertex is the vertex with only one incoming link, \etc.  So the
simplex in  Fig. \ref{mir}a has the following vertex ordering: $0,1,2,3$.

The branching structure also gives the simplex (and its sub-simplices) a
canonical orientation.  Fig. \ref{mir} illustrates two $3$-simplices with
opposite canonical orientations compared with the 3-dimension space in which
they are embedded.  The blue arrows indicate the canonical orientations of the
$2$-simplices.  The black arrows indicate the canonical orientations of the
$1$-simplices.

Given an Abelian group $(\M, +)$, an $n$-cochain $f_n$ is an assignment of
values in $\M$ to each $n$-simplex, for example a value $f_{n;i,j,\cdots,k}\in
\M$ is assigned to $n$-simplex $(i,j,\cdots,k)$.  So \emph{a cochain $f_n$ can
be viewed as a bosonic field on the spacetime lattice}. 

$\M$ can also be viewed a $\Z$-module (\ie a vector space with
integer coefficient) that also allows scaling by an integer:  
\begin{align}
	 x+y &= z,\ \ \ \ x*y=z, \ \ \ \ mx=y,
	\nonumber\\
	x,y,z & \in \M,\ \ \ m \in \Z.
\end{align}
The direct sum of two modules
$\M_1\oplus \M_2$ (as vector spaces) is equal to the direct product of the two
modules (as sets):
\begin{align}
 \M_1\oplus \M_2 \stackrel{\text{as set}}{=} \M_1\times \M_2
\end{align}

\begin{figure}[tb]
\begin{center}
\includegraphics[scale=0.5]{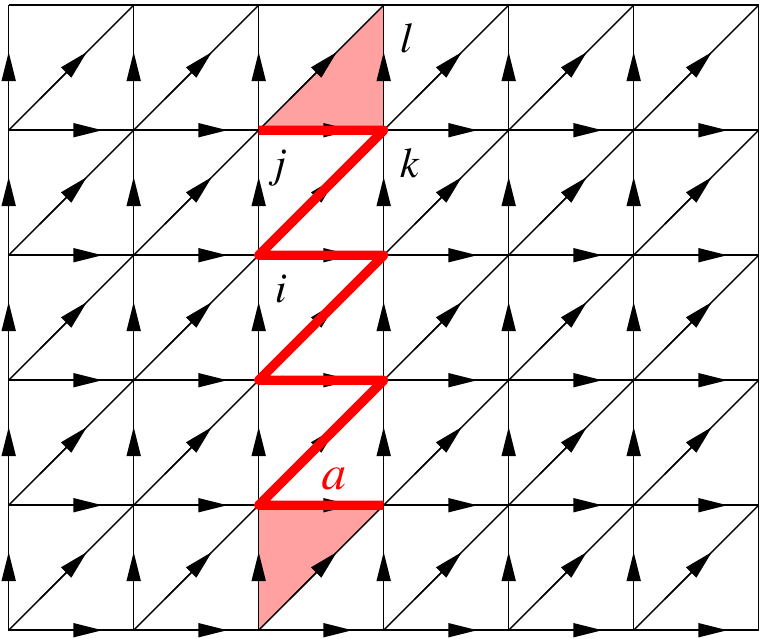} \end{center}
\caption{ (Color online)
A 1-cochain $a$ has a value $1$ on the red links: $ a_{ik}=a_{jk}= 1$ and a
value $0$ on other links: $ a_{ij}=a_{kl}=0 $.  $\dd a$ is non-zero on the
shaded triangles: $(\dd a)_{jkl} = a_{jk} + a_{kl} - a_{jl}$.  For such
1-cohain, we also have $a\smile a=0$.  So when viewed as a $\Z_2$-valued cochain,
$\Bs_2 a \neq a\smile a$ mod 2.
}
\label{dcochain}
\end{figure}

We like to remark that a simplex $(i,j,\cdots,k)$ can have two different
orientations. We can use $(i,j,\cdots,k)$ and $(j,i,\cdots,k)=-(i,j,\cdots,k)$
to denote the same simplex with opposite orientations.  The value
$f_{n;i,j,\cdots,k}$ assigned to the simplex with opposite  orientations should
differ by a sign: $f_{n;i,j,\cdots,k}=-f_{n;j,i,\cdots,k}$.  So to be more
precise $f_n$ is a linear map $f_n: n\text{-simplex} \to \M$. We can denote the
linear map as $\<f_n, n\text{-simplex}\>$, or
\begin{align}
 \<f_n, (i,j,\cdots,k)\> = f_{n;i,j,\cdots,k} \in \M.
\end{align}
More generally, a \emph{cochain} $f_n$ is a linear map
of $n$-chains:
\begin{align}
	f_n:  n\text{-chains} \to \M,
\end{align}
or (see Fig. \ref{dcochain})
\begin{align}
 \<f_n, n\text{-chain}\> \in \M,
\end{align}
where a \emph{chain} is a composition of simplices. For example, a 2-chain can
be a 2-simplex: $(i,j,k)$, a sum of two 2-simplices: $(i,j,k)+(j,k,l)$, a more
general composition of 2-simplices: $(i,j,k)-2(j,k,l)$, \etc.  The map $f_n$ is
linear respect to such a composition.  For example, if a chain is $m$ copies of
a simplex, then its assigned value will be $m$ times that of the simplex.
$m=-1$ correspond to an opposite orientation.  

We will use $C^n(\cM^D;\M)$ to denote the set of all
$n$-cochains on $\cM^D$.  $C^n(\cM^D;\M)$ can also be
viewed as a set all $\M$-valued fields (or paths) on  $\cM^D$.  Note
that $C^n(\cM^D;\M)$ is an Abelian group under the $+$-operation.

The total spacetime lattice $\cM^D$ correspond to a $D$-chain.  We
will use the same $\cM^D$ to denote it.  Viewing $f_D$ as a linear
map of $D$-chains, we can define an ``integral'' over $\cM^D$:
\begin{align}
 \int_{\cM^D} f_D &\equiv \<f_D,\cM^D\>
\\
&=\sum_{(i_0,i_1,\cdots,i_D)}
s_{i_0i_1\cdots i_D} (f_D)_{i_0,i_1,\cdots,i_D}
.
\nonumber 
\end{align}
Here $s_{i_0i_1\cdots i_D}=\pm 1$, such that a $D$-simplex in the $D$-chain
$\cM^D$ is given by $s_{i_0i_1\cdots i_D} (i_0,i_1,\cdots,i_D)$.

\begin{figure}[tb]
\begin{center}
\includegraphics[scale=0.5]{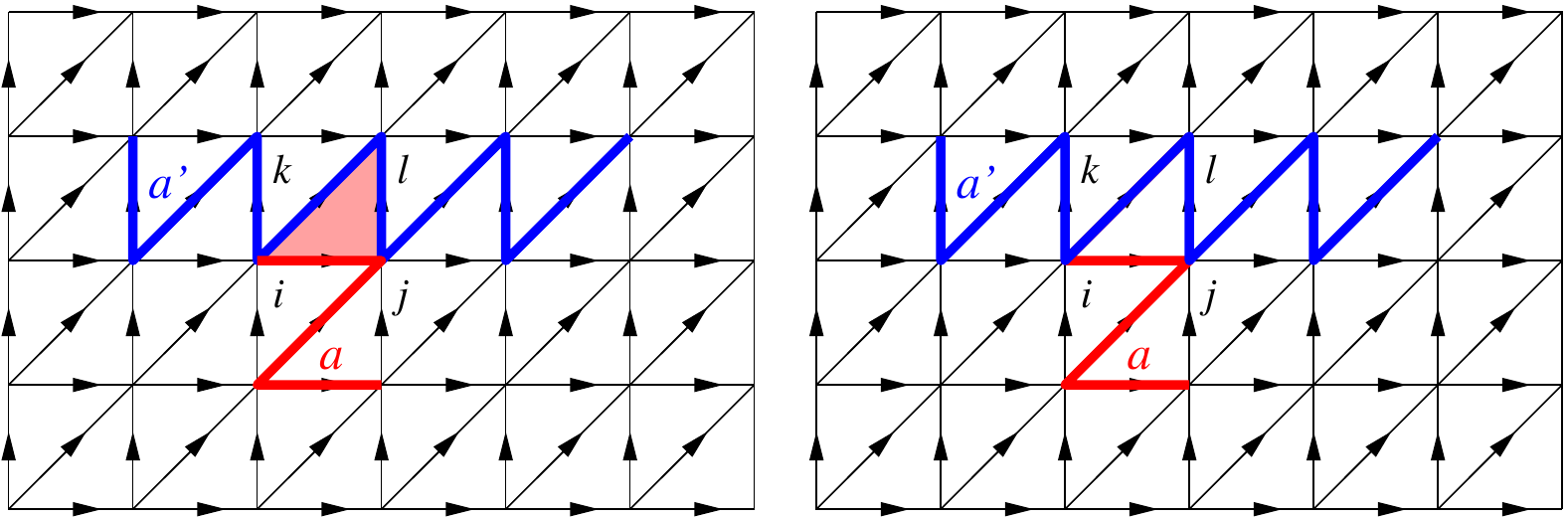} \end{center}
\caption{ (Color online)
A 1-cochain $a$ has a value $1$ on the red links, Another
1-cochain $a'$ has a value $1$ on the blue links.
On the left, $a\smile a'$ is non-zero on the shade triangles:
$(a\smile a')_{ijl}=a_{ij}a'_{jl}=1$.
On the right, $a'\smile a$ is zero on every triangle.
Thus $a\smile a'+a'\smile a$ is not a coboundary.
}
\label{cupcom}
\end{figure}

We can define a derivative operator $\dd$ acting on an $n$-cochain $f_n$, which
give us an $(n+1)$-cochain (see Fig. \ref{dcochain}):
\begin{align} 
\label{eq:differential}
&\ \ \ \ \<\dd f_n, (i_0i_1i_2\cdots i_{n+1})\>
\nonumber\\
&=\sum_{m=0}^{n+1} (-)^m 
\<f_n, (i_0i_1i_2\cdots\hat i_m\cdots i_{n+1})\>
\end{align}
where $i_0i_1i_2\cdots \hat i_m \cdots i_{n+1}$ is the sequence
$i_0 i_1 i_2 \cdots i_{n+1}$ with $i_m$ removed, and
$i_0, i_1,i_2 \cdots i_{n+1}$ are the ordered vertices of the $(n+1)$-simplex
$(i_0 i_1 i_2 \cdots i_{n+1})$.

A cochain $f_n \in C^n(\cM^D;\M)$ is called a \emph{cocycle} if $\dd
f_n=0$.   The set of cocycles is denoted by $Z^n(\cM^D;\M)$.  A
cochain $f_n$ is called a \emph{coboundary} if there exist a cochain $f_{n-1}$
such that $\dd f_{n-1}=f_n$.  The set of coboundaries is denoted by
$B^n(\cM^D;\M)$.  Both $Z^n(\cM^D;\M)$ and
$B^n(\cM^D;\M)$ are Abelian groups as well.  Since $\dd^2=0$, a
coboundary is always a cocycle: $B^n(\cM^D;\M) \subset
Z^n(\cM^D;\M)$.  We may view two  cocycles differ by a coboundary as
equivalent.  The equivalence classes of cocycles, $[f_n]$, form the so called
cohomology group denoted by \begin{align} H^n(\cM^D;\M)=
Z^n(\cM^D;\M)/ B^n(\cM^D;\M), \end{align}
$H^n(\cM^D;\M)$, as a group quotient of $Z^n(\cM^D;\M)$ by
$B^n(\cM^D;\M)$, is also an Abelian group.

For the $\Z_N$-valued cocycle $x_n$, $\dd x_n \se{N} 0$. Thus 
\begin{align}
\label{BsDef}
 \Bs_N x_n \equiv \frac1N \dd x_n 
\end{align}
is a $\Z$-valued cocycle. Here $\Bs_N$ is Bockstein homomorphism.

We notice the above definition for cochains still makes sense if we
have a non-Abelian group $(G, \cdot)$ instead of an Abelian group $(\M,
+)$, however the differential $\dd$ defined by \eqn{eq:differential}
will not satisfy $\dd \circ \dd = 1$, except for the first two
$\dd$'s. That is, one may still make sense of 0-cocycle and 1-cocycle,
but no more further naively by formula
\eqn{eq:differential}. For us, we only use non-Abelian 1-cocycle in
this article. Thus it is ok.  Non-Abelian cohomology is then thoroughly
studied in mathematics motivating concepts such as gerbes to enter.

From two cochains $f_m$  and $h_n$, we can construct a third cochain
$p_{m+n}$ via the cup product (see Fig. \ref{cupcom}):
\begin{align}
p_{m+n} &= f_m \smile h_n ,
\nonumber\\
\<p_{m+n}, (0 \to {m+n})\> 
&= 
\<f_m, (0 \to m)\> \times
\nonumber\\
&\ \ \ \ 
\<h_n,(m \to {m+n}) \>,
\end{align}
where $i\to j$ is the consecutive sequence from $i$ to $j$: 
\begin{align}
i\to j\equiv i,i+1,\cdots,j-1,j. 
\end{align}
Note that the above definition applies to cochains with global.

The cup product has the following property 
\begin{align}
\label{cupprop}
 \dd(h_n \smile f_m) &= (\dd h_n) \smile f_m + (-)^n h_n \smile (\dd f_m) 
\end{align}
for  cochains with global or local values.  
We see that $h_n \smile f_m $ is a
cocycle if both $f_m$ and $h_n$ are cocycles.  If both $f_m$ and $h_n$ are
cocycles, then $f_m \smile h_n$ is a coboundary if one of $f_m$ and $h_n$ is a
coboundary.  So the cup product is also an operation on cohomology groups
$\hcup{} : H^m(M^D;\M)\times H^n(M^D;\M) \to H^{m+n}(M^D;\M)$.  The cup product
of two \emph{cocycles} has the following property (see Fig. \ref{cupcom}) 
\begin{align}
 f_m \smile h_n &= (-)^{mn} h_n \smile f_m + \text{coboundary}
\end{align}

We can also define higher cup product $f_m \hcup{k} h_n$ which gives rise to a
$(m+n-k)$-cochain \cite{S4790}:
\begin{align}
\label{hcupdef}
&\ \ \ \
 \<f_m \hcup{k} h_n, (0,1,\cdots,m+n-k)\> 
\nonumber\\
&
 = 
\hskip -1em 
\sum_{0\leq i_0<\cdots< i_k \leq n+m-k} 
\hskip -3em  
(-)^p
\<f_m,(0 \to i_0, i_1\to i_2, \cdots)\>\times
\nonumber\\
&
\ \ \ \ \ \ \ \ \ \
\ \ \ \ \ \ \ \ \ \
\<h_n,(i_0\to i_1, i_2\to i_3, \cdots)\>,
\end{align} 
and $f_m \hcup{k} h_n =0$ for  $k<0$ or for $k>m \text{ or } n$.  Here $i\to j$
is the sequence $i,i+1,\cdots,j-1,j$, and $p$ is the number of permutations to
bring the sequence
\begin{align}
 0 \to i_0, i_1\to i_2, \cdots; i_0+1\to i_1-1, i_2+1\to i_3-1,\cdots
\end{align}
to the sequence
\begin{align}
 0 \to m+n-k.
\end{align}
For example
\begin{align}
&
 \<f_m \hcup1 h_n, (0\to m+n-1)\> 
 = \sum_{i=0}^{m-1} (-)^{(m-i)(n+1)}\times
\nonumber\\
&
\<f_m,(0 \to i, i+n\to m+n-1)\>
\<h_n,(i\to i+n)\>.
\end{align} 
We can see that $\hcup0 =\smile$.  
Unlike cup product at $k=0$, the higher cup product of two
cocycles may not be a cocycle. For cochains $f_m, h_n$, we have
\begin{align}
\label{cupkrel}
& \dd( f_m \hcup{k} h_n)=
\dd f_m \hcup{k} h_n +
(-)^{m} f_m \hcup{k} \dd h_n+
\\
& \ \ \
(-)^{m+n-k} f_m \hcup{k-1} h_n +
(-)^{mn+m+n} h_n \hcup{k-1} f_m 
\nonumber 
\end{align}

Let $f_m$ and $h_n$ be cocycles and $c_l$ be a chain, from \eqn{cupkrel} we
can obtain
\begin{align}
\label{cupkrel1}
 & \dd (f_m \hcup{k} h_n) = (-)^{m+n-k} f_m \hcup{k-1} h_n 
\nonumber\\
&
\ \ \ \ \ \ \ \ \ \
 \ \ \ \ \ \ \
+ (-)^{mn+m+n}  h_n \hcup{k-1} f_m,
\nonumber\\
 & \dd (f_m \hcup{k} f_m) = [(-)^k+(-)^m] f_m \hcup{k-1} f_m,
\nonumber\\
& \dd (c_l\hcup{k-1} c_l + c_l\hcup{k} \dd c_l)
= \dd c_l\hcup{k} \dd c_l 
\nonumber\\
&\ \ \ -[(-)^k-(-)^l]
(c_l\hcup{k-2} c_l + c_l\hcup{k-1} \dd c_l) .
\end{align}

From \eqn{cupkrel1}, we see that, for $\Z_2$-valued cocycles $z_n$,
\begin{align}
 \Sq^{n-k}(z_n) \equiv z_n\hcup{k} z_n
\end{align}
is always a cocycle.  Here $\Sq$ is called the Steenrod square.  More generally
$h_n \hcup{k} h_n$ is a cocycle if $n+k =$ odd and $h_n$ is a cocycle.
Usually, the Steenrod square is defined only for $\Z_2$-valued cocycles or
cohomology classes.  Here, we like to define a generalized
Steenrod square for $\M$-valued
cochains $c_n$:
\begin{align}
\label{Sqdef}
 \gSq^{n-k} c_n \equiv c_n\hcup{k} c_n +  c_n\hcup{k+1} \dd c_n .
\end{align}
From \eqn{cupkrel1}, we see that
\begin{align}
\label{Sqd1}
 \dd \gSq^{k} c_n &= \dd(
c_n\hcup{n-k} c_n +  c_n\hcup{n-k+1} \dd c_n )
\\
&= \gSq^k \dd c_n +(-)^{n}
\begin{cases}
0, & k=\text{odd} \\ 
2  \gSq^{k+1} c_n  & k=\text{even} \\ 
\end{cases}
.
\nonumber 
\end{align}
In particular, when $c_n$ is a $\Z_2$-valued cochain, we have
\begin{align}
\label{Sqd}
  \dd \gSq^{k} c_n \se{2} \gSq^k \dd c_n.
\end{align}

Next, let us consider the action of $\gSq^k$ on the sum of two
 $\M$-valued cochains $c_n$ and $c_n'$:
\begin{align}
& \gSq^{k} (c_n+c_n')
 = \gSq^{k} c_n + \gSq^k c_n' +
\nonumber\\
&\ \ \
 c_n \hcup{n-k} c_n' + c_n' \hcup{n-k} c_n 
+ c_n \hcup{n-k+1} \dd c_n' + c_n' \hcup{n-k+1} \dd c_n 
\nonumber\\
&=\gSq^{k} c_n + \gSq^k c_n' 
+[1 + (-)^k]c_n \hcup{n-k} c_n'
\nonumber\\
&\ \ \
-(-)^{n-k} [ - (-)^{n-k} c_n' \hcup{n-k} c_n + (-)^n c_n \hcup{n-k} c_n']
\nonumber\\
&\ \ \
+ c_n \hcup{n-k+1} \dd c_n' + c_n' \hcup{n-k+1} \dd c_n
\end{align}
Notice that (see \eqn{cupkrel})
\begin{align}
&\ \ \ \
- (-)^{n-k} c_n' \hcup{n-k} c_n + (-)^n c_n \hcup{n-k} c_n' 
\\
&= \dd(c_n'\hcup{n-k+1}c_n) 
- \dd c_n' \hcup{n-k+1} c_n 
-(-)^n c_n' \hcup{n-k+1} \dd c_n ,
\nonumber 
\end{align}
we see that
\begin{align}
& \gSq^{k} (c_n+c_n')
 = 
\gSq^{k} c_n + \gSq^k c_n' 
+[1 + (-)^k]c_n \hcup{n-k} c_n'
\nonumber\\
&
+(-)^{n-k} [ \dd c_n' \hcup{n-k+1} c_n +(-)^n c_n' \hcup{n-k+1} \dd c_n
]
\nonumber\\
&
-(-)^{n-k} 
\dd (c_n'\hcup{n-k+1}c_n) 
+c_n \hcup{n-k+1} \dd c_n'+ c_n' \hcup{n-k+1} \dd c_n
\nonumber\\
&=
\gSq^{k} c_n + \gSq^k c_n'  
+[1 + (-)^k]c_n \hcup{n-k} c_n'
\nonumber \\
&\  \ \
+[1+(-)^{k}]c_n' \hcup{n-k+1} \dd c_n 
-(-)^{n-k} \dd (c_n'\hcup{n-k+1}c_n)
\nonumber\\
&\ \ \
-[(-)^{n-k+1}\dd c_n' \hcup{n-k+1} c_n
- c_n \hcup{n-k+1} \dd c_n'] .
\end{align}
Notice that (see \eqn{cupkrel})
\begin{align}
&\ \ \ \ 
 (-)^{n-k+1}\dd c_n' \hcup{n-k+1} c_n - c_n \hcup{n-k+1} \dd c_n'
\nonumber\\
&= \dd(\dd c_n' \hcup{n-k+2} c_n) +(-)^n \dd c_n' \hcup{n-k+2} \dd c_n  ,
\end{align}
we find
\begin{align}
& \gSq^{k} (c_n+c_n')
=
\gSq^{k} c_n + \gSq^k c_n'  
+[1 + (-)^k]c_n \hcup{n-k} c_n'
\nonumber \\
&\  \ \
+[1+(-)^{k}]c_n' \hcup{n-k+1} \dd c_n 
-(-)^{n-k} \dd (c_n'\hcup{n-k+1}c_n)
\nonumber\\
&\ \ \
-\dd (\dd c_n'\hcup{n-k+2} c_n )
-(-)^{n} \dd c_n'\hcup{n-k+2} \dd c_n 
\nonumber\\
&=
\gSq^{k} c_n + \gSq^k c_n'  
-(-)^{n} \dd c_n'\hcup{n-k+2} \dd c_n 
\nonumber \\
&\ \ \
+[1+(-)^{k}][c_n \hcup{n-k} c_n'+ c_n' \hcup{n-k+1} \dd c_n] 
\nonumber\\
&\ \ \
-(-)^{n-k} \dd (c_n'\hcup{n-k+1}c_n)
-\dd (\dd c_n'\hcup{n-k+2} c_n )
.
\label{Sqplus1}
\end{align}
We see that, if one of the $c_n$ and $c_n'$ is a cocycle,
\begin{align}
\label{Sqplus}
  \gSq^{k} (c_n+c_n') \se{2,\dd} \gSq^{k} c_n + \gSq^k c_n' .
\end{align}
We also see that
\begin{align}
\label{Sqgauge}
&\ \ \ \
 \gSq^{k} (c_n+\dd f_{n-1})
\\
& = \gSq^{k} c_n + \gSq^k \dd f_{n-1} +
[1+(-)^k] \dd f_{n-1}\hcup{n-k} c_n
\nonumber\\
&\ \ \
-(-)^{n-k} \dd (c_n\hcup{n-k+1}\dd f_{n-1})
-\dd (\dd c_n\hcup{n-k+2} \dd f_{n-1} )
\nonumber\\
& = \gSq^{k} c_n 
+ [1+(-)^k] [\dd f_{n-1}\hcup{n-k} c_n +(-)^n \gSq^{k+1}f_{n-1}]
\nonumber\\
&
+\dd [\gSq^k  f_{n-1}
-(-)^{n-k} c_n \hskip -0.5em \hcup{n-k+1} \hskip -0.5em \dd f_{n-1}
-\dd c_n \hskip -0.5em \hcup{n-k+2}  \hskip -0.5em \dd f_{n-1} ]
.
\nonumber 
\end{align}

Using \eqn{Sqplus1}, we can also obtain the following result
if $\dd c_n =  \text{even}$
\begin{align}
\label{Sqplus2}
& \ \ \ \
 \gSq^k (c_n+2c_n')
\nonumber\\
& \se{4} \gSq^k c_n+2 \dd (c_n\hcup{n-k+1} c_n') +2 \dd c_n\hcup{n-k+1} c_n'
\nonumber\\
& \se{4} \gSq^k c_n+2 \dd (c_n\hcup{n-k+1} c_n') 
\end{align}

As another application, we note that, for a $\Q$-valued cochain $m_d$ and using
\eqn{cupkrel},
\begin{align}
\label{Sq1Bs}
& \gSq^1(m_{d}) = m_{d}\hcup{d-1} m_{d} + m_{d}\hcup{d} \dd m_{d}
\nonumber\\
&=\frac12 (-)^{d} 
[\dd (m_{d}\hcup{d} m_{d}) 
-\dd m_{d} \hcup{d} m_{d}] 
+\frac12  m_{d} \hcup{d} \dd m_{d} 
\nonumber\\
&=
(-)^{d} \Bs_2 (m_{d}\hcup{d} m_{d}) -(-)^d \Bs_2 m_{d} \hcup{d} m_{d}
+  m_{d} \hcup{d} \Bs_2 m_{d}
\nonumber\\
&=
(-)^{d} \Bs_2  \gSq^0 m_{d} 
-2 (-)^d \Bs_2 m_{d} \hcup{d+1} \Bs_2 m_{d}
\nonumber\\
&=
(-)^{d} \Bs_2 \gSq^0 m_{d} 
-2 (-)^d \gSq^0 \Bs_2 m_{d} 
\end{align}
This way, we obtain a relation between Steenrod square and Bockstein
homomorphism, when $m_d$ is a $\Z_2$-valued cochain
\begin{align}
\label{Sq1Bs2}
  \gSq^1(m_{d}) \se{2} \Bs_2 m_{d} ,
\end{align}
where we have used $\gSq^0 m_{d}= m_d$ for $\Z_2$-valued cochain.

For a $k$-cochain $a_k$, $k=\text{odd}$, we find that
\begin{align}
&\ \ \ \
 \gSq^k a_k = a_ka_k+a_k\hcup{1}\dd a_k
\\
&=
\frac12 [\dd a_k \hcup{1} a_k -a_k \hcup{1}\dd a_k -\dd(a_k\hcup{1}a_k) ]
+a_k\hcup{1}\dd a_k
\nonumber\\
&= \frac12 [\dd a_k \hcup{2}\dd a_k-\dd (\dd a_k\hcup{2} a_k)] 
-\frac12 \dd(a_k\hcup{1}a_k)
\nonumber\\
&= \frac14 \dd (\dd a_k \hcup{3}\dd a_k) 
-\frac12 \dd(a_k\hcup{1}a_k+\dd a_k\hcup{2} a_k)
\nonumber 
\end{align}
Thus $\gSq^k a_k$ is always a $\Q$-valued coboundary, when $k$ is odd.

\section{Procedure for deriving Hamiltonian from topological partition function}\label{apdx:h}

We briefly review the procedure for writing down local commuting projection Hamiltonians from the topological action. The reader may refer to Ref. \cite{Mesaros2013,CGL1172} for details.

\subsection{Zero background gauge field case}

Suppose $\cM^4 = \cM^3 \times I$ for some closed 3-manifold $\cM^3$ and $I$ is an interval parameterized by $t\in[0,T]$, to be regarded as the time direction. The space-time has boundaries at $t = 0,T$, where the field configurations are given by $\{a_0\}$ and $\{a_T\}$. The transfer matrix is given by
\begin{align}
&\bra{\{a_T\}}\ee^{-T \hat{H}^{\infty}}\ket{\{a_0\}}=Z^\text{top}[\{a_T\},\{a_0\}]\\
&Z^\text{top}_{\cM^3 \times I}[\{a_T\},\{a_0\}]\\
&=\frac{1}{n^{N_{l,int}+(N_{l,0}+N_{l,T})/2}} \sum_{\{a_{int}\}} \ee^{2\pi \ii \int_{\cM^3 \times I} \omega_4} ,\label{eth}
\end{align}
where $\int_{\cM^3 \times I} \omega_4$ is evaluated with link configurations at its boundaries fixed to be $\{a_0\}$, $\{a_T\}$. Links not living on the boundary are called internal links. Their configuration is given by $\{a_{int}\}$. $N_{l,0}$, $N_{l,T}$ and $N_{l,int}$ are the number of links at the two boundaries and in the space-time bulk respectively. In the following we assume the two boundaries have the same triangulation so $N_{l,0}=N_{l,T}=N_{l,\cM^3}$.

We may represent the transfer matrix diagrammatically as a spacetime cylinder
\begin{eqnarray*}
\includegraphics{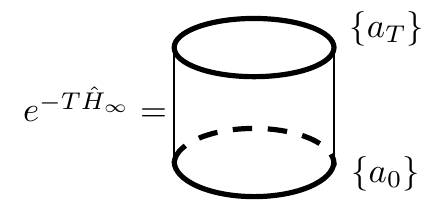}
\end{eqnarray*}
where the top and bottom ellipses represent the spatial closed manifold $\cM^3$ at $t=T,0$ respectively. They are the boundaries of the space-time cylinder and are drawn as bold lines. Note that although $\cM^3$ is a three-dimensional manifold, we draw it as a one-dimensional ellipse.

Recall from Ref.\cite{Mesaros2013,CGL1172} that under a local spacetime re-triangulation, the topological action $\int_{\cM^4} \omega_4$ changes by $\dd \om_4$. Hence the cocycle condition $\dd \om_4 \se{1} 0$ implies the action is invariant under re-triangulation mod 1. Moreover, during a re-triangulation, the boundary degrees of freedom cannot change, thus we can only conclude that the value of $\int_{\cM^3 \times I} \omega_4$ is independent of triangulations of the internal bulk, but it could depend on the boundary triangulation. Furthermore, $\int_{\cM^3 \times I} \omega_4$ is independent of the values of $a_{int}$. This is because during a re-triangulation, the internal link values are forgotten, which can be illustrated with the re-triangulation of a square:
\begin{eqnarray*}
\includegraphics{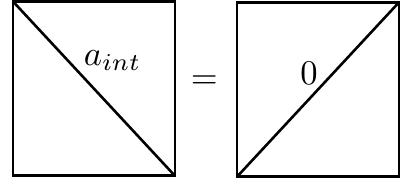}
\end{eqnarray*}

Thus $Z^\text{top}[\{a_T\},\{a_0\}]$ is independent of both the triangulation and field configuration of the internal bulk and only depends on the configuration at its boundaries.

We can show that the transfer matrix is a projection with a computation:
\begin{align*}
&\bra{\{a_{2T}\}}\ee^{-T \hat{H}^{\infty}}\ee^{-T \hat{H}^{\infty}}\ket{\{a_0\}}\nn
&=\sum_{\{a_T\}}Z^\text{top}[\{a_{2T}\},\{a_T\}]Z^\text{top}[\{a_T\},\{a_0\}]\\
&=\sum_{\{a_T,a_{int}\}}\frac{1}{n^{N_{l,int}+2N_{l,\cM^3}}} \ee^{2\pi \ii (\int_{\cM^3 \times [0,T]} \omega_4+\int_{\cM^3 \times [T,2T]} \omega_4)}\nn
&=\sum_{\{a_T,a_{int}\}}\frac{1}{n^{N_{l,int}+2N_{l,\cM^3}}} \ee^{2\pi \ii \int_{\cM^3 \times [0,2T]} \omega_4}\nn
&=\frac{1}{n^{N_{l,int'}+N_{l,\cM^3}}} \sum_{\{a_{int'}\}} \ee^{2\pi \ii \int_{\cM^3 \times [0,2T]} \omega_4}\nn
&=\bra{\{a_{2T}\}}\ee^{-T \hat{H}^{\infty}}\ket{\{a_0\}}
\end{align*}
where the label $int$ includes all the links not on the slices $t=0,T,2T$ and the label $int'$ includes all the links not on the slices $t=0,2T$. This computation can be expressed diagrammatically as
\begin{eqnarray*}
\includegraphics{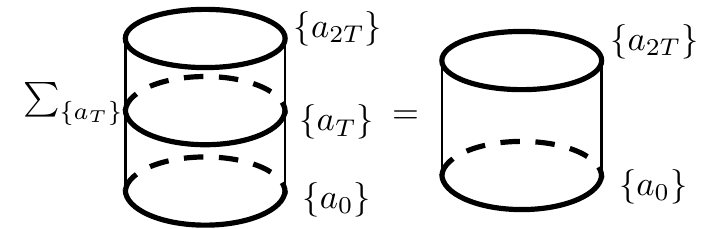}
\end{eqnarray*}

Since the eigenvalues of a projection is 1 or 0, correspondingly $\hat{H}_{\infty}$ has eigenvalues $0$ or $\infty$, \ie an infinite energy gap.

Moreover, the transfer matrix has trace 1. This is because $\Tr[\ee^{-T \hat{H}_{\infty}}]$ is evaluated by identifying the top and bottom link configurations of the cylinder and summing over them. With the two ends identified, $\cM^3\times I =\cM^3\times S^1$ becomes a closed manifold. As we showed in \eqref{Zclosedeq1}, on a closed manifold without any background gauge fields, $\int_{\cM^4}\om_4\se{1}0$. Thus we have
\begin{align}
\Tr[\ee^{-T \hat{H}_{\infty}}]=\frac{1}{n^{N_{l,int}+N_{l,\cM^3}}} \sum_{\{a_{int},a_{0}\}} 1=1\label{trp}
\end{align}
Diagrammatically, this is expressed as
\begin{eqnarray*}
\includegraphics{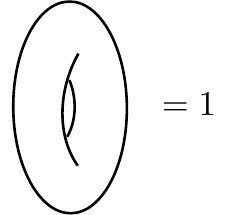}
\end{eqnarray*}
hence the ground state of $\hat{H}_{\infty}$ is unique.

Although the transfer matrix is a non-local operator, it can be decomposed into a product of local operators. Suppose we evaluate $\int_{\cM^3 \times I} \omega_4$ with a triangulation of the internal space-time, such that it consists of $N_{l,\cM^3}+1$ infinitesimal spatial slices, each slice having the same triangulation of the spatial slices at $t=0,T$. Between two adjacent slices, only a single link $ij$ is updated from $a_{0,ij}$ to $a_{T,ij}$,  while all other links remains the same. We have
\begin{align}
\ee^{-T \hat{H}^{\infty}} &= \prod_{ij}P_{ij} \label{pij}\\
\bra{\{a_T\}}P_{ij}\ket{\{a_0\}}&= n^{N_{l,\cM^3}-1} \prod_{i'j'\neq ij} \del_{a_{0,i'j'},a_{T,i'j'}} \times \nn
&\ \ \ \ Z^\text{top}_{\cM^3 \times I}[\{a_T\},\{a_0\}], \label{eq:Pelem}
\end{align}

In diagrams, this means
\begin{eqnarray*}
\includegraphics{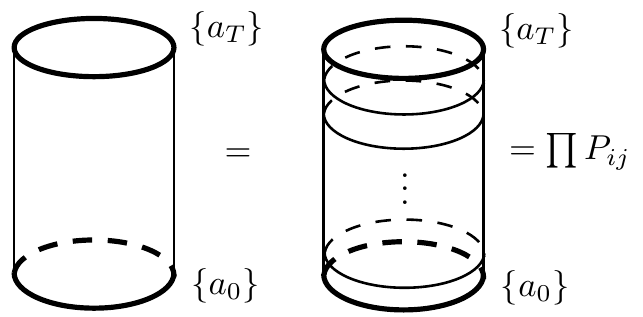}
\end{eqnarray*}
In \eqref{eq:Pelem}, it is not very clear that $P_{ij}$ is a local operator. The locality of $P_{ij}$ can be seen by examining the diagrammatic expression for $P_{ij}$,
\begin{eqnarray*}
\includegraphics{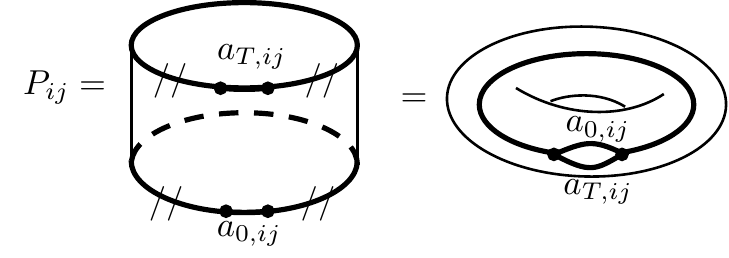}
\end{eqnarray*}
where double slash indicates the region in which field configurations on the top needs to be identified with that on the bottom. On the right hand side we see that $P_{ij}$ is associated with $\cM^3\times S^1$ with a slit at the link $ij$. This means that in $\cM^3$, the links far away from $ij$ become internal links in the non-zero matrix elements of $P_{ij}$, and hence the non-zero matrix elements of $P_{ij}$ are independent of the value of links far away from $ij$. Thus $P_{ij}$ is a local operator.

Using the same arguments as before, it can be shown that $P_{ij}$ is a projection operator with trace $n^{N_{l,\cM^3}-1}$. So each projection by $P_{ij}$ reduces the dimension of the ground state Hilbert space by a factor or $n$. Furthermore, in the following we will show that any two such operators $P_{ij}$, $P_{kl}$ commute. The two orderings $P_{ij}P_{kl}$ or $P_{kl}P_{ij}$ corresponds to triangulations shown below
\begin{eqnarray*}
\includegraphics{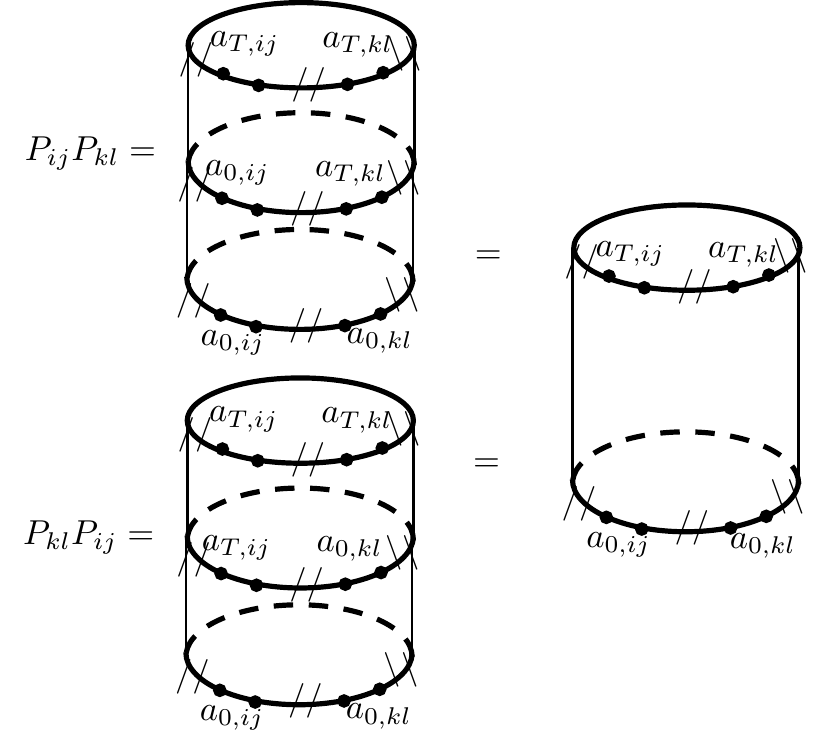}
\end{eqnarray*}
It is readily seen that the two diagrams only differs for the internal links. Thus $P_{ij}P_{kl}=P_{kl}P_{ij}$.

We note that the computation for $P_{ij}$ can be further simplified by setting $a_{int}=0$.
\begin{align}
&\bra{\{a_T\}}P_{ij}\ket{\{a_0\}}\nn
&=\prod_{i'j'\neq ij} \del_{a_{0,i'j'},a_{T,i'j'}}\frac{1}{n}\left. \ee^{2\pi \ii \int_{\cM^3 \times I} \omega_4}\right\rvert_{a_{int}=0}\label{eq:ZtopMI}
\end{align}

The ground state of $\hat{H}^{\infty}$ satisfies $P_{ij}\ket{\psi_0} = \ket{\psi_0}$. We can construct a Hamiltonian with finite gap but the same ground state as $\hat{H}^{\infty}$ by defining
\begin{align}
\hat{H} = - \sum_{ij} P_{ij}. \label{eq:Pij}
\end{align}

\subsection{Non-zero background gauge field case}\label{apdx:hnzB}
Suppose we are given a background gauge field on the spatial manifold $\cM^3$. In order to define the transfer matrix, we need to specify the background gauge field $\hat{B}$ on the spacetime $\cM^3\times I$. We propose that $\hat{B}$ should be \emph{static}, meaning that it should be invariant under time translation, \ie $\hat{B}$ is the same on every spatial slice. This is sensible because a non-static background gauge field actually correspond to the insertion of a 1-symmetry operator into the transfer matrix.

Such static background gauge field $\hat{B}$ on $\cM^3\times I$ can be constructed from a given flat $\hat{B}$ on $\cM^3$ as follows. We triangulate $\cM\times I$ such that any 2-cell $(ijk)$ in $\cM^3\times I$, when projected onto $\cM^3$, is either also a 2-cell $(i_0j_0k_0)$ in $\cM^3$, or a lower dimensional cell. Then we define 
\begin{align*}
\langle \hat{B},(ijk)\rangle :=\begin{cases} 
\langle \hat{B},(i_0j_0k_0)\rangle & \text{if $(ijk)$ projects to a 2-cell}\\
0 & \text{else}
\end{cases}
\end{align*}
it can be checked $\hat{B}=0$.

We then construct the transfer matrix with such static background gauge field. Diagrammatically, the transfer matrix is represented as follows:
\begin{eqnarray*}
\includegraphics{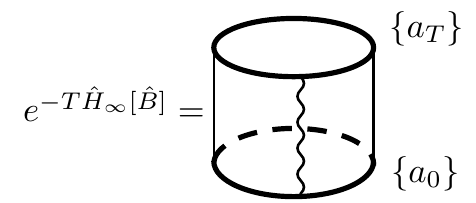}
\end{eqnarray*}
where the wiggly vertical line represents the static $\hat{B}$. We may repeat the same analysis as in the previous subsection, except that we include a wiggly vertical line in the diagrams. For example, in showing the transfer matrix is a projection, we have
\begin{eqnarray*}
\includegraphics{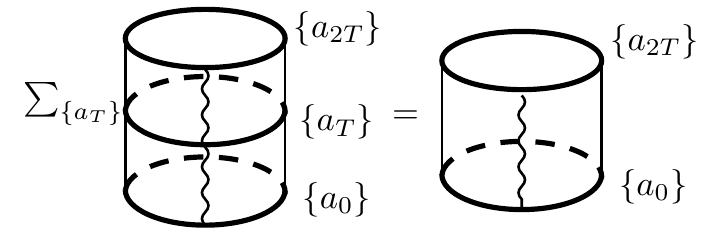}
\end{eqnarray*}

We need to be slightly careful about generalizing the argument that trace of transfer matrix is 1. Recall from the previous section at \eqref{trp}, we used the fact that on a closed manifold $\cM^4=\cM^3\times S^1$, we have
\begin{align*}
\int_{\cM^4}\om_4[\dd a]\se{1}\int_{\cM^4}\om_4[0]=0, 
\end{align*}
which is due to gauge invariance of the topological action \eqref{gaugeinv2}. In the present case we have 
\begin{align*}
\int_{\cM^4}\om_4[\hat{B}+\dd a]\se{1}\int_{\cM^4}\om_4[\hat{B}]
\end{align*}
To complete the argument, note that a ``static" background gauge field on
$\cM^3\times S^1$ may be extended into a higher dimensional manifold
$\cM^3\times D^2$, where $\p D^2=S^1$ with the same construction as before.
\footnote{The intuition is that the obstruction for such extension is due a
``symmetry twist'' in the time direction. For example, when the line dual to
$\hat{B}$ sweeps through a non-contractible surface in 3d, which is equivalent
to insertion of a 1-symmetry operator acting on the non-contractible surface.}
Thus
\begin{align}
&\int_{\cM^4}\om_4[\hat{B}]=\int_{\p(\cM^3\times D^2)}\om_4[\hat{B}]\nn
&=\int_{\cM^3\times D^2}\dd \om_4[\hat{B}]\se{1}0.\label{Bstatic}
\end{align}
using Stoke's theorem and the cocycle condition.

Therefore we have
\begin{eqnarray*}
\includegraphics{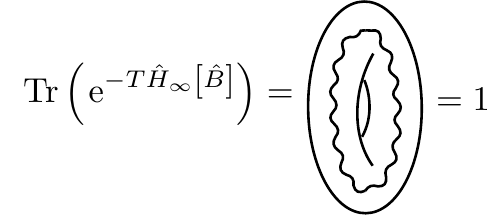}
\end{eqnarray*}
and the ground state is unique.

All the arguments in the previous section will follow through for the present case. We can construct commuting projections $P_{ij}[\hat{B}]$ which differs from the zero-gauge projections only when $ij$ is near the non-zero $\hat{B}$. Its corresponding diagram is
\begin{eqnarray*}
\includegraphics{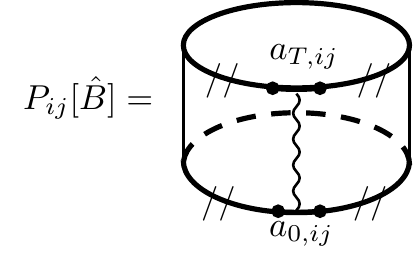}
\end{eqnarray*}

\section{Ground state wavefunction}\label{apdx:gs}
\subsection{Zero background gauge field case}
Suppose $\om_4 = d \phi_3$ for some 3-cochain $\phi_3$(which may not have 1-symmetry, so this does not mean $\om_4$ is a coboundary with 1-symmetry), then $\phi_3$ can be interpreted as the phase of a ground state wavefunction. Define $\ket{\psi_0} = \frac{1}{N_{\psi}}\sum_{\{a\}}\ee^{2\pi i \int_{\cM^3} \phi_3[a]} \ket{\{a\}}$ with normalization $N_{\psi}=\sqrt{n^{N_{l,\cM^3}}}$. Suppose the spatial manifold $\cM^3 = \p \cM^4_0$ is the boundary of some manifold $\cM^4_0$(such $\cM^4_0$ exists for any closed, oriented 3-manifold\cite{Kirby1989}). Then the amplitude is
\begin{align*}
&\int_{\cM^3} \phi_3[a]=\int_{\p \cM^4_0} \phi_3[a]=\int_{\cM^4_0} \dd \phi_3[a]=\int_{\cM^4_0} \om_4[a].
\end{align*}
So $\ket{\psi_0}$ may be represented diagrammatically as
\begin{eqnarray*}
\includegraphics{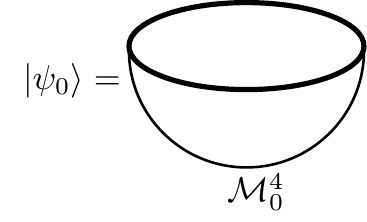}
\end{eqnarray*}

We check that $\ket{\psi_0}$ survives the $P_{ij}$ projection:
\begin{align}
&\bra{\{a_T\}}P_{ij} \ket{\psi_0}\nn
&=\frac{1}{N_{\psi}n}\sum_{\{a_0\}} \prod_{i'j'\neq ij} \del_{a_{0,i'j'},a_{T,i'j'}}\ee^{2\pi \ii [\int_{\cM^3 \times I} \omega_4+\int_{\cM^3}\phi_3[a_0]]}\nn
&=\frac{1}{N_{\psi}n}\sum_{\{a_0\}} \prod_{i'j'\neq ij} \del_{a_{0,i'j'},a_{T,i'j'}}\ee^{2\pi \ii \int_{\cM^3}\phi_3[a_T]}\nn
&=\frac{1}{N_{\psi}}\ee^{2\pi \ii \int_{\cM^3} \phi_3[a_T]}=\langle\{a_T\}\ket{\psi_0},\label{gsproj}
\end{align}
where in the second step we used Stoke's theorem $\int_{\cM^3 \times I} \omega_4 = \left. \int_{\cM^3} \phi_3 \right\vert^T_0$. The same result can also be derived diagrammatically as follows:
\begin{eqnarray*}
\includegraphics{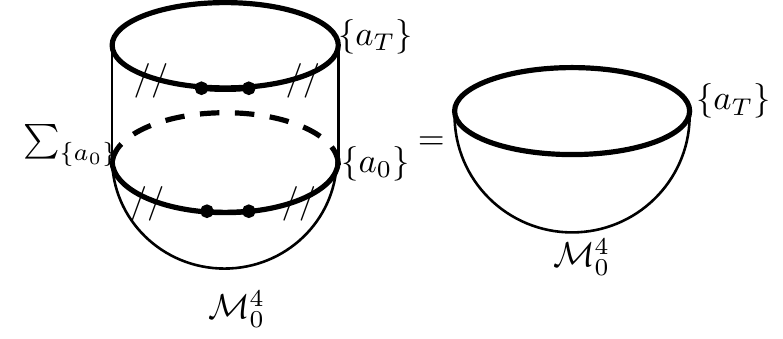}
\end{eqnarray*}

Therefore, the transfer matrix is \begin{align*}
\ee^{-T\hat{H}_{\infty}}=\ket{\psi_0}\bra{\psi_0},
\end{align*}
represented diagrammatically by
\begin{eqnarray*}
\includegraphics{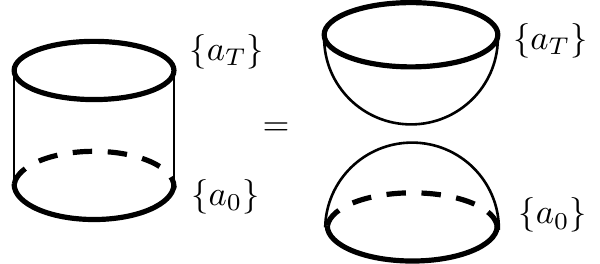}
\end{eqnarray*}
and the local projections $P_{ij}$ can be expressed in terms of $\phi_3$ as
\begin{align}
&\bra{\{a_T\}}P_{ij}\ket{\{a_0\}}\nn
&=\prod_{i'j'\neq ij} \del_{a_{0,i'j'},a_{T,i'j'}}\frac{1}{n} \ee^{2\pi \ii \int_{\cM^3} (\phi_3[a_T]-\phi_3[a_0])}
\end{align}
which is
\begin{eqnarray*}
\includegraphics{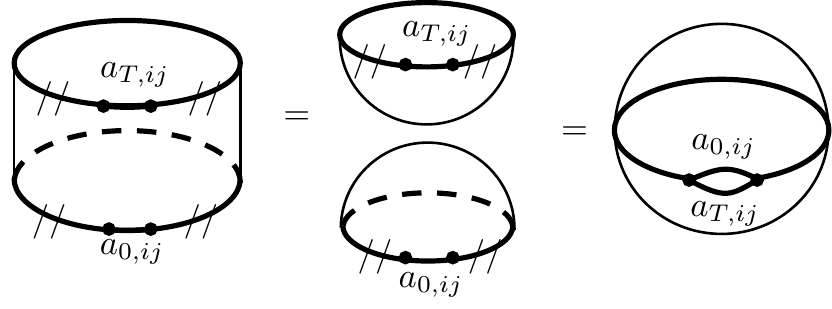}
\end{eqnarray*}

\subsection{Non-zero background gauge field case}\label{apdx:gsdzB}
Suppose $\om_4[\hat{B}+\dd a]=\om_4[\hat{B}]+\dd \phi_3[a,\hat{B}]$. While it is still true that $\cM^3=\p \cM^4_0$ for some manifold $\cM^4_0$, there may be obstructions in $\cM^4_0$ that forbids the extension of the background gauge field into $\cM^4_0$, while respecting the flatness constraint $\dd\hat{B}=0$.

So we will instead take $\cM^4_0 = \cM^3\times I$, where $I=[-1,0]$ is an interval. The boundary now have two components $\p\cM^4_0 = \cM^3\times\{0\} \coprod \cM^3\times\{-1\}$. We take the first component to be the original spatial manifold and extend the field configurations such that on the other end $\cM^3 \times \{-1\}$, we fix $a=0$. The background gauge field is extended to be ``static" as in the previous section.

We define 
\begin{align}
\ket{\psi_0[\hat{B}]} = \frac{1}{N_{\psi}}\sum_{\{a\}}\ee^{2\pi i \int_{\cM^3} \phi_3[a,\hat{B}]-\phi_3[0,\hat{B}]} \ket{\{a\}}.\label{gswfnzB}
\end{align}
Thus we have
\begin{align*}
&\int_{\cM^3} \phi_3[a,\hat{B}]-\phi_3[0,\hat{B}]=\int_{\p \cM^4_0} \phi_3[a,\hat{B}] \nn
&=\int_{\cM^4_0} \dd \phi_3[a,\hat{B}]=\int_{\cM^4_0} \om_4[\hat{B}+\dd a] - \int_{\cM^4_0} \om_4[\hat{B}]\nn
&=\int_{\cM^4_0} \om_4[\hat{B}+\dd a].
\end{align*}
where in the last step the term $\int_{\cM^4_0} \om_4[\hat{B}]\se{1}0$ because its field configuration at $\cM^3\times\{0\}$ and $\cM^3\times\{-1\}$ are the same and the two ends can be glued together to form a closed manifold. The same arguments used in \eqref{Bstatic} can be applied.

In diagram, this means
\begin{eqnarray*}
\includegraphics{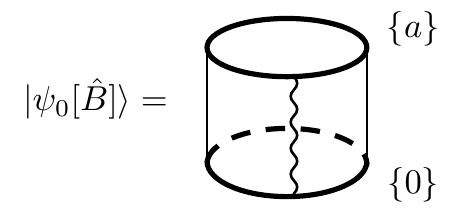}
\end{eqnarray*}
and it is the ground state for the projections $P_{ij}[\hat{B}]$:
\begin{eqnarray*}
\includegraphics{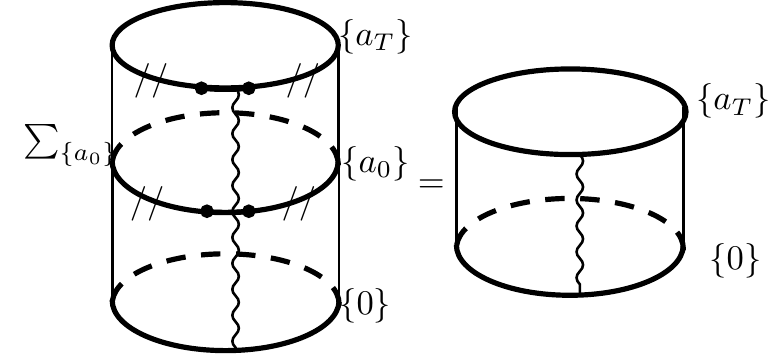}
\end{eqnarray*}
and the matrix elements of $P_{ij}[\hat{B}]$ can be expressed in terms of $\phi_3[a,\hat{B}]$:
\begin{align}
&\bra{\{a_T\}}P_{ij}[\hat{B}]\ket{\{a_0\}}\nn
&=\prod_{i'j'\neq ij} \del_{a_{0,i'j'},a_{T,i'j'}}\frac{1}{n} \ee^{2\pi \ii \int_{\cM^3} (\phi_3[a_T,\hat{B}]-\phi_3[a_0,\hat{B}])}\label{PelemnzB}
\end{align}

\section{Triangulation of hypercubic lattice}\label{apdx:lat}
$\R^d$ may be triangulated by first admitting a hypercubic lattice, and triangulating each hypercube $ I^d =\{(x_1,\dots,x_d):1\geq x_i\geq0~\forall i \}$ into $d!$ simplices $\Del_p$ labeled by $p$ in the permutation group $S_d$:
\begin{align*}
\Del_p=\{1\geq x_{p(1)}\geq \dots\geq x_{p(d)}\geq 0\}.
\end{align*}
The vertices and branching structure for each $\Delta_p$ are given by
\begin{align*}
&\vec{0}=(0,\dots,0) \\
&\ra \widehat{p(1)}\\
&\ra \widehat{p(1)}+\widehat{p(2)}\\
&\dots\\
&\ra \widehat{p(1)} + \dots + \widehat{p(d)}=(1,\dots,1)=\vec{1},
\end{align*}
where $\widehat{i}$ is the unit vector in the $x_i$ direction. The orientation of $\Delta_p$ is given by $\sigma(p)=\epsilon^{p(1)\dots p(d)}$. 

\section{Evaluation of $\int_{I^4} (b^{\Z_n})^2$ in a hypercube}\label{apdx:cube}
Let $b^{\Z_n}$ be a 2-cocycle. Under the triangulation in Appendix \ref{apdx:lat} for $d=4$, we have
\begin{align}
&\int_{I^4} (b^{\Z_n})^2 \nn
&=  \< (b^{\Z_n})^2, \sum_{p} \si(p)\Delta(p) \>= \< (b^{\Z_n})^2, \eps^{\mu\nu\rho\si} \Delta_{\{\mu\nu\rho\si\}} \> \nn
&=\eps^{\mu\nu\rho\si} \< b^{\Z_n}, (\vec{0},\widehat{\mu},\widehat{\mu}+\widehat{\nu}) \>\< b^{\Z_n}, (\widehat{\mu}+\widehat{\nu},\widehat{\mu}+\widehat{\nu}+\widehat{\rho},\vec{1}) \>\nn
&=\frac{1}{4}\eps^{\mu\nu\rho\si}\< b^{\Z_n}, (\vec{0},\widehat{\mu},\widehat{\mu}+\widehat{\nu})-(\nu\lra\mu) \>\nn
&\times \< b^{\Z_n}, (\widehat{\mu}+\widehat{\nu},\widehat{\mu}+\widehat{\nu}+\widehat{\rho},\vec{1}) - (\rho\lra\si) \>\nn
&=\frac{1}{4}\eps^{\mu\nu\rho\si}F_{\mu\nu}(\frac{\hat{\mu}}{2}+\frac{\hat{\nu}}{2})F_{\rho\si}(\hat{\mu}+\hat{\nu}+\frac{\hat{\rho}}{2}+\frac{\hat{\si}}{2}),\label{eq:cubeaction}
\end{align}
where 
\begin{align*}
F_{\mu\nu}(\vec{r})
:=& \<b^{\Z_n},(\vec{0},\hat{\mu},\hat{\mu}+\hat{\nu})_{\vec{r}-\frac{\hat{\mu}}{2}-\frac{\hat{\nu}}{2}} - (\mu\lra\nu)\>.
\end{align*}

\section{Evaluation of $P_{ij}$ in the $m$=even case}\label{apdx:Peven}

In this section we follow the procedure described in Appendix \ref{apdx:h} and write down the projections $P_{ij}$ in the $m$=even case for the topological action \eqref{eq:Ztopeven} with $\cM^3=\R^3$. The matrix elements are given by \eqref{eq:ZtopMI} and \eqref{om4even}:
\begin{align*}
\bra{\{a^{\Z_n}_T\}}P_{ij}\ket{\{a^{\Z_n}_0\}}&= \prod_{i'j'\neq ij} \del_{a^{\Z_n}_{0,i'j'},a^{\Z_n}_{T,i'j'}} \nn
&\ \ \ \ \times \frac{1}{n} \ee^{2\pi \ii \int_{\R^3 \times I} \frac{m}{2n} \gSq^2(\dd a^{\Z_n})},
\end{align*}
where
\begin{align*}
&\int_{\R^3 \times I} \gSq^2(\dd a^{\Z_n})= \int_{\R^3 \times I} \dd a^{\Z_n}\dd a^{\Z_n}\nn
&=\int_{\R^3 \times I} \dd (a^{\Z_n} \dd a^{\Z_n})=\left. \int_{\R^3} a^{\Z_n} \dd a^{\Z_n} \right\vert^{T}_{0}=\delta \left(\int_{\R^3} a^{\Z_n} \dd a^{\Z_n}\right)\nn
&= \int_{\R^3} \delta a^{\Z_n} \dd a^{\Z_n} +  a^{\Z_n} \dd \delta a^{\Z_n}+ \delta a^{\Z_n} \dd \delta a^{\Z_n} \nn
&\se{\dd}\int_{\R^3} \delta a^{\Z_n} \dd a^{\Z_n} + \dd a^{\Z_n}  \delta a^{\Z_n},
\end{align*}
where we have defined $\delta(x):=\left. x \right \vert^{T}_{0}$. In the last step we used integration by part and the fact that  $\delta a^{\Z_n} \dd \delta a^{\Z_n}=0$ because it is impossible for both factors of the cup product to be non-zero, since $\delta a^{\Z_n}$ is non-zero for only one link $ij$. So
Using the triangulation of Appendix \ref{apdx:lat} for $D=3$ space, we have
\begin{align}
&\int_{\R^3} \delta a^{\Z_n} \dd a^{\Z_n} + \dd a^{\Z_n}  \delta a^{\Z_n}\nn
&=\sum_{\vec{n}\in \Z^3} \eps^{\al \bt \ga} \<\delta a^{\Z_n} \dd a^{\Z_n} + \dd a^{\Z_n}  \delta a^{\Z_n},
(\vec{0},\hat{\al},\hat{\al}+\hat{\bt},\vec{1})_{\vec{n}}\>\nn
&=\sum_{\vec{n}\in \Z^3} \eps^{\al \bt \ga} 
\delta a^{\Z_n}(\vec{0},\hat{\al})_{\vec{n}} 
\dd a^{\Z_n}(\hat{\al},\hat{\al}+\hat{\bt},\vec{1})_{\vec{n}} \nn
&+\dd a^{\Z_n}(\vec{0},\hat{\al},\hat{\al}+\hat{\bt})_{\vec{n}} \delta a^{\Z_n}(\hat{\al}+\hat{\bt},\vec{1})_{\vec{n}}\nn
&=\sum_{\vec{n}\in \Z^3} \eps^{\al \bt \ga} 
\delta a^{\Z_n}(\vec{0},\hat{\al})_{\vec{n}} 
\dd a^{\Z_n}(\hat{\al},\hat{\al}+\hat{\bt},\vec{1})_{\vec{n}} \nn
&+\dd a^{\Z_n}(\vec{0},\hat{\bt},\hat{\bt}+\hat{\ga})_{\vec{n}-\hat{\bt}-\hat{\ga}} \delta a^{\Z_n}(\vec{0},\hat{\al})_{\vec{n}}\nn
&=\sum_{\vec{n}\in \Z^3} \frac{\eps^{\al \bt \ga} }{2}
\delta a^{\Z_n}(\vec{0},\hat{\al})_{\vec{n}}\nn
&\times
 [F_{\bt \ga}(\frac{\hat{\bt}}{2}+\frac{\hat{\ga}}{2})_{\vec{n}+\hat{\al}} + F_{\bt \ga}(\frac{\hat{\bt}}{2}+\frac{\hat{\ga}}{2})_{\vec{n}-\hat{\bt}-\hat{\ga}} ]\nn
&=\sum_{\vec{n}\in \Z^3} \frac{\eps^{\al \bt \ga} }{2}
\delta a^{\Z_n}(\vec{0},\hat{\al})_{\vec{n}} [F_{\bt \ga}(\vec{r}_{ij}+\vec{\frac 12}) + F_{\bt \ga}(\vec{r}_{ij}-\vec{\frac 12}) ], \label{eq:F}
\end{align}
where $(\vec{a},\vec{b},\dots,\vec{c})_{\vec{n}}$ is a shorthand for $(\vec{a}+\vec{n},\vec{b}+\vec{n},\dots,\vec{c}+\vec{n})$, and we have dropped the $\<\>$ brackets for pairing cochains and chains. $\vec{r}_{ij} = \vec{n}+\frac{\hat{\al}}{2}$ denote the mid-point of $ij=(\vec{0},\hat{\al})_{\vec{n}}$, $\vec{\frac 12}:=(\frac 12, \frac 12,\frac 12)$, and
\begin{align*}
F_{\bt\ga}(\vec{r})&:= \dd a^{\Z_n}(\vec{0},\hat{\bt},\hat{\bt}+\hat{\ga})_{\vec{r}-\frac{\hat{\bt}}{2}-\frac{\hat{\ga}}{2}} - (\bt\lra\ga). 
\end{align*}


Note that the final expression \ref{eq:F} only depends on links which are 1-diagonal. So the 2-diagonal and 3-diagonal links simply form decoupled product states. We may henceforth neglect all these links for the current analysis.

We may now write down the expression for $P_{ij}$
\begin{align}
P_{ij} =  \frac{1}{n} \sum_{k=0}^{n} \widehat{X}_{ij}^{k} \ee^{2\pi \ii \frac{m k}{2 n} \frac{\eps^{\al \bt \ga}}2 [F_{\bt \ga}(\vec{r}_{ij}+\vec{\frac 12})+F_{\bt \ga}(\vec{r}_{ij}-\vec{\frac 12})]}, \label{eq:Peven}
\end{align}
summed only over 1-diagonal links $ij=(\vec{n},\vec{n}+\hat{\al})$, $\widehat{X}_{ij}$ increments $a^{\Z_n}_{ij}$ by 1. It can be checked that $[P_{ij},P_{i'j'}]=0$ for distinct 1-diagonal links $ij$ and $i'j'$.

\section{Evaluation of $P_{ij}$ for general $m$}\label{apdx:Podd}
As in the $m$=even case, the matrix elements of the projections $P_{ij}$ are given by \eqref{eq:ZtopMI}:
\begin{align*}
\bra{\{a^{\Z_n}_T\}}P_{ij}\ket{\{a^{\Z_n}_0\}}&= \prod_{i'j'\neq ij} \del_{a^{\Z_n}_{0,i'j'},a^{\Z_n}_{T,i'j'}}  \nn
&\ \ \ \ \times\frac{1}{n} \ee^{2\pi \ii \int_{\R^3 \times I} \frac{m}{2n} \gSq^2(\dd a^{\Z_n}) +\dd \xi_3[a^{\Z_n}]},
\end{align*}
where the exponent is
\begin{align*}
&\int_{\R^3 \times I} \frac{m}{2n} \gSq^2\dd a^{\Z_n} +\dd \xi_3[a^{\Z_n}] \nn
&=\int_{\R^3 \times I} \dd\phi_3(a^{\Z_n})=\int_{\R^3} \del \phi_3[a^{\Z_n}],
\end{align*}
where $\phi_3$ is given in \eqref{phi3}. Again $\del a^{\Z_n} \dd \del a^{\Z_n}=0$ since we only change by one link.
\begin{align*}
\del \phi_3[a^{\Z_n}]&\se{\dd}  \frac{m}{2n}\left( \del a^{\Z_n} \dd a^{\Z_n} + \dd a^{\Z_n} \del a^{\Z_n} \right)\nn
&\ \ \ + \frac m2 \left[ (\dd a^{\Z_n}+ \dd \del a^{\Z_n})\hcup{1}\toZ{\frac {\dd a^{\Z_n}+ \dd \del a^{\Z_n}}n } \right] \nn
&\ \ \ -\frac m2 \left[ \dd a^{\Z_n}\hcup{1}\toZ{\frac {\dd a^{\Z_n}}n } \right]\nn
&\se{\dd,1} \frac{m}{2n}\left( \del a^{\Z_n} (\dd a)^{\Z_n} + (\dd a)^{\Z_n} \del a^{\Z_n} \right) \nn
&\ \ \ + \frac m2 \left[ \dd a^{\Z_n}\hcup{1} \del \toZ{\frac {\dd a^{\Z_n}}n } \right] \nn
&\ \ \ +\frac m2 \left[ \del a^{\Z_n}\hcup{1}\dd \toZ{\frac {\dd a^{\Z_n}+ \dd \del a^{\Z_n}}n } \right].
\end{align*}

Where we integrated by part in the last step and used $\del a^{\Z_n} \del\toZ{\frac{\dd  a^{\Z_n}}{n}}=\del\toZ{\frac{\dd  a^{\Z_n}}{n}}\del a^{\Z_n} =0$. Evaluating on the $d=3$ lattice triangulation described in Appendix \ref{apdx:lat}, we have
\begin{align*}
&\del \phi_3(\vec{0},\hat{\al},\hat{\al}+\hat{\bt},\vec{1})\nn
&=\frac{m}{2n}\big[\del a^{\Z_n}(\vec{0},\hat{\al})(\dd a)^{\Z_n}(\hat{\al},\hat{\al}+\hat{\bt},\vec{1}) \nn
&\ \ \ +(\dd a)^{\Z_n}(\vec{0},\hat{\al},\hat{\al}+\hat{\bt})  \del a^{\Z_n}(\hat{\al}+\hat{\bt},\vec{1})\big]\nn
&\ \ \ +\frac{m}{2}\big[\dd a^{\Z_n}(\vec{0},\hat{\al}+\hat{\bt},\vec{1})\del \toZ{\frac{\dd a^{\Z_n}}{n}}(\vec{0},\hat{\al},\hat{\al}+\hat{\bt})\nn
&\ \ \ +\dd a^{\Z_n}(\vec{0},\hat{\al},\vec{1})\del \toZ{\frac{\dd a^{\Z_n}}{n}}(\hat{\al},\hat{\al}+\hat{\bt},\vec{1})\nn
&\ \ \ +\del a^{\Z_n}(\vec{0},\vec{1})\dd \toZ{\frac {\dd a^{\Z_n}+ \dd \del a^{\Z_n}}n }(\vec{0},\hat{\al},\hat{\al}+\hat{\bt},\vec{1})  \big].
\end{align*}
The projections can be written as
\begin{align}
P_{ij} = & \frac{1}{n} \sum_{k=0}^{n}  \widehat{X}_{ij}^{k} \ee^{2\pi \ii \int_{\R^3}\del_k \phi_3[a^{\Z_n}]},\label{eq:Podd}
\end{align}
where $\delta_k a^{\Z_n} = (a_{ij}+k)^{\Z_n}-a_{ij}^{\Z_n}$ is non-zero for only one link $ij$.

There are three cases to consider: $ij$ can be 1-, 2- or 3-diagonal, as defined in subsection \ref{evenm} of the main text.

For the 3-diagonal links $ij=(\vec{n},\vec{n}+\vec{1})$, 
\begin{align*}
&\int_{\R^3}\del_k \phi_3[a^{\Z_n}]=\sum_{\vec{n}\in \Z^3} \eps^{\al \bt \ga}\del_k \phi_3(\vec{0},\hat{\al},\hat{\al}+\hat{\bt},\vec{1})_{\vec{n}}\nn
&=\sum_{\al,\bt,\ga} \eps^{\al \bt \ga}\frac m2 \del_k a^{\Z_n}(\vec{0},\vec{1})_{\vec{n}}\nn
&\ \ \ \times \dd \toZ{\frac {\dd a^{\Z_n}+ \dd \del_k a^{\Z_n}}n }(\vec{0},\hat{\al},\hat{\al}+\hat{\bt},\vec{1})_{\vec{n}}\nn
&\se{1}\frac m2 \del_k a^{\Z_n}(\vec{0},\vec{1})_{\vec{n}}\nn
&\ \ \ \times \sum_{\al,\bt,\ga} \eps^{\al \bt \ga} \toZ{\frac {\dd a^{\Z_n}}n }\big[(\hat{\al},\hat{\al}+\hat{\bt},\vec{1})_{\vec{n}}+(\vec{0},\hat{\al},\hat{\al}+\hat{\bt})_{\vec{n}}\big],
\end{align*}
where $(\vec{a},\vec{b},\vec{c})_{\vec{n}}$ is a shorthand for $(\vec{a}+\vec{n},\vec{b}+\vec{n},\vec{c}+\vec{n})$. We see that the 3-diagonal link $ij$ is coupled to $\toZ{\frac {\dd a^{\Z_n}}n}$ on twelve triangles making up the six faces of the cube whose diagonal is $ij$.

For the 2-diagonal links $ij=(\vec{n},\vec{n}+\hat{\al}+\hat{\bt})$,
\begin{align*}
&\int_{\R^3}\del_k \phi_3[a^{\Z_n}]=\sum_{\vec{n}\in \Z^3} \eps^{\al \bt \ga}\del_k \phi_3(\vec{0},\hat{\al},\hat{\al}+\hat{\bt},\vec{1})_{\vec{n}}\nn
&=\sum_{\ga} \eps^{\al \bt \ga}\frac m2 \Big\{ \dd a^{\Z_n}(\vec{0},\hat{\al}+\hat{\bt},\vec{1})_{\vec{n}}\nn
&\ \ \ \times \del_k \toZ{\frac{\dd a^{\Z_n}}{n}}\big[(\vec{0},\hat{\al},\hat{\al}+\hat{\bt})_{\vec{n}}-(\vec{0},\hat{\bt},\hat{\al}+\hat{\bt})_{\vec{n}} \big]\nn
&\ \ \ +\dd a^{\Z_n}(\vec{0},\hat{\ga},\vec{1})_{\vec{n}-\hat{\ga}}\nn
&\ \ \ \times \del_k \toZ{\frac{\dd a^{\Z_n}}{n}}\big[(\hat{\ga},\hat{\ga}+\hat{\al},\vec{1})_{\vec{n}-\hat{\ga}}-(\hat{\ga},\hat{\ga}+\hat{\bt},\vec{1})_{\vec{n}-\hat{\ga}}\big]\Big\}\nn
&=\frac m2\del_k \toZ{\frac{\dd a^{\Z_n}}{n}}\big[(\vec{0},\hat{\al},\hat{\al}+\hat{\bt})_{\vec{n}}-(\vec{0},\hat{\bt},\hat{\al}+\hat{\bt})_{\vec{n}}\big]\nn
&\ \ \ \times \sum_{\ga} \eps^{\al \bt \ga}\dd a^{\Z_n} \big[(\vec{0},\hat{\al}+\hat{\bt},\vec{1})_{\vec{n}} - (-\hat{\ga},\vec{0},\hat{\al}+\hat{\bt})_{\vec{n}}\big].
\end{align*}
For instance, if $\al,\bt$ = $x_1,x_2$, the link $ij$ is involved as $\del_k \toZ{\frac{\dd a^{\Z_n}}{n}}$ in two triangles making up the square in $x_1$--$x_2$ plane enclosing $ij$. Each of the triangles is coupled to $\dd a^{\Z_n}$ on two other faces in the $(x_1+x_2)$--$x_3$ plane. All four triangles intersect at $ij$.

For the 1-diagonal links $ij=(\vec{n},\vec{n}+\hat{\al})$,
\begin{align*}
&\int_{\R^3}\del_k \phi_3[a^{\Z_n}]=\sum_{\vec{n}\in \Z^3} \eps^{\al \bt \ga}\del_k \phi_3(\vec{0},\hat{\al},\hat{\al}+\hat{\bt},\vec{1})_{\vec{n}}\nn
&=\sum_{\bt,\ga} \eps^{\al \bt \ga}
\frac{m}{2n} \Big(\del a^{\Z_n}(\vec{0},\hat{\al})_{\vec{n}} (\dd a)^{\Z_n}(\hat{\al},\hat{\al}+\hat{\bt},\vec{1})_{\vec{n}}\nn
&+(\dd a)^{\Z_n} (\vec{0}, \hat{\bt}, \hat{\bt}+\hat{\ga})_{\vec{n}-\hat{\bt}-\hat{\ga}} \del a^{\Z_n} (\hat{\bt} + \hat{\ga}, \vec{1})_{\vec{n}-\hat{\bt}-\hat{\ga}}
\Big)\nn
&+\frac{m}{2}\Big(\dd a^{\Z_n} (\vec{0},\hat{\al}+\hat{\bt},\vec{1})_{\vec{n}} \del \toZ{\frac{\dd a^{\Z_n}}{n}}(\vec{0},\hat{\al},\hat{\al}+\hat{\bt})_{\vec{n}}\nn
&+\dd a^{\Z_n} (\vec{0},\hat{\ga}+\hat{\al},\vec{1})_{\vec{n}-\hat{\ga}} \del \toZ{\frac{\dd a^{\Z_n}}{n}}(\vec{0},\hat{\ga},\hat{\ga}+\hat{\al})_{\vec{n}-\hat{\ga}}\nn
&+\dd a^{\Z_n} (\vec{0},\hat{\ga},\vec{1})_{\vec{n}-\hat{\ga}} \del \toZ{\frac{\dd a^{\Z_n}}{n}}(\hat{\ga},\hat{\ga}+\hat{\al},\vec{1})_{\vec{n}-\hat{\ga}}\nn
&+\dd a^{\Z_n} (\vec{0},\hat{\bt},\vec{1})_{\vec{n}-\hat{\bt}-\hat{\ga}} \del \toZ{\frac{\dd a^{\Z_n}}{n}}(\hat{\bt},\hat{\bt}+\hat{\ga},\vec{1})_{\vec{n}-\hat{\bt}-\hat{\ga}}
\Big)\nn
&=\sum_{\bt,\ga} \eps^{\al \bt \ga}
\Big\{
\frac{m}{2n}\del a^{\Z_n}(\vec{0},\hat{\al})_{\vec{n}}\nn
&\ \ \ \times(\dd a)^{\Z_n}\big[(\hat{\al},\hat{\al}+\hat{\bt},\vec{1})_{\vec{n}}+(-\hat{\bt}-\hat{\ga},-\hat{\ga},\vec{0})_{\vec{n}}\big]\nn
&+\frac{m}{2}\Big(
\del \toZ{\frac{\dd a^{\Z_n}}{n}}(\vec{0},\hat{\al},\hat{\al}+\hat{\bt})_{\vec{n}}\nn
&\ \ \ \times \dd a^{\Z_n} \big[(\vec{0},\hat{\al}+\hat{\bt},\vec{1})_{\vec{n}}+ (-\hat{\ga},\vec{0},\hat{\al}+\hat{\bt})_{\vec{n}}\big]\nn
&+\del \toZ{\frac{\dd a^{\Z_n}}{n}}(-\hat{\ga},\vec{0},\hat{\al})_{\vec{n}} \nn
&\ \ \ \times \dd a^{\Z_n} \big[(-\hat{\ga},\hat{\al},\hat{\al}+\hat{\bt})_{\vec{n}}+ (-\hat{\bt}-\hat{\ga},-\hat{\ga},\hat{\al})_{\vec{n}}\big]
\Big)
\Big\}.
\end{align*}

In the case $n=2$, $\toZ{\frac{\dd a^{\Z_2}}{2}}\se{2}\gSq^1(a^{\Z_2})$.

\section{Calculation details for $\th_q$, $\th_{q_1q_2}$}\label{apdx:stat}
It turns out we only need to keep track of the two triangles and five links in the central square, shown in \Fig{fig:scatter}. This is slightly non-trivial, essentially due to $\phi_2[a,h]=0$ when $\dd h=0$. 
In this section we assume $a_i=a^{\Z_n}_i$ and $q=q^{\Z_n}$. Applying \eqref{bdstate}, we have
\begin{align*}
&W_i^q\ket{\{a^{\Z_n}\}} = \ee^{2\pi \ii \big[\phi_2(\{a_1,a_4,a_0\},h(W_i^q))-\phi_2(\{a_2,a_3,a_0\},h(W_i^q))\big]}\nn
&\times \ket{\{[a+\dd h(W_i^q)]^{\Z_n})\}},
\end{align*}
with $h(W_i^q)$ depicted in the bottom of \Fig{fig:scatter}.

Evaluating $\phi_2$ using \eqref{bdphaseodd}, we have
\begin{align*}
&\phi_2(\{a_1,a_4,a_0\},h(W_1^q))\se{1}\frac{m}{2}(a_4-a_0)\toZ{\frac{a_4+q}{n}}\nn
&\phi_2(\{a_2,a_3,a_0\},h(W_1^q))\nn
&\se{1}\frac{m}{2n}qa_3+\frac{m}{2}(a_2+a_3-a_0)\toZ{\frac{a_2+q}{n}}\nn
&\phi_2(\{a_1,a_4,a_0\},h(W_2^q))\nn
&\se{1}\frac{m}{2n}(-q)^{\Z_n}a_4+\frac{m}{2}(a_1+a_4-a_0)\toZ{\frac{a_1+(-q)^{\Z_n}}{n}}\nn
&\phi_2(\{a_2,a_3,a_0\},h(W_2^q))\se{1}\frac{m}{2}(a_3-a_0)\toZ{\frac{a_3+(-q)^{\Z_n}}{n}}\nn
&\phi_2(\{a_1,a_4,a_0\},h(W_3^q))=0\nn
&\phi_2(\{a_2,a_3,a_0\},h(W_3^q))\se{1}\frac{m}{2n}qa_3\nn
&\ \ \ +\frac{m}{2}\big(a_0(\frac{q+(-q)^{\Z_n}}{n})
+(a_2+q)\toZ{\frac{a_3+(-q)^{\Z_n}}{n}}\nn
&\ \ \ +(a_2+a_3-a_0)(\toZ{\frac{a_2+q}{n}}+\toZ{\frac{a_3+(-q)^{\Z_n}}{n}})\big)\nn
&\phi_2(\{a_1,a_4,a_0\},h(W_4^q))\se{1}\frac{m}{2n}(-q)^{\Z_n}a_4\nn
&\ \ \ +\frac{m}{2}\big(a_0(\frac{q+(-q)^{\Z_n}}{n})
+(a_1+(-q)^{\Z_n})\toZ{\frac{a_4+q}{n}}\nn
&\ \ \ +(a_1+a_4-a_0)(\toZ{\frac{a_1+(-q)^{\Z_n}}{n}}+\toZ{\frac{a_4+q}{n}})+q\toZ{\frac{-q}{n}}\big)\nn
&\phi_2(\{a_2,a_3,a_0\},h(W_4^q))=0.
\end{align*}
So for self-statistics \eqref{thq}, after some algebra, we are left with
\begin{align*}
\th_{q}&\se{1}
\phi_2(\{a_1,a_4,a_0\},h(W_2^{q}))-\phi_2(\{a_2,a_3,a_0\},h(W_2^{q}))\nn
&\ \ \ +\phi_2(\{(a_1-q)^{\Z_n},a_4,(a_0-q)^{\Z_n}\},h(W_1^{q}))\nn
&\ \ \ -\phi_2(\{a_2,(a_3-q)^{\Z_n},(a_0-q)^{\Z_n}\},h(W_1^{q}))\nn
&\ \ \ -\phi_2(\{a_1,a_4,a_0\},h(W_3^q))+\phi_2(\{a_2,a_3,a_0\},h(W_3^q))\nn
&\ \ \ -\phi_2(\{a_1,a_4,a_0\},h(W_4^q))\nn
&\ \ \ +\phi_2(\{(a_2+q)^{\Z_n},(a_3-q)^{\Z_n},a_0\},h(W_4^q))\nn
&\se{1}q^2\frac{m}{2n}.
\end{align*}
Whereas for mutual-statistics \eqref{thq1q2}, we have
\begin{align*}
\th_{q_1q_2}&\se{1}
\phi_2(\{a_1,a_4,a_0\},h(W_2^{q_2}))-\phi_2(\{a_2,a_3,a_0\},h(W_2^{q_2}))\nn
&\ \ \ +\phi_2(\{(a_1-q_2)^{\Z_n},a_4,(a_0-q_2)^{\Z_n}\},h(W_1^{q_1}))\nn
&\ \ \ -\phi_2(\{a_2,(a_3-q_2)^{\Z_n},(a_0-q_2)^{\Z_n}\},h(W_1^{q_1}))\nn
&\ \ \ -\phi_2(\{a_1,a_4,a_0\},h(W_1^{q_1}))+\phi_2(\{a_2,a_3,a_0\},h(W_1^{q_1}))\nn
&\ \ \ -\phi_2(\{a_1,(a_4+q_1)^{\Z_n},(a_0+q_1)^{\Z_n}\},h(W_2^{q_2}))\nn
&\ \ \ +\phi_2(\{(a_2+q_1)^{\Z_n},a_3,a_0\},h(W_2^{q_2}))\nn
&\se{1}q_1q_2\frac{m}{n}.
\end{align*}

\section{Evaluation of $W_{\bigodot i}$ for $(n,m)=(2,1)$}\label{wi}
In this section we derive \eqref{eq:wi}. We also assume $a=a^{\Z_n}$ for all initial link values in this section.  Restricting to $(n,m)=(2,1)$ and enforcing ``no flux" rule $\dd a \se{2}0$, \eqref{bdphaseodd} is
\begin{align}
&\phi_2[a,h^{\Z_2}]\se{1}\frac{1}{4} \al^{\Z_2} a+\frac{1}{2} \big(a \hcup{1} \frac{\dd \al^{\Z_2}}{2}   \nn
&\ \ \ +(a+\al^{\Z_2})\toZ{\frac{a+\al^{\Z_2}}{2}}+h^{\Z_2}\dd \toZ{\frac{\dd h^{\Z_2}}{2}}\big).\label{phi2nf}
\end{align}
Applying \eqref{bdstate}, we have
\begin{align*}
&\left.W_{\bigodot i}\ket{\{a_{ij},a_{jj'}\}}\right\vert_{\dd a^{\Z_2}\se{2}0}\nn
&=\ee^{2\pi \ii \Phi[a]}\ket{\{(a_{ij}+1)^{\Z_2},a_{jj'}\}},
\end{align*}
where
\begin{align*}
\Phi[a]&=\<\phi_2,(1,2,i)\>-\<\phi_2,(2,i,3)\>+\<\phi_2,(i,3,4)\>\nn
&\ \ \ -\<\phi_2,(i,5,4)\>+\<\phi_2,(6,i,5)\>-\<\phi_2,(1,6,i)\>.
\end{align*}
Applying \eqref{phi2nf} for each 2-simplex in \Fig{fig:wi}, we get
\begin{align*}
\<\phi_2,(1,2,i)\>&=\frac{1}{2}\big(a_{12}\toZ{\frac{a_{2i}+1}{2}}\big)\nn
\<\phi_2,(2,i,3)\>&=\frac{1}{4}a_{i3}+\frac{1}{2}\big(a_{23}+(a_{2i}+1)\toZ{\frac{a_{i3}+1}{2}}\big)\nn
\<\phi_2,(i,3,4)\>&=\frac{1}{4}a_{34}\nn
\<\phi_2,(i,5,4)\>&=\frac{1}{4}a_{54}\nn
\<\phi_2,(6,i,5)\>&=\frac{1}{4}a_{i5}+\frac{1}{2}\big(a_{65}+(a_{6i}+1)\toZ{\frac{a_{i5}+1}{2}}\big)\nn
\<\phi_2,(1,6,i)\>&=\frac{1}{2}\big(a_{16}\toZ{\frac{a_{6i}+1}{2}}\big).
\end{align*}
Note for $a=a^{\Z_2}$ and $a'=a'^{\Z_2}$, we have $\toZ{\frac{a+a'}{2}}=aa'$. Also for any simplex $(i,j,k)$, the ``no flux" constraint means
\begin{align*}
a_{jk}&=(a_{ij}+a_{ik})^{\Z_2}=a_{ij}+a_{ik}-2\toZ{\frac{a_{ij}+a_{ik}}{2}}\nn
&=a_{ij}+a_{ik}-2a_{ij}a_{ik}.
\end{align*}
After a bit of algebra, simplifying using the above identities, we finally arrive at
\begin{align*}
\Phi[a]\se{1}\frac{1}{2}\sum_{\<jj'\>}a_{ij}a_{ij'}.
\end{align*}

\subsection{DS projection Hamiltonian}
For completeness, we supplement this section by briefly explaining the projection Hamiltonian for DS topological order from the action (up to a volume term)
\begin{align*}
Z_{DS} = \sum_{da\se{2}0}\ee^{2\pi \ii \int_{M^3} \frac{1}{2} aaa}.
\end{align*}
The construction was well-studied in the literature, see \eg. \Ref{Mesaros2013}. It is similar to that described in Appendix \ref{apdx:h}, except that six links connecting to the same site is updated. We have
\begin{align*}
\hat{H} = - \sum_{i} P_{i}\prod_{\Delta_i}\delta_{\<da,\Delta\>,0} -\sum_{\Delta}\delta_{\<da,\Delta\>,0},
\end{align*}
where $\Delta$ is summed over all 2-simplices, $\Delta_i$ are product over all 2-simplices having $i$ as a vertex.
\begin{align*}
P_i\ket{\{a_{ij},a_{jj'}\}}=\ee^{2\pi \ii \Phi_{DS}[a]}\ket{\{(a_{ij}+1)^{\Z_2},a_{jj'}\}},
\end{align*}
and $\Phi_{DS}[a]$ is evaluating the cocycles on the six tetrahedrons involved when a site is updated. Using \Fig{fig:wi} and updating $i$ to $i'$ with $i'$ out of paper, where $a_{i'j}=(a_{ij}+1)^{\Z_2}$ and $a_{ii'}=1$, the result is
\begin{align*}
\Phi_{DS}[a]&\se{1}\frac{1}{2}\big[a_{12}a_{2i}+a_{2i}(a_{i3}+1)+(a_{i3}+1)a_{34}\nn
&\ \ \ +(a_{i5}+1)a_{54}+a_{6i}(a_{i5}+1)+a_{16}a_{6i}\big]\nn
&\se{1}\frac{1}{2}\sum_{\<jj'\>}a_{ij}a_{ij'}.
\end{align*}
We see it describes the same phase as $H_{\p}$ in \eqref{hp}.

\section{$\omega_4$, $\phi_3$ and $\phi_2$}\label{apdx:gen}
In the main text, we find that for $\Z_n$-1-SPT, the 4-cocycle $\omega_4$, the ground state wavefunction amplitude $\phi_3$, and the boundary transform anomalous phase $\phi_2$ are related  via \eqref{om4da} and \eqref{delphi3}:
\begin{align*}
\om_4[\dd a^{\Z_n}] &=\dd \phi_3 [a]\nn
-\del_{\al} \phi_3[a^{\Z_n}] &=\dd \phi_2[a,h].
\end{align*}

In general, given $\om_4$ satisfying $\dd \om_4 =0$. We can define the 3-cochain $\phi_3^{\str{0}}$ as follows:
\begin{align*}
\<\phi_3^{\str{0}},(1234)\>:=\<\om_4,(\str{0}1234)\>,
\end{align*}
where we have introduced an extra ``reference" vertex $\str{0}$. A heuristic way to interpret $\str{0}$ is that it is located at $t=-\infty$ whereas the other vertices $i=1,2,3,4$ are located at a spatial slice at $t=0$. So $a_{ii'}$ are ``spatial" links and $a_{\str{0}i}$ are ``temporal" links. We may choose the links $a_{\str{0}i}=0$, $i=1,2,3,4$ as a convention. The dependence of $\phi_3^{\str{0}}$ on $\str{0}$ is the choice of such convention. For arbitrary 4-chain $(01234)$, we have
\begin{align*}
&\<\dd \phi_3^{\str{0}},(01234)\>\nn
&=\sum_{m=0}^4 (-)^m \<\phi_3^{\str{0}},(0\dots\hat{m}\dots4)\>\nn
&=\sum_{m=0}^4 (-)^m \<\om_4,(\str{0}0\dots\hat{m}\dots4)\>\nn
&=\<\om_4,(01234)\> -\<\dd \om_4,(\str{0}01234)\>\nn
&=\<\om_4,(01234)\>,
\end{align*}
so $\om_4 = \dd \phi_3^{\str{0}}$.

To generalize \eqref{delphi3}, note that if we have a 1-symmetry $\al=\dd h$ only on the spatial links, then we can use the invariance of $\om_4$ under space-time 1-symmetry to undo $h$ from the spatial links and act $(-h)$ on the temporal links instead, \ie
\begin{align*}
\<\phi_3^{\str{0}}[a+\al],(1234)\> &= \<\phi_3^{\str{0}}[a+\dd h],(1234)\>\nn
&=\<\om_4[a+(\dd h)_{spatial}],(\str{0}1234)\>\nn
&=\<\om_4[a],(\str{1}1234)\>\nn
&=\<\phi_3^{\str{1}}[a],(1234)\>.
\end{align*}
So $\del_{\al}\phi_3^{\str{0}}=\phi_3^{\str{1}}-\phi_3^{\str{0}}$. Here $(\dd h)_{spatial}$ means it only exists on spatial links $a_{ii'}$, and we have introduced a new vertex $\str{1}$ where
\begin{align*}
a_{\str{1}i}:=a_{\str{0}i}-h_i=-h_i.
\end{align*}
 If we define
\begin{align*}
\<\phi_2^{\str{0}\str{1}},(234)\>:=\<\om_4,(\str{0}\str{1}234)\>,
\end{align*}
it can then be checked that for arbitrary 3-chain $(1234)$, we have
\begin{align*}
&\<\dd \phi_2^{\str{0}\str{1}},(1234)\>\nn
&=\sum_{m=1}^4 -(-)^m \<\phi_2^{\str{0}\str{1}},(1\dots\hat{m}\dots4)\>\nn
&=\sum_{m=1}^4 -(-)^m \<\om_4,(\str{0}\str{1}1\dots\hat{m}\dots4)\>\nn
&=\sum_{\str{m}=\str{0}}^{\str{1}} -(-)^{\str{m}} \<\om_4,(\str{0}\dots\hat{\str{m}}\dots\str{1}1234)\>  +\<\dd \om_4,(\str{0}\str{1}1234)\>\nn
&=-\<\om_4,(\str{1}1234)\>+\<\om_4,(\str{0}1234)\>\nn
&=-\<\phi_3^{\str{1}},(1234)\>+\<\phi_3^{\str{0}},(1234)\>.
\end{align*}
So $\del_{\al}\phi_3^{\str{0}}=-\dd \phi_2^{\str{0}\str{1}}$.

In general we may define 
\begin{align*}
\<\phi^{\str{0}\dots\str{(4-k-1)}}_k,(01234)\>:=\<\om,(\str{0}\dots\str{(4-k-1)}(4-k)\dots4)\>
\end{align*}
for $k=3,2,1,0,-1$. They represent the anomaly in the boundary transformation in $k$-dimensional sub-manifolds in the boundary. $k=-1$ means dimension 0 in the bulk. They satisfy
\begin{align*}
\str{\dd} \phi_k = (-)^{k}\dd\phi_{k-1},
\end{align*}
where 
\begin{align*}
(\str{\dd} \phi_k)^{\str{0}\dots\str{(4-k)}}:=\sum_{\str{m}=0}^{4-k}(-)^{\str{m}}\phi_k^{\str{0}\dots\hat{\str{m}}\dots\str{(4-k)}}.
\end{align*}

\section{Generalization of 
\eqref{selfstat} and \eqref{mutstat} 
to $G$-protected 1-SPT for finite unitary groups}\label{apdx:gen2}

In general, we can carry through the calculations for self-statistics and mutual-statistics for transformation strings, for a $G$-protected 1-SPT in 3+1D as well, where $G$ is any unitary group. Note $G$ is Abelian since it is a 1-symmetry. In this section we will only present the final results.

Following similar strategies for deriving self- and mutual-statistics in the $\Z_{n}$ case, it can be shown that for general unitary group $G$, the self- and mutual- statistics of transformation strings are given by
\begin{align}
\th_{q}&=-\om_4(-q,-q,0,-q,0,q)+ \om_4(-q,-q,-q,-q,0,0)\nn
&\ \ \ -\om_4(0,-q,0,-q,-q,0) + \om_4(0,0,0,0,-q,0)\nn
&\ \ \ + \om_4(0,0,q,0,0,0)-\om_4(0,0,-q,0,-q,-q)\label{selfstat2} \\
\th_{q_1 q_2}&=\Big\{\big[\om_4(-q_1,0,-q_1,q_1,0,-q_1-q_2)\nn
&\ \ \ +\om_4(0,0,-q_1,-q_2,-q_1-q_2,-q_1)\nn
&\ \ \ -\om_4(q_1,0,0,-q_1,-q_1-q_2,-q_2)\nn
&\ \ \ -(q_1\ra 0)\big] -(q_2\ra 0)\Big\} +(q_1\lra q_2),\label{mutstat2}
\end{align}
where $q,q_1,q_2\in G$ labels the group element associated with the transformation string, $\om_4[\cB]=\om_4(\cB_{012},\cB_{013},\cB_{014},\cB_{023},\cB_{024},\cB_{034})$ where $\dd\cB=0$. It can be checked \eqref{selfstat2} and \eqref{mutstat2} are topological invariants, namely, they are unchanged under $\om_4\ra \om_4 + \dd \bt_3$ for any 1-symmetric 3-cochain $\bt_3$.

We will check that \eqref{selfstat2} and \eqref{mutstat2} recovers \eqref{selfstat} and \eqref{mutstat} in the case $G=\Z_{n}$.
The $\Z_{n}$ 4-cocycle \eqref{om4} is
\begin{align}
&\omega_4[\cB]=\frac{m}{2n} \gSq^2\cB^{\Z_n}\nn
&=\frac{m}{2n}\big(\cB^{\Z_n}_{012}\cB^{\Z_n}_{234} +\cB^{\Z_n}_{034}(\dd \cB^{\Z_n})_{0123}+\cB^{\Z_n}_{014}(\dd \cB^{\Z_n})_{1234}\big),\label{om42}
\end{align}
where
\begin{align*}
(\dd \cB^{\Z_n})_{0123}&=\cB^{\Z_n}_{123}-\cB^{\Z_n}_{023}+\cB^{\Z_n}_{013}-\cB^{\Z_n}_{012}\nn
(\dd\cB^{\Z_n})_{1234}&=\cB^{\Z_n}_{234}-\cB^{\Z_n}_{134}+\cB^{\Z_n}_{124}-\cB^{\Z_n}_{123}\nn
\cB^{\Z_n}_{ijk}&=(\cB^{\Z_n}_{0jk}-\cB^{\Z_n}_{0ik}+\cB^{\Z_n}_{0ij})^{\Z_n}~\text{for}~i\neq0,
\end{align*}
so \eqref{selfstat2} and \eqref{mutstat2} are
\begin{align*}
\th_{q}&= -\om_4(-q,-q,0,-q,0,q) + \om_4(-q,-q,-q,-q,0,0)\nn
&\ \ \ - \om_4(0,-q,0,-q,-q,0) + \om_4(0,0,0,0,-q,0)\nn
&\ \ \ + \om_4(0,0,q,0,0,0) - \om_4(0,0,-q,0,-q,-q)\nn
&\se{1}-0 + \big(\frac{m}{2n}q^2)-0 +0 +0 -0=\frac{m}{2n}q^2.\\
\th_{q_1 q_2}&=\Big\{\big[\om_4(-q_1,0,-q_1,q_1,0,-q_1-q_2)\nn
&\ \ \ +\om_4(0,0,-q_1,-q_2,-q_1-q_2,-q_1)\nn
&\ \ \ -\om_4(q_1,0,0,-q_1,-q_1-q_2,-q_2)\nn
&\ \ \ -(q_1\ra 0)\big] -(q_2\ra 0)\Big\} +(q_1\lra q_2)\nn
&=\Big\{\big[(\frac{m}{2n}(-q_1)^{\Z_n}(-q_2)^{\Z_n}\nn
&\ \ \ +\frac{m}{2}(-q_1-q_2)^{\Z_n}\frac{-q_1^{\Z_n}-(-q_1)^{\Z_n}}{n})\nn
&\ \ \ -\frac{m}{2}(-q_2)^{\Z_n}\frac{-(-q_1)^{\Z_n}-q_1^{\Z_n}}{n}\nn
&\ \ \ -(q_1\ra 0)\big]-(q_2\ra 0)\Big\}+(q_1\lra q_2)\nn
&\ \ \ \nn
&\se{1}\Big\{\big[\big(\frac{m}{2n}q_1q_2+\frac{m}{2}[q_1\toZ{\frac{-q_2}{n}}+q_2\toZ{\frac{-q_1}{n}}\nn
&\ \ \ +(q_1+q_2)(\toZ{\frac{q_1}{n}}+\toZ{\frac{-q_1}{n}})]\big) \nn
&\ \ \ -\frac{m}{2}q_2(\toZ{\frac{q_1}{n}}+\toZ{\frac{-q_1}{n}})\nn
&\ \ \ -(q_1\ra 0)\big]-(q_2\ra 0)\Big\} +(q_1\lra q_2)\nn
&\se{1}\Big\{\big[\big(\frac{m}{2n}q_1q_2+\frac{m}{2}[q_1\toZ{\frac{-q_2}{n}}+q_2\toZ{\frac{-q_1}{n}}\nn
&\ \ \ +q_1(\toZ{\frac{q_1}{n}}+\toZ{\frac{-q_1}{n}})]\big)\big]-(q_2\ra 0)\Big\}+(q_1\lra q_2)\nn
&\se{1}\big(\frac{m}{2n}q_1q_2+\frac{m}{2}[q_1\toZ{\frac{-q_2}{n}}+q_2\toZ{\frac{-q_1}{n}}]\big)+(q_1\lra q_2)\nn
&\se{1}\frac{m}{n}q_1q_2.
\end{align*}
Thus \eqref{selfstat} and \eqref{mutstat} are recovered.

\bibliography{bib,../../bib/all,../../bib/publst} 

\end{document}